\documentclass[graybox, envcountchap]{svmult}

\usepackage{mathptmx}        % selects Times Roman as basic font
\usepackage{amsmath}
\usepackage{amssymb}
\usepackage{color}
\usepackage{helvet}          % selects Helvetica as sans-serif font
\usepackage{courier}         % selects Courier as typewriter font
\usepackage{dirtree}
\usepackage{makeidx}        % allows index generation
\usepackage{graphicx}        % standard LaTeX graphics tool
\usepackage{subfig}

\usepackage{multicol}        % used for the two-column index
\usepackage[bottom]{footmisc}% places footnotes at page bottom

\usepackage{hyperref}        %for hyperlinks
\hypersetup{colorlinks=true,urlcolor=blue}

\usepackage[misc]{ifsym}

\makeindex             % used for the subject index
\def\Msun{\ifmmode {\rm M}_{\odot} \else M$_{\odot}$\fi}

\def\ovii{O\,{\sc vii}}
\def\oviii{O\,{\sc viii}}
\def\oviiviii{O\,{\sc vii-viii}}

\def\nex{Ne\,{\sc x}}
\def\neixx{Ne\,{\sc ix-x}}

\def\fexvii{Fe\,{\sc xvii}}
\def\fexxv{Fe\,{\sc xxv}}

\def\chandra{{\it Chandra}}
\def\xmm{{\it XMM-Newton}}

\def\nustar{{\it NUSTAR}}
\def\extp{{\it eXTP}}

\begin{document}

%%%%%%%%%%%%%%%%%%%%%%%%%%%%%%%%%%%%%%%%%%%%%%%%%%%%%%%%%%%%%%%%%

\title{Ultra-luminous X-ray sources: \\extreme accretion and feedback}
\titlerunning{Ultra-luminous X-ray sources} 
\author{Ciro Pinto and Dominic J. Walton}  
\institute{C. Pinto (\Letter) \at Istituto Nazionale di Astrofisica (INAF) – IASF Palermo, Via U. La Malfa 153, I-90146 Palermo, Italy, \email{ciro.pinto@inaf.it}
\and D. J. Walton (\Letter) \at Centre for Astrophysics Research, University of Hertfordshire, College Lane, Hatfield AL10 9AB, UK \email{d.walton4@herts.ac.uk}
}
\maketitle

\abstract{Ultra-luminous X-ray sources (ULXs) are the most extreme members of the X-ray binary population, exhibiting X-ray luminosities that can surpass the $10^{39}$ erg\,s$^{-1}$ threshold (by orders of magnitude). They are mainly seen in external galaxies and are most preferentially found in star-forming galaxies with lower metallicities. The vast majority of these systems are now understood to be powered by super-Eddington accretion of matter onto stellar-mass compact objects (black holes and neutron stars). This is driven by the discovery of coherent pulsations, cyclotron lines and powerful winds in members of the ULX population. The latter was possible thanks to high-resolution X-ray spectrometers such as those aboard \textit{XMM-Newton}. ULX winds carry a huge amount of power owing to their relativistic speeds (0.1-0.3\,$c$) and are likely responsible for the $\sim$100\,pc superbubbles observed around many ULXs. The winds also regulate the amount of matter that can reach the central accretor. Their study is, therefore, essential to understanding how quickly compact objects can grow and how strong their feedback onto the surrounding medium can be. This may also be relevant to understand supermassive black hole growth, particularly in the early Universe. Here we provide an overview on ULX phenomenology, highlight some recent exciting results, and show how future missions such as \textit{XRISM} and \textit{ATHENA} will drive further significant progress in this field.}

%%%%%%%%%%%%%%%%%%%%%%%%%%%%%%%%%%%%%%%%%%%%%%%%%%%%%%%%%

\section{Introduction} \label{sec:intro}

Ultra-luminous X-ray sources (ULXs) are non-nuclear astronomical objects mainly found in external galaxies. Under the assumption of isotropic emission, their X-ray luminosities surpass the Eddington limit for a standard 10\,{\Msun} stellar-remnant black hole (BH), or $10^{39}$\,erg\,s$^{-1}$, and in rare cases reaching a few times $10^{41}$\,erg\,s$^{-1}$. This makes them the most extreme amongst X-ray binaries (XRBs) and excellent targets to study super-Eddington accretion or to search for unusually massive black holes. In both cases, it is clear that ULXs provide an important workbench to study the origin of the early supermassive black holes that have been discovered at high redshifts when the Universe was young \cite{Fan2003, Banados2018}.

\textcolor{black}{In this section, we highlight important results obtained with X-ray and multi-wavelength observations of ULXs. In Sect.\,\ref{sec:timing} and \ref{sec:broadband} we report some relevant achievements from broadband timing and spectroscopy studies of ULXs. In Sect.\,\ref{sec:hrxs} we detail the groundbreaking discoveries obtained through high-resolution X-ray spectroscopy of ULXs. We discuss the implications of the discoveries and their comparison with theoretical models in Sect.\,\ref{sec:implications}. Current limitations and future prospects on possible advances in this research field are discussed in Sect.\,\ref{sec:prospects}.}

\subsection{X-ray surveys} \label{sec:surveys}

ULXs were discovered about 40 years ago with the \textit{Einstein} observatory \cite{Long1981,Fabbiano1989}. Dedicated studies using the ROSAT all-sky survey have underlined the strong connection between ULXs and star formation based on the fact that ULXs are preferentially seen in late-type galaxies, especially in star-forming regions such as spiral arms \cite{Liu2005}. Typically, ULX candidates are identified by searching the sky area\footnote{The D25 isophote -- the best elliptical fit to the area over which the B-band surface brightness exceeds 25 mag arcsec$^{-2}$ -- is usually used to represent the extent of the galaxy in these searches.} subtended by known galaxies for X-ray sources (excluding their nuclear regions), and then using the known galaxy distances to pick out sources that have luminosities that exceed $10^{39}$\,erg\,s$^{-1}$.

After the launch of more powerful telescopes with the high effective area and spatial resolution, such as \textit{Chandra} and \textit{XMM-Newton}, it was possible to significantly improve the detection and characterisation of ULXs. ULX catalogues were built bearing details on their X-ray luminosity, spectral hardness (normally calculated as the ratio between the 2-10 keV and the 0.3-2 keV energy bands), and more properties
\cite{Swartz2004,Walton2011}. It was soon clear that a large fraction ($>20\%$) of the ULXs have \textcolor{black}{steep spectral indices (e.g. $\Gamma > 2$)} indicative of soft, and likely thermal spectra.
About $10\%$ of the ULX sample shows significant X-ray flux variability but this value is likely a lower limit due to the sparse sampling and limited amount of sources with multiple observations \cite{Robba2022}.  
From the study of the X-ray luminosity function (XLF) it became clear that ULXs in elliptical galaxies may be considered as the high-luminosity end of the low-mass X-ray binary population (LMXBs) whilst in star-forming galaxies, there is a clear contribution from younger high-mass X-ray binary (HMXB) systems \cite{Mineo2012, Lehmer2019}.
The combination of a larger number of ULXs in star-forming galaxies and star-forming regions suggests that most ULXs represent a young but short-lived population, such as the high-mass X-ray binaries.

Two decades of observations with \textit{Chandra}, \textit{XMM-Newton} and \textit{Swift} have allowed a relatively sizeable population of ULX candidates to be compiled, resulting in the current sample of 1843 ULX candidates in 951 host galaxies (\cite{Walton2022}; see also \cite{Kovlakas2020, Inoue2021, Bernadich2022}). The number of variable ULXs keeps increasing as further observations are obtained, with some sources showing 2-3 orders of magnitude of changes in flux (e.g. \cite{Walton2015, Earnshaw2018, Song2020}).

\subsection{Brief summary of ULX main properties as seen in X-rays} \label{sec:xray-summary}

In the early 2000s, it was speculated that the extreme luminosities exhibited by ULXs were produced by intermediate-mass $(10^{2-4} M_{\odot})$ black holes (IMBHs) accreting at relatively normal (i.e. sub-Eddington) rates. This was related to the apparent detection of relatively cool ($\sim0.1$ keV or $10^6$ K) accretion disc components in ULX spectra, temperatures which were in between the UV-dominated ($\sim0.01$ keV) spectra of supermassive black holes (SMBHs) powering active galactic nuclei (AGN) and the X-ray peaked ($\sim1$ keV) spectra in the stellar-mass black holes (BHs) powering X-ray binaries. An accretion disc with $kT\sim0.1$ keV and a bolometric luminosity of $10^{40}$ erg\,s$^{-1}$ would correspond to BHs of several hundred Solar masses accreting at relatively low Eddington-normalised rates, which are expected to produce hard X-ray spectra through comparison with Galactic XRBs \cite{Kaaret2001,Miller2003}. Amongst the best candidates for IMBHs there is the hyper-luminous source (HLX-1) in ESO 243-49 \cite{Farrell2009}. Its outburst and spectral states are similar to the sub-Eddington Galactic BHs. With a peak luminosity of $10^{42}$ erg\,s$^{-1}$, it likely has a mass of $10^{4-5} M_{\odot}$, and could be a failed tidal disruption event owing to a recent delay in the outbursts \cite{Lin2020}.

Two alternative scenarios invoked either extreme geometrical beaming in Eddington limited stellar-remnant black holes \cite{King2001} or super-Eddington accretion onto such objects (black holes were typically assumed in this scenario), similarly to the Galactic microquasar SS 433 \cite{Begelman2002} but potentially viewed at a more favourable angle (our view of SS433 is almost perfectly edge-on).

The ULX field was revolutionised after the discoveries of X-ray pulsations from the extreme ULX M82 X-2 with \textit{NuSTAR} (peak luminosity of $L_{\rm{X}} \sim 2 \times 10^{40}$ erg\,s$^{-1}$ \cite{Bachetti2014}), clearly demonstrating this source is powered by a highly super-Eddington (and strongly magnetised) neutron star. Since this discovery, a handful of other ULXs have been confirmed as neutron star accretors via the detection of pulsations with either \textit{XMM-Newton} or \textit{NuSTAR} \cite{Fuerst2016, Israel2017a, Israel2017b, Carpano2018, Sathyaprakash2019, Rodriguez2020}.
More detail on the properties of ULX variability at short- and long-term time scales, including fast-coherent and slow-(super)orbital modulations, is reported in Sect.\,\ref{sec:timing}.

Deep observations with \textit{XMM-Newton} and, later on, \textit{NuSTAR} have shown that most ULXs actually show X-ray spectra that differ from the sub-Eddington accretion states seen in Galactic black hole XRBs, further supporting the super-Eddington accretion scenario. ULX spectra typically show a strong curvature in the $\sim$2--10\,keV band before breaking to very steep spectra above $
\sim$10\,keV, with spectral slopes of $\Gamma \sim 3$ \cite{Bachetti2013, Walton2018a}, and the brightest sources are now often interpreted as exhibiting two thermal components that dominate below 10 keV (with typical temperatures $\sim$0.2--0.5 and $\sim$2--4\,keV).
It has therefore become largely obsolete to relate ULX spectra to the standard XRB accretion states.
Instead, ULX spectra are now more typically discussed in the context of the so-called `\textit{ultra-luminous state}' \cite{Gladstone2009}, a distinct accretion state related to super-Eddington accretion.
Observationally, the ultra-luminous state is further classified according to three main regimes depending on the effective spectral slope below 10\,keV: 
soft ultra-luminous (SUL, in which $\Gamma>2$, or equivalently the cooler thermal component dominates) or hard ultra-luminous (HUL, in which $\Gamma<2$, where the hotter thermal component dominates). In the latter, if the X-ray spectrum has a single peak and a blackbody-like shape, it is called broadened disc regime (BD, \cite{Sutton2013}). In addition to these classifications, there is also a rare population of extremely soft ULXs which have spectra that are primarily dominated by even cooler thermal emission ($kT \sim 0.1$\,keV) and are often referred to as `ultra-luminous supersoft sources' (ULSs).
More detail on X-ray broadband spectral properties of ULXs is given in Sect.\,\ref{sec:broadband}.

Atomic features provide key information on accretion flow geometry/dynamics and on the presence of any outflows (which are naturally predicted for super-Eddington accretion), but have been challenging to detect in ULX spectra.
Nevertheless, early studies spotted some spectral features around 1 keV where the resolving power of CCD spectrometers is rather low \cite{Stobbart2006}, and more recent studies have suggested that the shape and variability of these features imply that they are associated with the ULX itself rather than to the hot interstellar medium (ISM) of the host galaxy \cite{Sutton2015,Middleton2015b}.
ULXs are generally difficult targets for high-resolution X-ray spectroscopy. Despite their high X-ray luminosities, the Mpc distances normally limit their flux in the canonical 0.3-10 keV band to less than a few $10^{-12}$ erg\,s$^{-1}$\,cm$^{-2}$. This makes it difficult to fill the thousands of energy channels in high-resolution grating spectrometers with sufficient photons. 
However, the more recent availability of deep observations (300-500 ks) with moderate effective area detectors, mainly the Reflection Grating Spectrometers (RGS) aboard \textit{XMM-Newton}, finally enabled the detection and identification of emission and absorption lines in ULX X-ray spectra \cite{Pinto2016}. The emission lines are commonly found at their laboratory wavelengths, with the exception of NGC 5204 (UL)X-1 where they are blueshifted by $\sim0.3c$ \cite{Kosec2018a}. Multiple observations show that the emission lines vary over time and have huge X-ray luminosities ($L_{\rm X}\sim10^{38}$ erg\,s$^{-1}$; \cite{Pinto2020b}) that are orders of magnitude brighter than in Eddington-limited Galactic XRBs \cite{Amato2021,Psaradaki2018}.

The absorption lines are instead almost always highly blueshifted (0.1-0.3\,$c$) and vary with the ULX regime and amongst ULXs with different spectral slope \cite{Pinto2017,Kosec2018b,Pinto2020b}. 
\textit{Chandra} gratings confirmed similar outflows in a Galactic transient ULX (\cite{vdEijnden2019}). Further work on moderate-resolution detectors confirmed the presence of blueshifted absorption and emission features \cite{Walton2016a,Wang2019}. These discoveries, particularly the relativistically blueshifted absorption lines, likely reveal the long-sought powerful winds predicted by the theoretical simulations of radiation pressure in super-Eddington accretion discs \cite{Ohsuga2005}.
Detail on the methods used in high-resolution X-ray spectroscopy and their applications onto ULX spectra is provided in Sect.\,\ref{sec:hrxs}.

\subsection{Brief summary of ULX multi-wavelength observations} \label{sec:multiwavelength}

Observations of Galactic XRBs in other energy bands improve our knowledge, especially on the nature of the donor, the mass of the compact object, the presence of collimated outflows (jets) and the local environment. For ULXs, this is complicated by their larger distances (at least two orders of magnitude than Galactic XRBs).

Starting in the early 2000s, it was found that many bright ULXs are spatially associated with an optical \cite{Pakull2003} and/or radio nebula \cite{Kaaret2004}. Originally they were thought to be powered by radiation and being mainly photoionised owing to their strong He~{\sc ii} (4686\,{\AA}), [Ne~{\sc v}] (3426\,{\AA}) and [O {\sc iv}] (25.89 $\mu$m) line emission. The He~{\sc ii} line requires photons with energies in excess of 54.4~eV, which implies strong X-ray illumination of the ISM by the ULX and rather low beaming, even accounting for precession. For Holmberg II X-1 a lower limit on the luminosity required to illuminate its surrounding nebula of $1.1 \times 10^{40} \rm \, erg \, s^{-1}$ was obtained \cite{Berghea2010}, which - given the long recombination time of He {\sc iii} - indicates that the average luminosity has been at least $10^{40} \rm \, erg \, s^{-1}$ for several thousand years \cite{Kaaret2017}.

The discovery of large velocity broadening (up to 150 km s$^{-1}$), i.e. outflows, and the presence of shocks and lines from collisional excitation (e.g. [O~{\sc i}] at 6300\,{\AA}) indicated that in many cases mechanical power due, e.g., to winds or jets is actually inflating them. A clear example is the optical bubble, previously thought to be a nebula, around NGC 1313 (UL)X-1. Recent, accurate work with the \textit{Very Large Telescope (VLT)} / MUSE has confirmed the collisional nature of this interstellar cavity \cite{Gurpide2022}. The bubbles are young ($10^{5-6}$ yr) and their H\,$\alpha$ luminosities require a mechanical power of $10^{39-40}$\,erg\,s$^{-1}$ \cite{Pakull2003}, which agrees with early estimates of the mechanical power associated with the X-ray winds in ULXs \cite{Pinto2020a}. More detail on the comparison between the properties of ULX X-ray winds and multi-wavelength observations is provided in Sect.\,\ref{sec:implications}.

High-quality optical spectra of ULXs also revealed the presence of winds in the form of broad (500-1500 km s$^{-1}$) emission lines from the Balmer series and from He~{\sc ii} \cite{Fabrika2015}. This implies that caution is needed when attempting to use the emission lines to measure radial velocity curves and the mass of the compact object. The supersoft ULX M81 ULS-1 shows a blueshifted H\,$\alpha$ line, with the blueshift varying with time and corresponding to a range of projected velocities between 0.14-0.17\,$c$ \cite{Liu2015}. This indicates that it originates in either a relativistic baryonic jet like in SS 433 (see below), or a conical wind similar to those observed in X-rays \cite{Pinto2016}.

In some cases, jets could play a role in powering ULX bubbles. There is indeed evidence of jets through radio observations of a few nearby ULXs such as Holmberg II X-1, M\,31 ULX, and NGC 7793 S26 \cite{Cseh2014, Middleton2013, Pakull2010}. The latter is particularly interesting as the jet mechanical power of a few $10^{40} \rm \, erg \, s^{-1}$ is orders of magnitude higher than the X-ray luminosity, such that this system strongly resembles the case of the Galactic microquasar SS 433 \cite{Marshall2002,Fabrika2004}. There is a large consensus that SS 433 is a compact object accreting at a super-Eddington rate of $\sim 10^{-4} M_{\odot} \rm \, yr^{-1}$ via Roche lobe overflow from a massive donor, likely an evolved A supergiant, and it is X-ray weak due to strong circumstellar obscuration owing to an edge-on viewing angle. Should it be observed face-on, i.e. closer to the accretion disc axis, it would appear as a ULX \cite{Waisberg2019,Middleton2021}. The M31 ULX was detected in the radio whilst exhibiting a thermal spectrum, similar to the XRB soft X-ray states. The radio emission was likely due to discrete ejecta similar to the Galactic BH GRS 1915+105. They also have a comparable X-ray peak luminosity of $1.3 \times 10^{39}$ erg s$^{-1}$, suggesting that M31 ULX is a BH accreting near its Eddington limit. For ESO 243-49 HLX-1 and NGC 2276-3c, instead, the detection of flaring radio emission and hard X-ray spectral states would suggest the presence of an Eddington-limited IMBH as the accretor.

Substantial work has been done in the last decade in order to search for the compact optical counterparts of ULXs, with a view to identifying their donor stars and, in turn, obtaining dynamical mass constraints for the accretors via radial velocity (RV) studies. ULX distances imply searches with current facilities are only sensitive to optical counterparts brighter than $\sim$20th magnitude, even for massive donors, which makes their detection difficult. 
Moreover, the high probability of ULXs being located in star-forming regions, i.e. crowded fields, complicates the situation even further. Here the best results have been obtained through the simultaneous use of the high-spatial-resolution images ($<0.3''$) from \textit{Chandra} and the \textit{Hubble Space Telescope (HST)}. For about 20 ULXs unique optical counterparts were found \cite{Tao2011,Gladstone2013}. The magnitudes and B-V colours suggest OB supergiant companion stars if these counterparts are really dominated by the stellar companion, but the optical emission could be significantly affected by X-ray reprocessing in the outer disc. Indeed, this is confirmed in some ULXs by the detection of strong optical variability \cite{Tao2011}.

These efforts have resulted in two initial claims for dynamical mass estimates that support the presence of stellar-remnant accretors (as opposed to IMBHs). The first is M 101 ULX-1, where evidence for a Wolf-Rayet companion is seen, and a probable mass of 20-30\,\Msun\ for the accretor is proposed (with a lower limit of 5\,\Msun, which would require the source to be powered by a black hole). However, caution is needed here, as the sampling of the RV data is very sparse, and the stellar velocity shifts are inferred from emission lines, which may be problematic in the presence of strong outflows from the ULX (as noted above). The second is NGC 7793 P13, where a B9Ia supergiant companion is seen, and an upper limit to the accretor mass of 15\,\Msun\ for the accretor is claimed based on the stellar companion type and the detection of a $\sim$64-day optical photometric period (interpreted to be the binary orbit \cite{Motch2014}). As it turns out, NGC 7793 P13 is the second source confirmed to host a neutron star through the detection of X-ray pulsations \cite{Fuerst2016}, consistent with the dynamical mass constraint.

\textit{VLT} / X-shooter spectra of the pulsating NS NGC 300 ULX-1, covering the wavelength range 3,500-23,000\,{\AA} clearly showed the presence of a red supergiant (RSG) donor star that is best matched by a stellar atmosphere with $T_{\rm eff}$ = 3,650-3,900 K and log(L$_{\rm bol} / L_{\odot}$) = 4.25 $\pm$ 0.10, which yields a stellar radius $R / R_{\odot}$ = 310 $\pm$ 70 \cite{Heida2019}. Given the large donor-to-compact object mass ratio, orbital modulations of the radial velocity of the RSG are likely undetectable. Other ULXs have an RSG as a donor \cite{Heida2016} with NGC 253 ULX favouring a massive BH ($\gtrsim$ 50 M$_{\odot}$) as the compact object \cite{Heida2015}. In these cases, long-term monitoring may be used to search for line modulations and allow for further dynamical mass estimates in the future.

The community of experts in stellar population synthesis has been actively working on the predictions of ULX binary systems properties.
For instance, it was shown that extremely high, super-Eddington accretion rates, of $10^{-3} ~ M_{\odot} \rm \, yr^{-1}$ or above  can be easily sustained for $\sim 10^4$ yr in BHs and NSs with companion stars of similar masses \cite{Rappaport2005,Wiktorowicz2015}.
Exploring the different formation channels of merging double compact objects (DCOs: BH-BH/BH-NS/NS-NS) using the STARTRACK code, it was found that in the local Universe typically 50\,\% of merging BH-BH progenitor binaries would have evolved through a phase of super-Eddington accretion (when only one of the binary members had evolved into a compact object) and would have potentially appeared as a ULX \cite{Mondal2020}. This indicates that ULXs can be used to study the origin of gravitational wave (GW) sources. The fraction of observed ULXs that will form merging DCOs in future varies between 5 and 40\,\%. More detail on the accretion rates and the comparison with observations is provided in Sect.\,\ref{sec:implications}.

%%%%%%%%%%%%%%%%%%%%%%%%%%%%%%%%%%%%%%%%%%%%%%%%%%%%%%%%%

\section{X-ray timing properties}
\label{sec:timing}

One of the key observational aspects of accretion-powered sources is temporal variability, and ULXs are no different. These variability properties are often characterised via Fourier techniques and have revealed a wide range of phenomenology in Galactic XRBs, including broadband noise, quasi-periodic oscillations and coherent pulsations (which can reveal the nature of the compact object) on short timescales, as well as potentially orbital or super-orbital variations on (typically) longer timescales (e.g. \cite{Remillard2006}). In addition to these flux variations, time lags between different energy bands may unveil photon reprocessing or disc perturbations, and help to reconstruct the accretion geometry. Here we provide a brief overview on ULX variability properties focusing first on short-term and then long-term variability timescales.

\subsection{Pulsations}
\label{sec:timing-pulsations}

Undeniably the most important recent discovery on X-ray timing properties of ULXs is the coherent pulsations in a growing sample of ULXs. To date, there are 6 `traditional' ULXs that are known to be pulsars which show sustained luminosities comfortably in excess of $10^{39}$\,erg s$^{-1}$, although there is plenty of speculation that more of the ULX population will reveal themselves to be pulsars over time. In addition to these sources, there are a similar number of transient Be X-ray binaries with neutron star accretors in the local group that occasionally peak just above $10^{39}$\,erg s$^{-1}$ whilst in outburst, and so could also formally be considered ULX pulsars (e.g. \cite{Skinner1982, Wilson-Hodge2018, Vasilopoulos2020}).

\begin{figure}[h] %%%[b]
%\sidecaption
% For example, with the graphicx style use
\includegraphics[scale=.65]{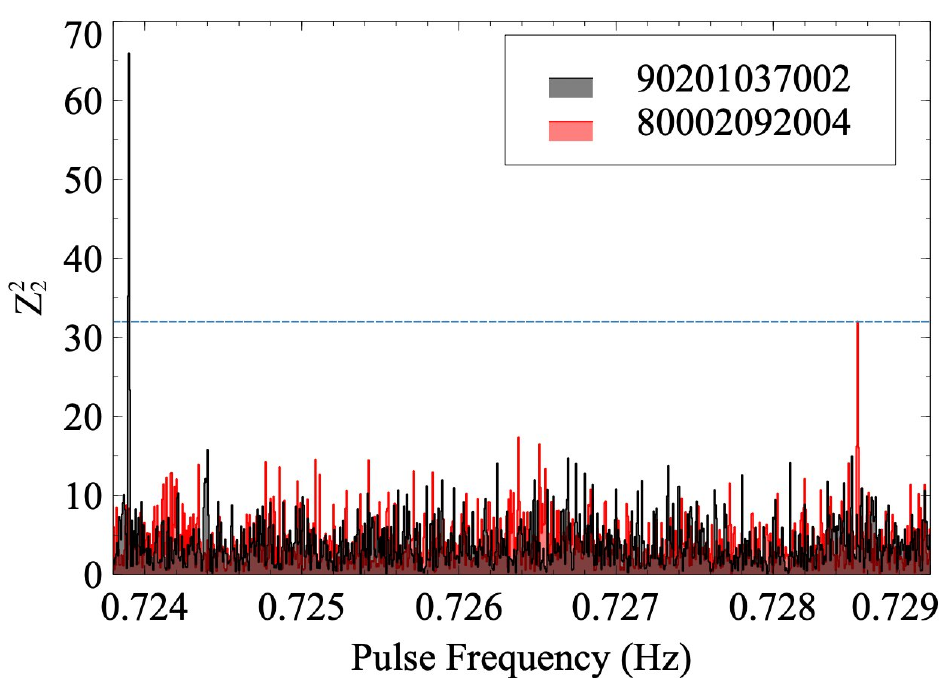}
\includegraphics[scale=.255]{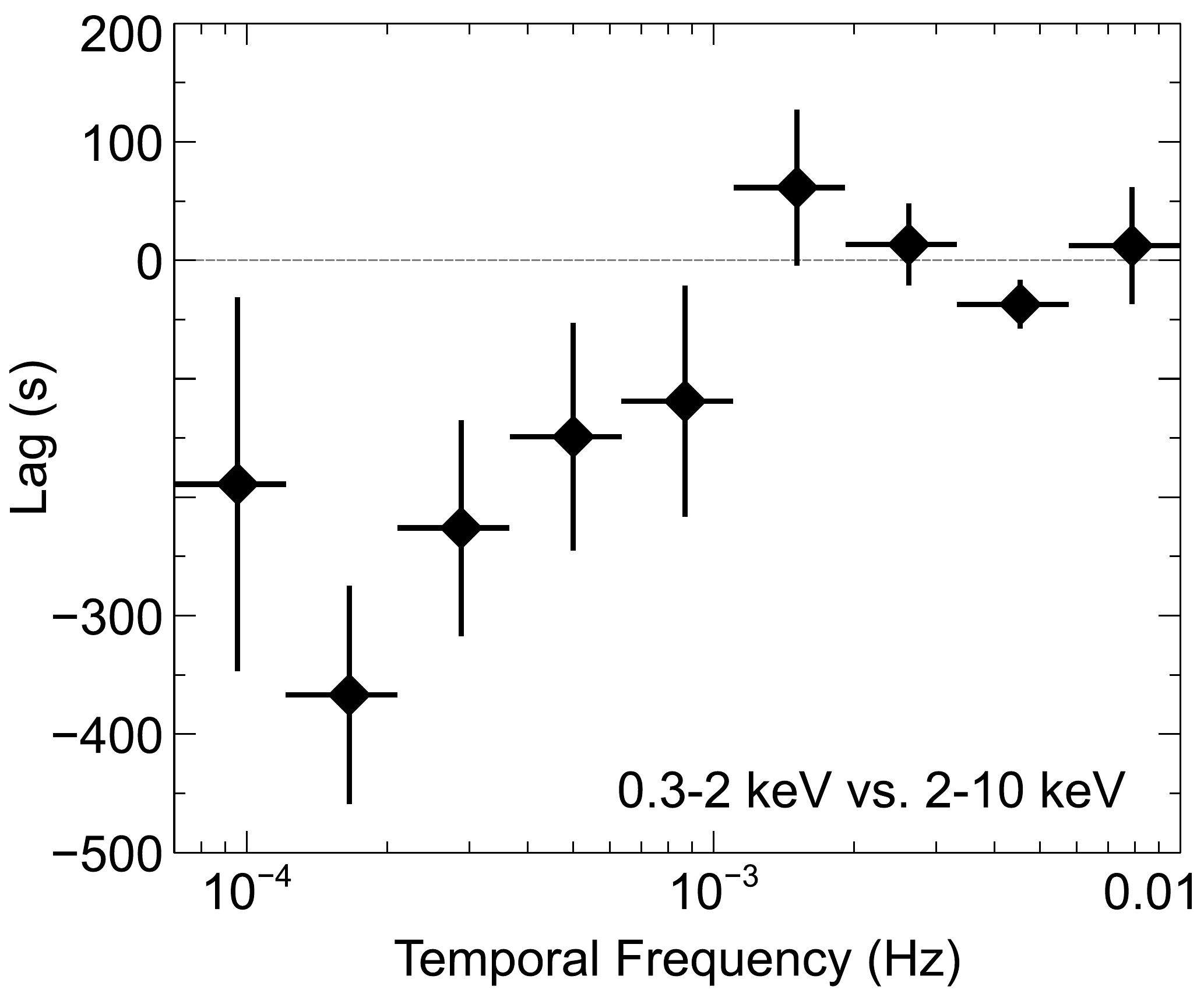}
%\includegraphics[scale=.625]{fig_Fuerst20162a}
%
% If no graphics program available, insert a blank space i.e. use
%\picplace{5cm}{2cm} % Give the correct figure height and width in cm
%
\caption{\textcolor{black}{Left: Detection of pulsations in M 82 X-2 with \textit{NUSTAR} in 2014 (red) and 2016 (black). The horizontal line indicates the 3\,$\sigma$ level for the given number of trials \cite{Bachetti2020}. 
Right: soft (0.3--2~keV) vs. hard (2--10~keV) lags as a function of temporal frequency for NGC 1313 X-1 (\textit{XMM-Newton} / EPIC-pn data). Negative lags are soft band lags. The variability on long timescales show a soft band lag of $\sim150$~s \cite{Kara2020}. }}
%%%\caption{Left: Detection of pulsations in M 82 X-2 with \textit{NUSTAR} in 2014 (red) and 2016 (black). The horizontal line indicates the 3\,$\sigma$ level for the given number of trials \cite{Bachetti2020}. Right: Power spectral density (PSD) of the 2016 \textit{XMM-Newton} EPIC-pn (0.3–10 keV; top) and NuSTAR \textit{FPMA} (3–78 keV; bottom) data between 0.00124 and 6.812 Hz for NGC 7793 P13. The red dashed line shows the 5\,$\sigma$ detection limit assuming white noise and accounting for the number of trials \cite{Fuerst2016}.}
\label{fig:pulsations}       % Give a unique label
\end{figure}

M 82 X-2 was the first ULX to show pulsations \cite{Bachetti2014} with a spin period of 1.37\,s. The evolution of the pulsations also revealed a 2.5\,day orbital modulation and a continuous spin-up of $\dot{P} = -2 \times 10^{-10} \, \rm s \, s^{-1}$ (see Fig.\,\ref{fig:pulsations}, left panel). Coherent pulsations are the robust signature of a rotating neutron star, which indicates that the observed luminosity surpasses the Eddington limit by orders of magnitude. Long-term monitoring of the pulse period has shown that the neutron star exhibits periods of both secular spin-up and spin-down, potentially indicating the neutron star is close to spin equilibrium, and also that the orbital period is evolving owing to the extreme mass transfer, confirming accretion rates that significantly exceed the Eddington limit \cite{Bachetti2020, Bachetti2022}.

After M82 X-2, the best-studied ULX pulsars are NGC\,7793 P13 and NGC\,5907 ULX-1.
NGC 7793 P13 is the second pulsating ULX (PULX) discovered \cite{Fuerst2016,Israel2017b}. It shows pulsations at 0.43\,s with a secular spin-up of $\dot{P} = -3 \times 10^{-11} \, \rm s \, s^{-1}$. In this system, the X-ray pulsations reveal a much longer orbital period of 67 days, similar to but formally slightly longer than the optical photometric period (which now appears to be a \textit{sub}-orbital periodicity \cite{Fuerst2021}). The most extreme PULX is NGC 5907 (UL)X-1, which reaches a peak luminosity  $\sim10^{41} \rm \, erg \, s^{-1}$, corresponding to $\sim500$ times the Eddington limit of a neutron star \cite{Israel2017a, Fuerst2017}. This sources also exhibit pulsations on timescales of $\sim1$\,s and a strong $\dot{P} = -8 \times 10^{-10} \, \rm s \, s^{-1}$. The orbital solution is not as well determined in this case, but the best estimates from the evolution of the X-ray pulsations put the orbital period in the range $\sim5-30$ days \cite{Israel2017a}.

PULXs typically show nearly sinusoidal pulse profiles which would argue against the presence of strong beaming, in broad agreement with the moderate beaming implied by the presence of the photoionised components of ULX bubbles (Sect.\,\ref{sec:multiwavelength}). The pulsed fraction is typically $\sim$10\,\% in the \textit{XMM-Newton} band (0.3--10\,keV), but rises with the energy such that typical pulsed fractions above 10\,keV are $\sim$20--30\% (although the most extreme case, NGC 300 ULX-1, shows an extreme pulsed fraction of $\sim$80\% at these energies \cite{Carpano2018}). PULX spectra are discussed in Sect.\,\ref{sec:broadband}, but tend to fit in the hard end of ULX spectra. Whilst the traditional PULXs do spend extended periods at their extreme luminosities, almost all also show very strong long-term flux variations, and the most extreme cases can also drop down to fluxes below $10^{38} \rm \, erg \, s^{-1}$ \cite{Belfiore2020}. Sect.\,\ref{sec:implications} reports some theoretical interpretations.

\subsection{Quasi-periodic oscillations}
\label{sec:timing-qpo}

High-frequency quasi-periodic oscillations (HF-QPOs) can be identified with their main harmonics above 100\,Hz thanks to their 3:2 frequency ratio \cite{Remillard2006}. They are often used to probe the inner edge of the accretion disc owing to their high coherence and dependence on the BH mass and spin. They tend to be found mainly during BH hard states.
\textit{RXTE} observations of M 82 X-1 provided evidence for a pair of QPOs at $3.32 \pm 0.06$ and $5.07 \pm 0.06$ Hz, i.e. with a harmonic ratio compatible with HF-QPOs seen in Galactic BHs. Each QPO has a rather low significance ($3.7\sigma$ and $2.7\sigma$), which combined yields $4.7\,\sigma$. Assuming that the QPOs are of the same nature, scaling according to the frequency-mass relation, leads to a mass estimate of $428 \pm 105 M_\odot$ for the compact object mass in M 82 X-1, i.e. an IMBH \cite{Pasham2014}.

In Galactic BHs, low frequency (LF-) QPOs are divided into three types, A, B, and C \cite{Casella2005}. Type-C frequency is correlated with the spectral index and the disk flux. The QPOs detected in M82 X-1 and NGC 5408 X-1 are similar to BH type-C QPOs. In M82 X-1, however, the QPO frequency is uncorrelated with the spectral index, which cautions against its use as a BH mass indicator. This was confirmed in NGC 5408 X-1 where the QPO frequencies are not correlated with the continuum noise break frequencies and the spectral parameters \cite{Middleton2011}.

Quasi-periodicities can also be produced by other phenomena such as dips. The \textit{XMM-Newton} light curve of the supersoft source in NGC 247 shows strong flux dips on time scales of 200 s or longer. These cannot be explained just by an increase of \textcolor{black}{photoelectric} absorption \cite{Feng2016}. Power spectra and coherence reveal the dipping preferentially occurs on 5 ks and 10 ks timescales \cite{Alston2021}. A likely explanation is therefore occultation of the central X-ray source by an optically thick structure, such as warping of the accretion disc, or obscuration by a wind launched from the accretion disc, or both. This supports the idea that supersoft ULXs are viewed close to edge-on (see Sect. \ref{sec:implications}). An alternative explanation invokes the occurrence of the propeller effect. \textcolor{black}{Assuming this as an origin for} the jump in flux and, thereby, in luminosity from $2.5 \times 10^{39} \rm \, erg \, s^{-1}$ to $2.3 \times 10^{38} \rm \, erg \, s^{-1}$, a magnetic field of $2\times10^{11}$\,G was constrained \cite{DAi2021} which is in line with estimates in other ULXs \cite{King2020}.

\subsection{Time delays between energy bands}
\label{sec:time-lags}

An advanced technique to decompose, in a model-independent way, the spectrum of a variable source and, in particular, the structure of an accretion disc consists of searching for correlated variability and time delays between time series of different energy bands. The low count rates seen from ULXs combined with the lower levels of short-timescale variability they exhibit in comparison to Galactic XRBs, make it difficult with the current instruments. In the last decade, dedicated work has found evidence for delays in the soft X-ray band at various frequencies: from $\sim 10$~mHz in NGC 5408 X-1 \cite{Heil2010}) down to much lower frequencies, $f\sim0.1-1$~mHz in other ULXs \cite{Pinto2017, Kara2020, Pintore2021} (Fig.\,\ref{fig:pulsations}, right panel). Given the two orders of magnitude difference in their time lags, it is not obvious that these are due to the same mechanism. The covariance spectra suggest that emission contributing to the lags is largely associated with the hotter of the two thermal-like components, likely originating from the inner accretion flow (see Sect.\,\ref{sec:broadband}). All these soft lags are however $\sim5-20$\,\% of the corresponding characteristic variability timescales. If these soft lags can be understood in the context of a unified picture of ULXs, then lag timescales may provide constraints on the density and extent of the accretion flow (see Sect.\,\ref{sec:implications}).

%\begin{figure}[h] %%%[b]
%%\sidecaption
%% Use the relevant command for your figure-insertion program
%% to insert the figure file.
%% For example, with the graphicx style use
%\includegraphics[scale=.22]{fig_Kara20202}
%\includegraphics[scale=.42]{fig_Pinto201710}
%%
%% If no graphics program available, insert a blank space i.e. use
%%\picplace{5cm}{2cm} % Give the correct figure height and width in cm
%%
%\caption{Left: soft (0.3--2~keV) vs. hard (2--10~keV) lags as a function of temporal frequency for NGC 1313 X-1 (\textit{XMM-Newton} / EPIC-pn data). Here, negative lags are soft band lags. The variability on long timescales show a soft band lag of $\sim150$~s \cite{Kara2020}. Right: Lag-energy spectra at low frequencies for NGC 55 ULX \cite{Pinto2017}. The dotted red lines show the magnitude expected from the Poisson noise. Both ULXs show a soft X-ray lag, which increases by a factor 10 in NGC 55 ULX.}
%\label{fig:lags}       % Give a unique label
%\end{figure}

\subsection{Long-term modulations}
\label{sec:timing-long-term}

Since the launch of the {\it Rossi X-ray Timing Explorer (RXTE)} and, later on, the \textit{Neil Gehrels Swift Observatory} it was possible to obtain long and regularly sampled light curves of X-ray sources. The brighest, persistent, ULXs with fluence $\gtrsim$ 10$^{-12}$ erg s$^{-1}$ cm$^{-2}$ make very good {\it Swift} targets. This enables the search for long-term (days to months) periodicities which could be associated with orbital or super-orbital (e.g. precession) variability. This is common practice in Galactic XRBs where periodic cycles can identify disc instabilities, enhanced accretion rates at specific orbital phases (e.g. periastron) and warps in the accretion disc \cite{Priedhorsky1987,Wijers1999}.

The first evidence for a ULX long-term (62-day) periodicity in a ULX was a 62-day period reported for the M82 galaxy based on data from \textit{RXTE} \cite{Kaaret2006}. Although initially assumed to be associated with M82 X-1, as this is typically the brightest X-ray source in the M82 galaxy, it is now understood that this period is actually associated with the ULX pulsar M82 X-2 \cite{Qiu2015, Brightman2019}. As such, given that the orbital period for this system is known to be $\sim$2.5 days, the 62-day variations must be super-orbital in nature.
\textit{Swift} monitoring significantly enlarged the detections of long-term periodicities, including a longer period of 115 days found in NGC 5408 X-1 \cite{Strohmayer2009}, which then shifted to 136 days \cite{An2016} suggesting that the modulation is super-orbital.

Long-term X-ray periods are relatively common amongst the PULX population, although they appear to have a range of origins. \cite{Walton2016b} report a 78-day X-ray period in NGC\,5907 ULX-1, which is likely super-orbital in nature given the current best estimates for the orbital period (between 5--30 days \cite{Israel2017a}). Similarly, M 51 ULX-7 shows super-orbital X-ray variations with a period that drifts from $\sim$38--44 days \cite{Vasilopoulos2020, Brightman2022}; here the orbital period is known to be $\sim$2 days based again on the evolution of the X-ray pulsations \cite{Rodriguez2020}. Although NGC\,7793 P13 shows a long-term X-ray periodicity with a 67-day timescale, the best current solution for the binary orbit based on the X-ray pulsations suggests that the variations are orbital in nature. 

At the other end of the mass scale, the IMBH candidate ESO 243-49 HLX-1 regularly goes into outburst (as mentioned in Sect.\,\ref{sec:xray-summary}). Initially it appeared as though these outbursts had a periodic recurrence time of $\sim$300 days \cite{Godet2014}, but more recently the period between outbursts has been increasing dramatically \cite{Lin2020}.
Amongst the several interpretations here, one potential option is an IMBH that is fed by winds from a giant star with a tidally stripped envelope \cite{Miller2014}.

%%%%%%%%%%%%%%%%%%%%%%%%%%%%%%%%%%%%%%%%%%%%%%%%%%%%%%%%%

\section{Broadband X-ray spectroscopy}
\label{sec:broadband}

\begin{figure}[h] %%%[b]
%\sidecaption
% For example, with the graphicx style use
\includegraphics[scale=.385]{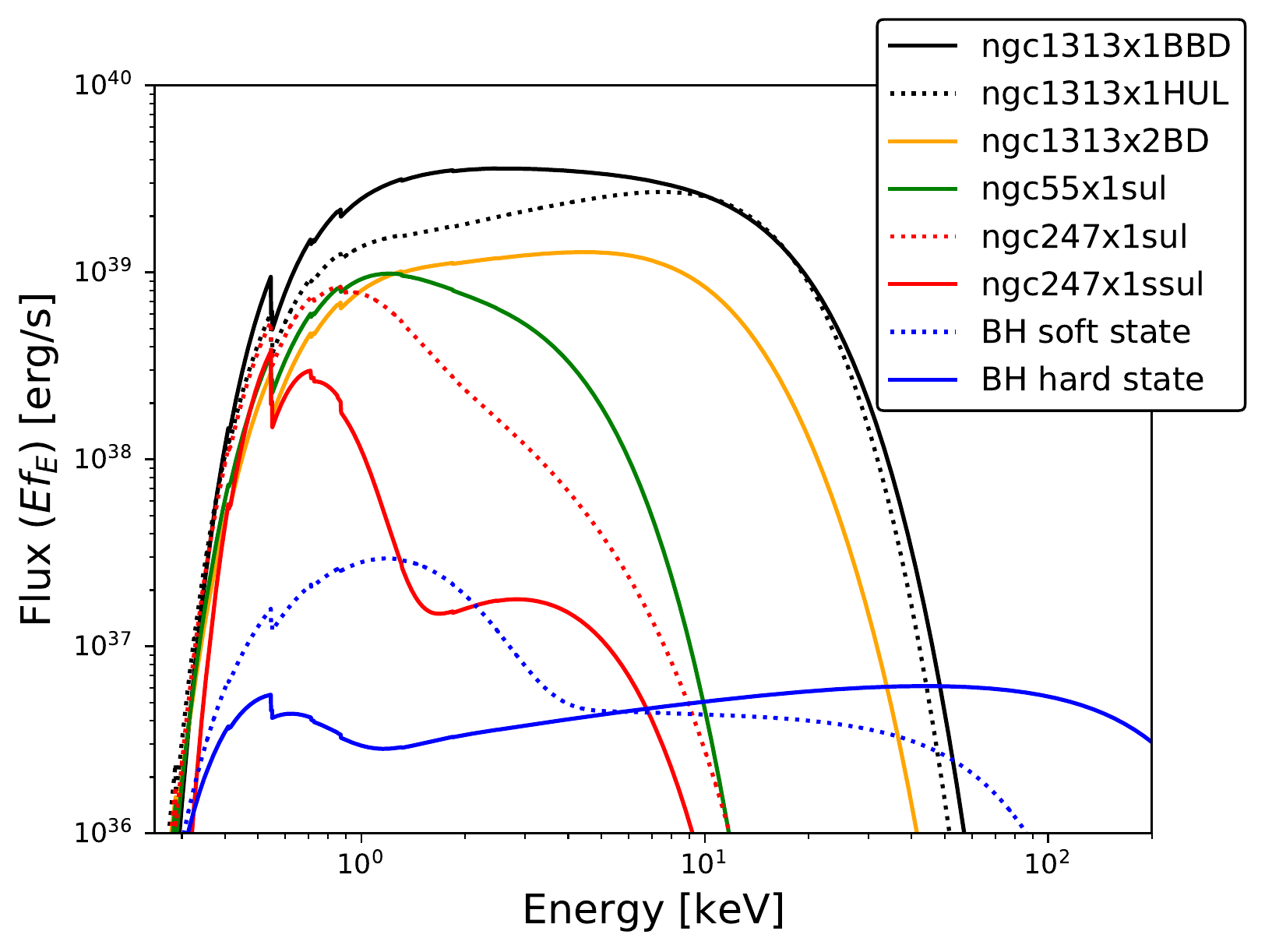}
\includegraphics[scale=.36]{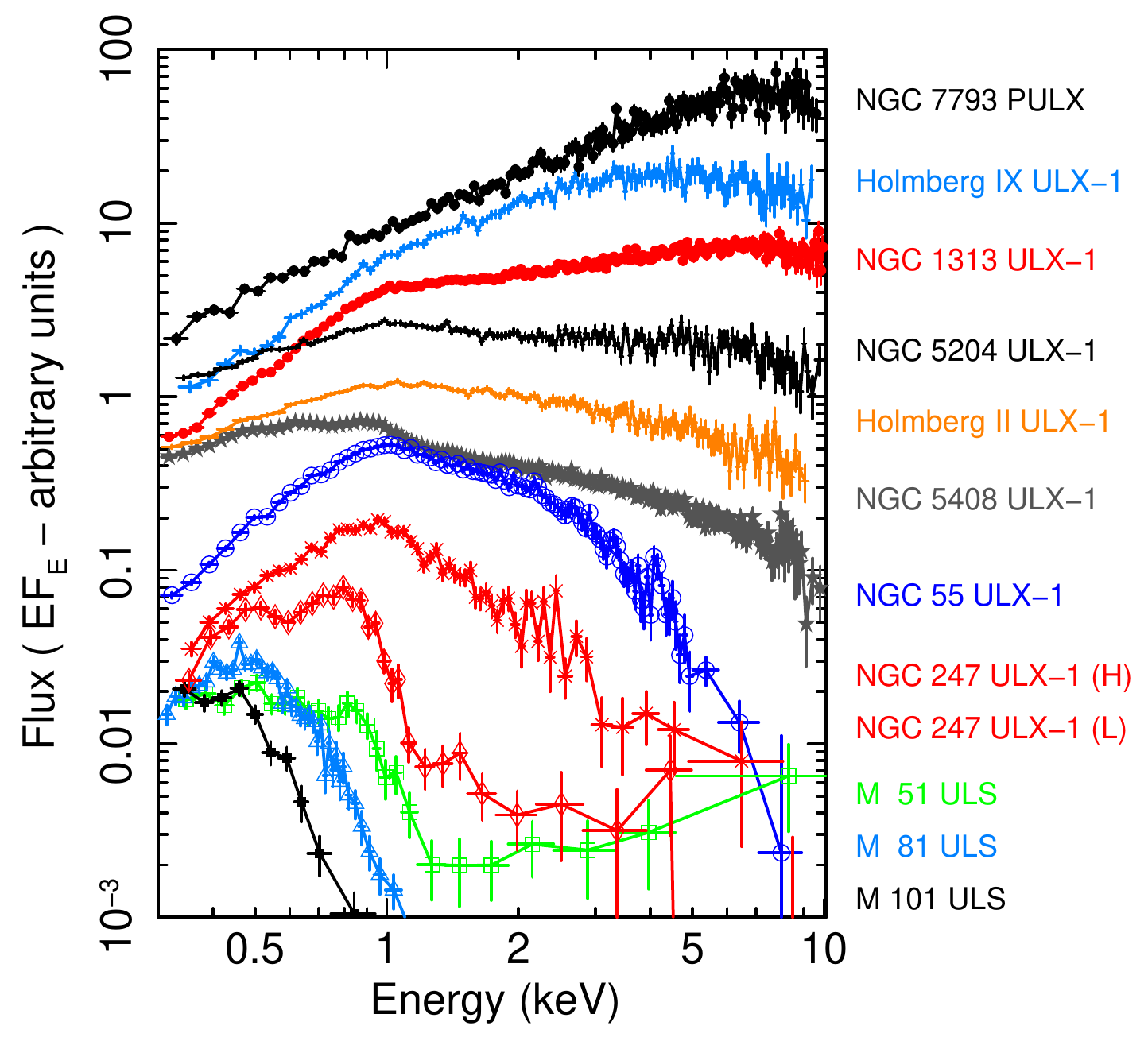}
%
% If no graphics program available, insert a blank space i.e. use
%\picplace{5cm}{2cm} % Give the correct figure height and width in cm
%
\caption{Left: Comparison of typical spectra of ULXs and Galactic BH binaries. ULX spectra are classified as supersoft (SSUL or ULS), soft (SUL) and hard (HUL) ultra-luminous and broadened disc (BD), and the brighter broadened disc regime (BBD). Right: X-ray spectra of some amongst the brightest ULXs with hardness increasing from bottom to top taken with \textit{XMM-Newton} (mainly) and \textit{Chandra} (ULS). Y-axis units are $E \times F_E$ in an arbitrary scale for displaying purposes \cite{Pinto2020a,Barra2022}.}
\label{fig:SEDs}       % Give a unique label
\end{figure}

Historically speaking, ULX spectra have typically been compared to the accretion states seen in Galactic black hole XRBs, which are broadly characterised by the relative contributions of the accretion disc (which provides thermal blackbody emission at lower energies, often characterised by temperatures $kT \sim 1$\,keV) and the up-scattering `corona' (which provides non-thermal Comptonised emission at higher energies, often characterised by electron temperatures $kT_{\rm{e}} \sim 50-100$\,keV). Most of these sources are transient LMXBs that accrete via Roche lobe overflow. When they go through an outburst, during which they can reach luminosities comparable to the Eddington limit, in addition to the strong flux variability they also exhibit strong spectral variability (i.e. strong changes in their spectral `hardness'), with most sources tending to follow a relatively well-defined pattern of behaviour. Sources generally rise through the `hard' state (in which the corona dominates the observed X-ray spectrum, and persistent radio jets are seen), before transitioning to the `soft' state (in which the accretion disc dominates, and winds are often observed) when they reach higher luminosities. As the sources fade, they transition back to the hard state and eventually return to quiescence (note that there is typically some hysteresis seen, such that the hard-to-soft and soft-to-hard transitions during the rise and decay, respectively, do not occur at the same luminosities (see Fig.\,\ref{fig:SEDs}, left panel \cite{Russell2022}).
There are also a small number of more persistent black hole HMXBs (most notably Cygnus X-1) which broadly show similar states to their LMXB cousins but do not fade to quiescence on observable timescales (as they primarily accrete from the stellar wind of their companion).

Early observations of ULXs with ASCA yielded the first high-counts spectra in the 0.5--10\,keV band \cite{Colbert1999}. Some spectra appeared thermal in nature, albeit with a slightly broader profile and higher disc temperatures than typically seen in classic Galactic XRB soft states, whilst others appeared power-law-like and showed evidence for short-timescale variability analogous to the hard state (e.g. \cite{Makishima2000, Kubota2002}). There was therefore speculation that ULXs were showing the same accretion states as seen in Galactic black hole XRBs, but at higher luminosities, thereby implying the presence of higher-mass black holes.

\subsection{The golden age of ULX X-ray spectroscopy} \label{sec:golden-age}

The launch of \textit{Chandra} \cite{Weisskopf2002} and \textit{XMM-Newton} \cite{Jansen2001} brought X-ray detectors with improved effective area, facilitating more detailed studies of ULX spectra. Early observations with these facilities revealed spectra that required two continuum components. These could be well explained by a disc--corona model, which initially seemed to further support the connection to sub-Eddington accretion states. The disc temperatures obtained were relatively cool ($kT \sim 0.3$\,keV), which would correspond to an IMBH with mass $\sim$1,000\,\Msun\ \cite{Miller2003}, as the inner disc temperature should scale with the mass as $T_{\rm in} \propto M^{-0.25}$.

However, sub-Eddington thin discs are expected to show a relationship between their bolometric luminosity and inner temperature $L_{\rm disc} \propto T_{\rm in}^4$. As multi-epoch spectra were available it became clear that ULXs exhibit spectral variability that disagreed with this trend, instead showing flatter relations \cite{Poutanen2007,Barra2022}. Moreover, the discovery of an inverted $L_{\rm disc} \propto T_{\rm in}^{-4}$ trend in the most luminous ($L_{\rm X} > 10^{40}$ erg s$^{-1}$) ULXs with multiple high-quality spectra \cite{Kajava2009,Robba2021} has been interpreted as due to beaming caused by a large-scale cone inflated by strong radiation in a super-Eddington accretion regime \cite{King2009}.
More discussion on super-Eddington accretion and geometrical beaming is provided in Sect.\,\ref{sec:SEdd}.

In addition, as deep observations became available, evidence started to emerge that the higher energy component was not power-law-like (e.g. \cite{Gladstone2009}), instead seeming to show some subtle curvature, casting doubt on the standard disc-corona models that had been considered previously. A key development in broadband X-ray spectroscopy came with the launch of \textit{NuSTAR} in 2012 \cite{Harrison2013}, covering a higher energy bandpass (3--78\,keV) than \textit{XMM-Newton} and \textit{Chandra}, and carrying the first imaging X-ray optics to extend above 10\,keV. \textit{NuSTAR} observations (often in coordination with \textit{XMM-Newton} to provide genuine broadband observations) have provided spectra of ULXs up to $\sim$30-40\,keV (in the best cases), and unambiguously confirmed that the higher energy emission seen by \textit{XMM-Newton} and \textit{Chandra} is not power-law-like, and the curvature indicated by these missions represents a genuine spectral cutoff (e.g. \cite{Bachetti2013,Walton2014}; see Fig. \ref{fig:SEDs}). The broadband spectra revealed were not consistent with any of the standard sub-Eddington accretion states, strongly arguing that ULXs are primarily a population of super-Eddington accretors (instead of sub-Eddington IMBHs).
The existence of ULX pulsars was also discovered by \textit{NuSTAR} around this time, further confirming these conclusions.

Owing to their distinct spectral appearance, comparisons between ULX spectra and the standard sub-Eddington accretion states have become largely obsolete. Instead, ULXs are now mostly interpreted as representing a new, super-Eddington accretion state, dubbed the \textit{ultra-luminous state} \cite{Gladstone2009}. However, a rich diversity of spectra are seen from the ULX population (see Figure \ref{fig:SEDs}), and so the ultra-luminous state has itself been empirically divided into 4 main sub-classifications, depending on the different spectral properties that ULXs exhibit in the 0.3--10\,keV band \cite{Sutton2013}:

\begin{itemize}
\item{\textbf{Hard ultra-luminous} (HUL) -- the 0.3-10\,keV spectrum shows two distinct continuum components. If modelled with a cool accretion disc and a higher energy power-law, the latter is brighter and yields slopes $\Gamma<2$. Several ULXs (and all pulsating ULXs) show HUL spectra \cite{Pintore2017,Walton2018a}. Some notable examples are: NGC 7793 (p)ULX, Holmberg IX X-1 and NGC 1313 X-1 in Fig. \ref{fig:SEDs} (right).}
\item{\textbf{Soft ultra-luminous} (SUL) -- the spectrum is similar to the HUL in the sense that two continuum components are required, but here the thermal component is brighter and $\Gamma>2$. SUL spectra yield kT$_{\rm BB} \sim 0.1-0.3$ keV, whilst the value can be slightly higher in the HUL regime. Notable examples of SUL sources are: NGC 55 ULX-1, NGC 5408 X-1, NGC 247 X-1 (H) and Holmberg II X-1.}
\item{\textbf{Supersoft ultra-luminous} (SSUL or ULS) -- these spectra are almost entirely dominated by a cool blackbody-like component, with $kT \sim 0.1$~keV. They have a much fainter (a factor ten) hard tail with $\Gamma = 2-4$. SSUL spectra are broadly similar to those during the supersoft (SS) phases of novae, although ULSs are persistent and up to 100 times brighter. Some notable examples are shown in Fig. \ref{fig:SEDs} including a low-flux epoch of M 101 ULX-1 and NGC 247 X-1 (L).}
\item{\textbf{Broadened disc} (BD) -- these spectra also seem to be dominated by a single thermal component, but it is hot and has a broader profile than expected from a standard thin accretion disc \cite{Shakura1973}. In the 0.3-10 keV band, they can be described with an advection-dominated disc model with $p \sim 0.5-0.6$, where $T(R) \propto R^{-p}$ ($p=0.75$ for a standard thin disc), and $kT_{\rm in} \sim 1-2$~keV. Notable examples are NGC 1313 X-1 (high-BD or BBD) and X-2 (low-BD), and M 33 X-8.}
\end{itemize}

The diversity of these spectra is proposed to be primarily related to the fact that at super-Eddington accretion rates, the inner regions of the accretion disc are expected to be both geometrically and optically thick, and to launch powerful outflows \cite{Shakura1973, Abramowicz1988, Poutanen2007}. For such a geometry there is a natural expectation that the observed properties of a given source will depend on the angle it is viewed at \cite{Middleton2015a}. The general picture often now invoked is that the innermost (and hottest) regions of the accretion disc are the primary origin of the hotter of the two emission components that dominate below 10\,keV, but are most readily viewed when the disc is observed close to face-on owing to their geometrically thick nature, resulting in a HUL source. As the viewing angle is increased these hotter regions are eventually blocked from view, meaning that cooler thermal emission from larger radii and possibly the outflow (which may itself be optically thick) dominate the observed spectra, resulting first in a SUL source, and at even higher inclinations a SSUL source. Broadened disc sources stand out as something of an exception as they are not necessarily related to this viewing angle dependence, but are often thought to represent the transitional regime around the Eddington threshold in which the disc has started to deviate from a normal thin disc. \textcolor{black}{There are exceptions to this, where very luminous examples of BD spectra have been seen. Therefore, the full geometric effects associated with strongly super-Eddington accretion have not yet been fully established.} It is important to note that many ULXs switch between different regimes as they vary with time. For instance, NGC 1313 X-1 is known to show SUL, BD and HUL spectra at different epochs with some BD spectra fainter and others brighter than its HUL spectrum (see Fig. \ref{fig:SEDs}). NGC 247 X-1 instead shows notable  transitions between SUL and SSUL spectra. This could be due to the precession of the accretion flow, or to changes in the scale height of the funnel formed by the inner disc/wind.

\subsection{Moving Beyond Simple Spectral Models}

Whilst the discovery of ULX pulsars and the non-standard broadband spectra observed have demonstrated that the majority of ULXs are super-Eddington accretors, their broadband X-ray spectra are not necessarily solely described by thermal emission from a super-Eddington accretion disc. 
In addition to the lower-energy thermal components, a further relevant ingredient is a discovery in \textit{XMM-Newton} and \textit{NuSTAR} spectra of a third, harder continuum component that extends beyond the Wien tail from these thermal components. This third component seems to be present in all ULXs observed by \textit{NuSTAR} to date with sufficient statistics above 10 keV \cite{Walton2018a}. The nature of this component is not well understood, although it is thought to be produced either by scattering in an optically-thin corona or in the accretion column onto a magnetised neutron star (phase-resolved spectroscopy of the known ULX pulsars has found that the pulsed emission from the accretion columns dominates the highest energy emission in those sources \cite{Brightman2016b, Walton2018a, Walton2018b}). 

Detailed spectral modelling of ULXs is also often complicated by the presence of spectral residuals such as those around 1 keV (most likely from winds, see Sect. \,\ref{sec:hrxs}) and at lower energies possibly due to variability in the line-of-sight neutral column density. The latter is known to occur for some Galactic XRBs, particularly in HMXBs. Some absorption is certainly produced by the Galactic ISM with abundances close to Solar (albeit with a gradient towards the centre of the Milky Way \cite{Pinto2013}). In many cases, there is, however, excess absorption in ULX spectra from a few $10^{21} \rm \, cm^{-2}$ due to either the host galaxy ISM, the circumstellar medium (CSM) around the ULX \cite{Winter2007}, or even the ULX wind \cite{Middleton2015b}.
Further modelling complications might arise from non-Solar metallicity in the host galaxy (which is often sub-Solar for ULX hosts) or in the ULX CSM \cite{Winter2007}.

\subsection{Insights on the $L-T$ relation and spectral evolution}

The last decade has seen significant interest in attempts to understand how ULX spectra evolve, and to determine the physical origin of their spectral components. This was enabled by long-term monitoring primarily with \textit{Swift} and multiple deep observations with \textit{XMM-Newton}, \textit{NuSTAR} and, in a few cases, \textit{Chandra} and \textit{Suzaku}.
First attempts of accounting for the wind effects and the dependence on both mass accretion rate and line of sight showed encouraging results in the description of the trends between spectral hardness and variability power. The covariance spectra provided evidence on correlated variability mainly in the hard X-ray band with a shape similar to that of the hot ($\sim$ 1-3 keV) component \cite{Middleton2015a}.

\begin{figure}[h] %%%[b]
%\sidecaption
% For example, with the graphicx style use
\includegraphics[scale=.475]{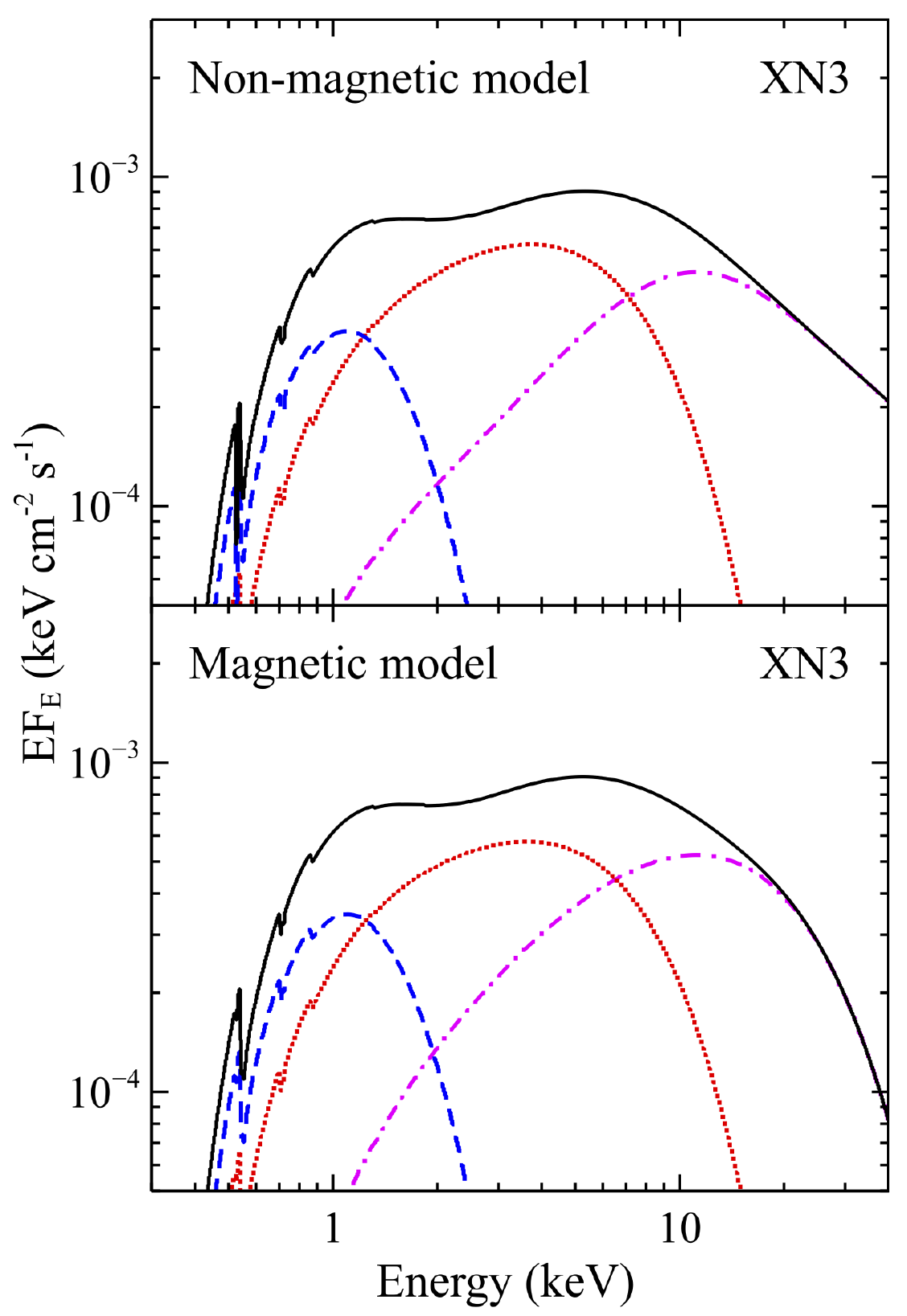}
\includegraphics[scale=.265]{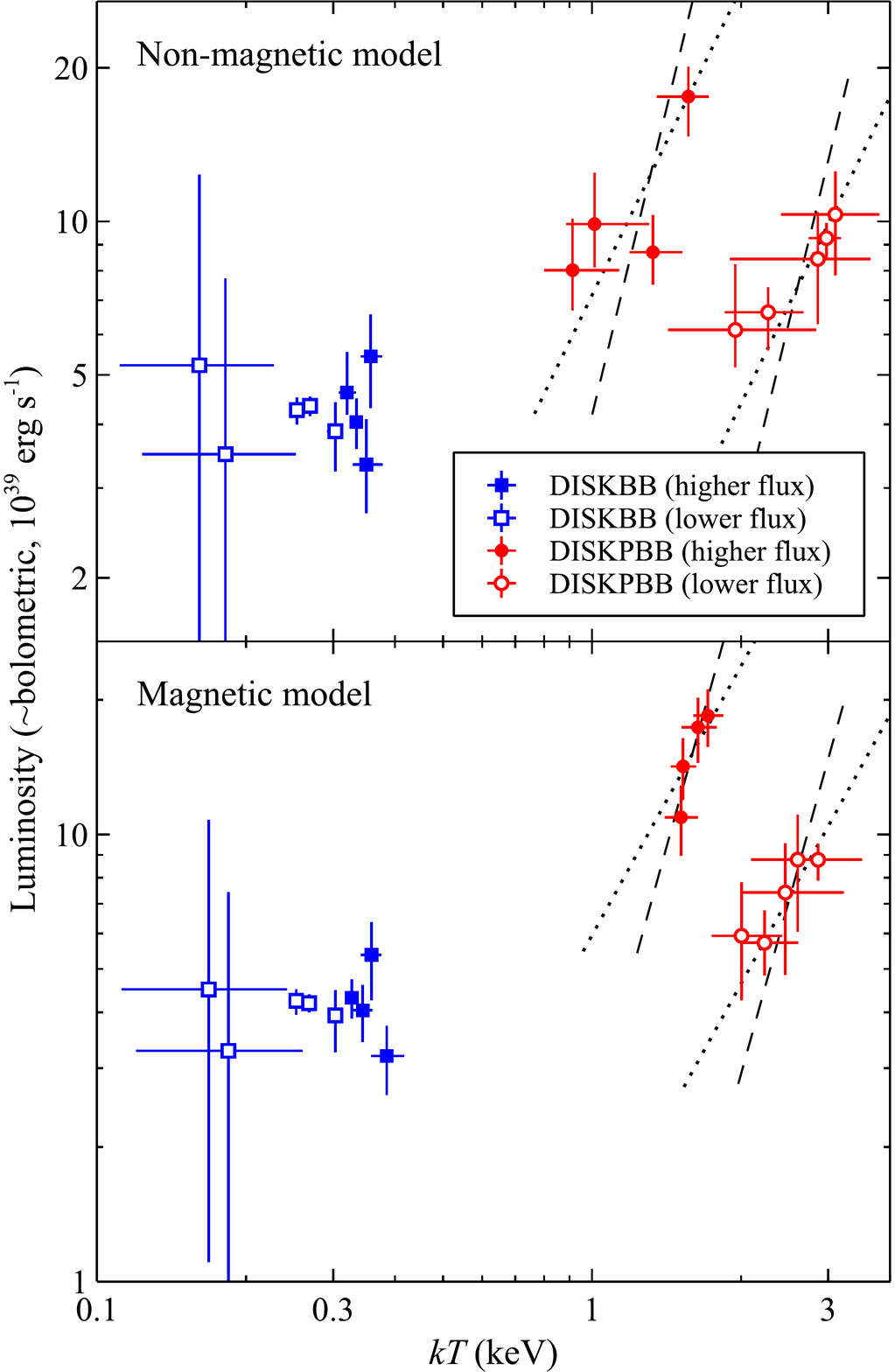}
%
% If no graphics program available, insert a blank space i.e. use
%\picplace{5cm}{2cm} % Give the correct figure height and width in cm
%
\caption{Left: Example model fits to the broadband spectra of NGC 1313 X-1 (a low-flux epoch) for two alternative accretor models. The total model is shown in solid black, the cooler {\scriptsize{DISKBB}} and the hotter {\scriptsize{DISKPBB}} components are shown in dashed blue and dotted red, and the highest energy component (either {\scriptsize{SIMPL}} or {\scriptsize{CUTOFFPL}}) is shown in dash-dotted magenta. Right: corresponding $L-T$ trends for 9 epochs. Epochs of low-flux (HUL) and high-flux (BBD) regimes are labelled. The {\scriptsize{DISKPBB}} component shows distinct tracks at high and low luminosities. Each track broadly agrees with the $L \propto T^2$ (thick disc) and $L \propto T^4$ (thin disc) predictions \cite{Walton2020}.}
\label{fig:broadband}       % Give a unique label
\end{figure}

In 2017, the deepest broadband view to date of a bright ULX, NGC 1313 X-1, was performed, combining substantial investments from \textit{XMM-Newton} (750 ks), \textit{NUSTAR} (500 ks) and \textit{Chandra} (500 ks). This allowed a detailed study of the evolution of each individual spectral component over a range of timescales spanning weeks to years. As noted above, the spectrum below 10\,keV was modelled with two thermal components (i.e. a \textsc{diskbb}+\textsc{diskpbb} combination in {\sc{xspec}}; both are multi-colour blackbody accretion disc models, but the latter also includes the radial temperature index as a free parameter). The softer thermal component (0.1-0.3 keV) component and the high-energy emission above 10 keV did not show strong variability. In contrast, the hotter (and dominant) thermal component ($\sim$ 1-3 keV) showed a surprising variability pattern which could not be described with a simple trend. Instead the data seemed to exhibit two correlated $L-T$ relationships which split above $\sim$10$^{40}$ erg s$^{-1}$, or alternatively at temperatures above $\sim$1.9\,keV \cite{Walton2020}. In Fig.\,\ref{fig:broadband} (left panel) two different modelling approaches for the high-energy continuum are shown: one assumes the source is a non-magnetised accretor (e.g. a BH) and that the highest energy emission arises in an up-scattering corona (with a power-law spectral profile), whilst the other assumes a magnetised accretor and that the highest energy emission arises in a ULX-pulsar-like accretion column\footnote{Here the emission from the accretion columns is treated empirically, assuming a \textsc{cutoffpl} model with parameters $\Gamma = 0.59$ and $E_{\rm{cut}} = 7.9$\,keV; this is based on the average parameters found for the pulsed emission from the known ULX pulsars via phase-resolved spectroscopy.}. Qualitatively similar results are seen, but the cause of this behaviour remains poorly understood.

The campaign on NGC 1313 was particularly productive since it also enabled the discovery of time lags (see Sect. \ref{sec:time-lags}) and variable winds (see Sect. \ref{sec:line-detections}) in X-1, as well as pulsations in X-2 \cite{Sathyaprakash2019}. A similar, follow-up, spectroscopic work on NGC 1313 X-2 focused on the full set of available observations with \textit{XMM-Newton} -- spanning a 17-year baseline -- in order to extend the variability range and constrain the $L-T$ relationship (NGC\,1313 X-2 is typically both softer and fainter than X-1, making it harder to study with other facilities). As X-2 is now known to be a pulsating ULX, a magnetic accretor model was then adopted (formally a \textsc{diskbb}+\textsc{diskbb}+\textsc{cutoffpl} model, with the parameters of the latter component fixed again to the values highlighted above). Trends of $L \propto T^{-4}$ and $L \propto T^{3}$ were obtained for the cooler and hotter thermal components, respectively, suggesting that the inner disc is not geometrically thin and that the cooler component is perhaps tracing an expanding disc photosphere or soft X-ray emission from the wind \cite{Robba2021}.

Another deep X-ray campaign was performed on the supersoft / SUL source NGC 247 ULX-1 with \textit{XMM-Newton} (800 ks). Starting in December 2019 and consisting of 8 uninterrupted exposures, it yielded the clearest view of an extremely soft ULX. These data revealed clear evidence of winds (see Sect. \ref{sec:physical-scan-absorption}) and quasi-periodicities (see Sect. \ref{sec:timing-qpo}). A detailed X-ray spectroscopic analysis was performed to study the spectral evolution over time and within various ranges of spectral hardness. The results show a clear anti-correlation between the temperature and radius of the cool blackbody component \cite{DAi2021} in agreement with an outflowing photosphere model \cite{Soria2016}, as well as with early results with different facilities \cite{Feng2016} and results for other ULSs \cite{Urquhart2016}.

More recent work performed on a sample of 17 ULXs observed with a range of X-ray facilities has suggested that the hardest sources might be powered by strongly-magnetised neutron stars with the high-energy emission likely dominated by the accretion column. This is supported by the strong variability of such components in pulsating ULXs. Softer sources may be explained by weakly magnetised neutron stars or black holes with strong outflows producing Compton down-scattering \cite{Gurpide2021a}.

%%%%%%%%%%%%%%%%%%%%%%%%%%%%%%%%%%%%%%%%%%%%%%%%%%%%%%%%%

\section{High-resolution X-ray spectroscopy}
\label{sec:hrxs}

Early analyses of the low-to-moderate resolution CCD spectra (particularly from \textit{XMM-Newton} / EPIC) with the continuum models discussed above showed evidence for atomic spectral features in the soft energy band ($<$ 2 keV), particularly near 1 keV \cite{Cropper2004,Goad2006,Stobbart2006}. Below 2 keV the resolving power of CCD spectra, $R = E / \Delta E \lesssim 20$, was not sufficient to resolve them. This led to several explanations including incorrect modelling of the metallicity of the ISM absorption component, the supernova remnants and a hot component of the ISM of the host galaxy. Detailed work on spatially-resolved X-ray spectra extracted with the high-spatial-resolution \textit{Chandra} / ACIS detector showed that the soft X-ray features seen in NGC 5408 X-1 must be associated with the ULX itself \cite{Sutton2015}.

Particularly striking is the similar shape of such residuals for different ULXs with high signal-to-noise \textit{XMM-Newton} / EPIC spectra. Later work showed that these can be modelled by absorption from a partly ionised medium, outflowing at $v \approx 0.1c$ \cite{Middleton2014}. This interpretation was further supported by the discovery of an anti-correlation between the strength of these features with the ULX spectral hardness, which hinted at a connection with the ULX accretion regime \cite{Middleton2015b}.

\subsection{Atomic lines as probes of winds}
\label{sec:wind-lines}

It is not an easy task to resolve and identify these spectral features with the current instruments. The grating spectrometers currently operational have a very good resolving power: the first order spectra of \textit{XMM-Newton} / RGS achieve $R \sim 100-600$ in the 0.3-2 keV band, whilst the gratings aboard \textit{Chandra}, especially HETGS, yield up to 1,000 around 1 keV and cover a broader (0.4-10 keV) band. However, owing to their thousands of energy channels, the counts per bin in grating spectra are an order of magnitude lower than in CCD spectra, thereby requiring exposures 10 times longer. Bright ULXs typically exhibit a fluency of a few $10^{-12}$ erg\,s$^{-1}$\,cm$^{-2}$ in the 0.3-10 keV band, which requires exposure times of at least 100\,ks with RGS and 3-4 times longer with \textit{Chandra} gratings to achieve a $3\,\sigma$ detection of a line with an equivalent width (EW) of a few eV and a flux $\sim 10^{-15}$ erg\,s$^{-1}$\,cm$^{-2}$ \cite{Kosec2021}. Moreover, whilst CCD spectra blend neighbouring lines (such as the whole Fe\,{\sc xx-xiv} / Ne\,{\sc ix-x} complex between 0.9-1.0 keV) the gratings spectra would show excess at much narrower energy ranges. Finally, a possible low-metallicity environment which is common amongst star-forming regions where ULXs are often observed might further dim any lines since they are mainly produced by heavy elements.

Searching for, resolving, and modelling individual lines in high-resolution X-ray spectra is a challenging task that requires knowledge of atomic physics, of the properties of the ISM along the line of sight and expertise in statistics and instrumental calibration. In Galactic XRBs, this is easier than in ULXs thanks to their proximity (at least three orders of magnitude smaller distances), leading to much higher quality spectra. Furthermore, winds can be blown by various phenomena in XRBs and, therefore, can exhibit different patterns of emission or absorption lines:

\begin{itemize}
\item{\textbf{Thermal winds} -- The hard X-ray photons from the inner disc heat the cooler gas in the outer discs of accreting compact objects, which can reach velocities above 100 km s$^{-1}$ and escape. The plasma is generally photoionised by the hard X-ray radiation field and exhibits a high ionisation parameter: log $\xi=L_{\rm ion}/n_{\rm H}R^2 \sim 4-5$ (where $L_{\rm ion}$ is the ionising luminosity, $n_{\rm H}$ is the number density and $R$ is the distance from the ionising source). They have been mainly studied through their Fe {\sc xxv-xxvi} lines at 6.6-7 keV with \textit{Chandra} / HETGS \cite{Ponti2012}. Similar, albeit cooler, winds are found in AGN and are typically referred to as \textit{warm absorbers}; these are thought to be produced by evaporation of the dusty torus \cite{Kallman2019}.}
\item{\textbf{Radiative winds} -- Here the momentum and/or energy of UV and soft X-ray photons is absorbed by the plasma, which is then pushed away. A well-known subsample are \textbf{stellar winds}, seen in hot stars where the momentum of photospheric photons is absorbed by thousands of spectral lines and, thus, transferred to the atmospheric plasma. These \textit{line-driven} winds can also be of hybrid ionisation due to collisions with the companion star or its wind \cite{Amato2021}. In nova outbursts larger velocities ($> 3,000$ km s$^{-1}$) can be reached but here the plasma is cooler ($\xi \sim 1-3$) due to their supersoft SED \cite{Pinto2012a}.
%%%Stellar winds produce strong emission lines in the soft X-ray band whilst the SS phase of novae is characterised by strongly saturated absorption lines but in both cases \textit{XMM-Newton} / RGS and \textit{Chandra} / LETGS are ideal analysis instruments.
UV lines are known to provide a strong contribution to radiative winds in AGN, particularly, quasars where relativistic velocities ($\gtrsim0.1c$) can be reached \cite{Proga2000,QB2020}. In these cases, they are also known as \textit{ultra-fast outflows} or UFOs. At super-Eddington accretion rates Thomson scattering becomes important and absorption of X-ray photons can be sufficient to push plasmas at similar or higher velocities \cite{Ohsuga2005}. Historically the study of these winds focused on the Fe K band \cite{Tombesi2010} as they typically have an ionisation degree similar to XRB thermal winds (and have thus mainly with CCD spectrometers).
%%%, as AGN grating spectra typically have low S/N at these energies). 
However, \textcolor{black}{recent, deep, grating observations} have enabled the detection of features from these fast winds in the soft X-ray spectra of AGN as well (primarily with \textit{XMM-Newton} / RGS \cite{Pinto2018a,Kosec2018c}).}
\item{\textbf{Magnetic winds} -- the role of magnetic fields is less well understood than radiation pressure in accretion disc winds. Although it seems that magnetic pressure is not necessarily required to launch winds in stellar-mass or supermassive compact objects \cite{QB2020,Tomaru2022}, \textcolor{black}{magneto-hydrodynamic (MHD) simulations have indeed shown that magnetic fields may help in lifting the plasma away from the mid-plane, and accelerate it over a wide range of velocities for} a broad range of BH masses \cite{Fukumura2017}. Current spectrometers lack the necessary spectral resolution to distinguish different line shapes (asymmetric for MHD winds) which would provide the ideal means to differentiate between launching models. Nevertheless, an alternative diagnostic comes from studying the temporal response of the wind lines to variations in the continuum. Although the results are pointing towards a radiation / thermal origin of the winds \cite{Tomaru2022,Pinto2018a,Matzeu2017} a thorough work on a sample of sources is required to place stronger constraints, especially given the potential detections of fast winds ($0.04-0.05c$) in some sub-Eddington XRBs \cite{King2012,King2014}.}
\end{itemize}

\subsection{Spectral codes for the study of photoionised winds}
\label{sec:photoionisation}

Early searches for spectral lines in ULXs focused on the high-energy band and, particularly, the Fe K transitions from Fe {\sc xxv-xxvi} mentioned above. Spectral fits of individual observations typically yield 90\,\% upper limits for the EWs of 10-50\,eV. In a few cases, weak ($<3\,\sigma$) evidence of the Fe 6.4\,keV fluorescence line has been reported \cite{Walton2017b,Mondal2021} which might show irradiation of a dusty CSM by the ULX.

However, owing to the soft SED of ULXs \cite{Pinto2020a} and the low statistics in the hard band (2-10 keV), it is obvious that the Fe K is the wrong place to search for strong lines in ULX spectra. The plasma is likely mildly rather than fully ionised, which means that most features should be from the Fe L species and other lower-temperature ions such as {O \sc vii-viii} and {Ne \sc ix-x}, i.e. the dominant species of photoionised gas in the soft X-ray band ($< 2$ keV). 

It is possible to determine whether the emitting gas is in collisional or photoionisation equilibrium through the well-known diagnostics tools known as the $r$ and $g$ ratios for the He-like triplets. These are defined as follows: $r (n_{\rm e}) = {F}/{I}$ and $g (T_{\rm e}) = ({F+I})/{R}$, where $F$, $I$ and $R$ are the fluxes of the forbidden, intercombination and resonance lines. In X-ray emitting plasmas low values of $r$ (or very weak forbidden lines) indicate high-density gas (a strong UV radiation field could be an alternative cause), whilst low values of $g$ are typical of low-temperature plasmas. For example, a combination of high $g$ and $r$ is common for the low-density photoionised gas that is often found in clouds around AGN \cite{Porquet2000}, whilst low values of $r$ and high values of $g$ are commonly observed in high-density photoionised atmospheres of discs in X-ray binaries \cite{Psaradaki2018}. Simultaneous low $g$ and $r$ ratios are typical of low-density collisionally-ionised plasmas in the interstellar- and intra-cluster medium or shocked gas in jets, such as those observed in the Galactic microquasar SS 433 \cite{Marshall2002}. It is therefore more common to use photoionised plasma models for emission or absorption lines from plasmas in accretion discs around compact objects.

\begin{figure}[h] %%%[b]
\includegraphics[scale=.355]{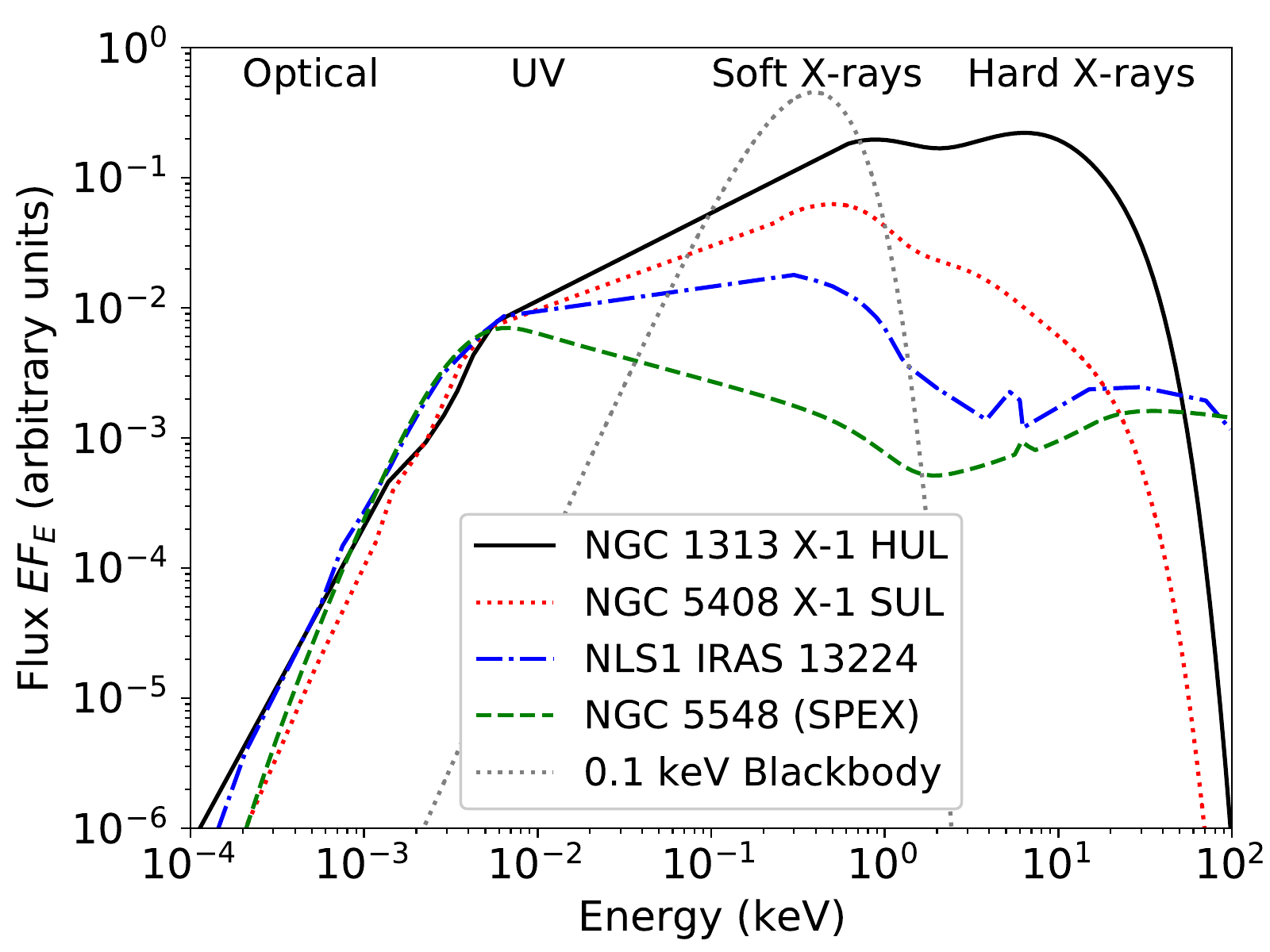}
\includegraphics[scale=.38]{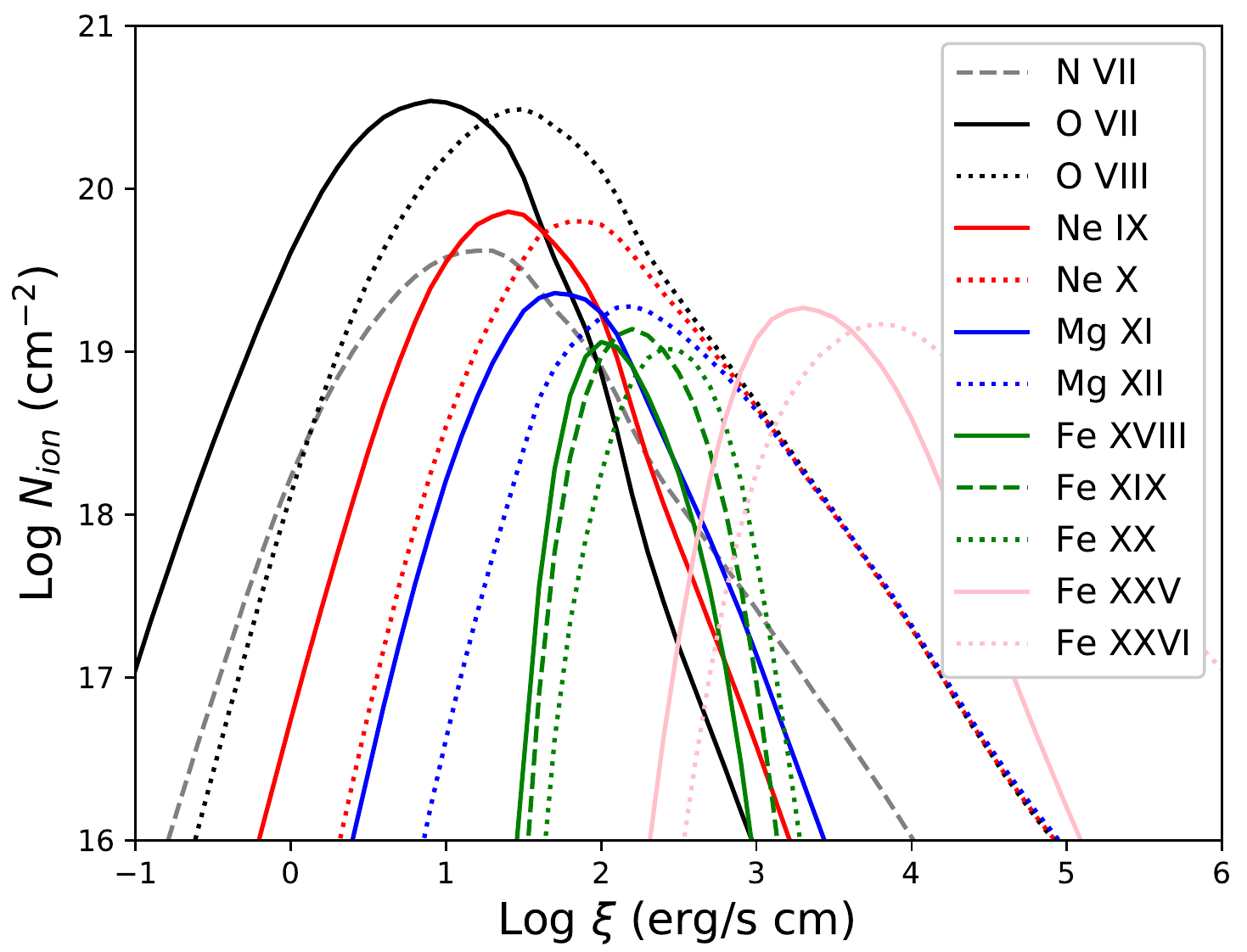}
\caption{Left panel: a comparison between the SEDs of two ultra-luminous state regimes (HUL and SUL) with those from the Sy1 NGC 5548, the NLSy1 IRAS 13224-3809 and a 0.1 keV blackbody model (the latter mimics the spectrum of a tidal-disruption event or a nova SS phase \cite{Pinto2020a}). Right panel: ionic column densities vs. ionisation parameter computed for NGC 1313 X-1 HUL spectrum.}
\label{fig:sed_columns}       % Give a unique label
\end{figure}

Fig. \ref{fig:sed_columns} (left panel) shows the spectral energy distributions (SEDs) of the hard and soft ultra-luminous spectra of the archetypal ULXs NGC 1313 X-1 and NGC 5408 X-1 compared with the SEDs of two different AGN, namely the Seyfert\,1 galaxy (Sy1) NGC 5548, and the high-accretion Narrow Line Seyfert\,1 (NLSy1) IRAS 13224-3809. We also show a 0.1 keV blackbody model mimicking the spectrum of a typical tidal-disruption event or that of a nova SS phase. ULXs are characterised by a strong soft X-ray excess and very little flux above 20-30 keV when compared to AGN. The SEDs of HUL, SUL and BD regimes (although the latter is not shown directly) are more similar to those of NLSy1s whilst SSUL SEDs are somewhere in between that of a NLSy1 and the 0.1 keV blackbody (see Fig. \ref{fig:SEDs}).

There are multiple currently available spectral codes that enable the computation of photoionisation balance. The most widely used are {\sc{spex}} \cite{Kaastra1996}, {\sc{cloudy}} \cite{Ferland1998} and {\sc{xstar}} \cite{Kallman2001} (see Chapter 7, `An Overview of Astrophysical Plasmas' by Timothy Kallman). {\sc{spex}}, in particular, is a full end-to-end package that includes atomic data, computation of different plasma conditions (e.g. photoionisation, collisional ionisation, out-of-equilibrium, etc.) and spectral modelling. Furthermore, the {\scriptsize{PION}} code available in {\sc{spex}} permits an instantaneous calculation of the balance and modelling of a given spectrum without the need to assume an SED \textit{a priori}. We have applied {\scriptsize{PION}} to the HUL spectrum of NGC 1313 X-1, assuming a column density $N_{\rm H} = 10^{24} \rm cm^{-2}$ and Solar abundances for illustration. Fig. \ref{fig:sed_columns} (right panel) shows the corresponding column density of each ion for a broad range of ionisation parameters. The dominant species are {O \sc vii-viii} and {Ne \sc ix-x} for log $\xi$ up to 3. At log $\xi > 3$, the Fe ions are dominant \textcolor{black}{amidst those that are not fully ionised} but their column densities distribute amongst many ionic species which implies that their individual lines are likely weak \textcolor{black}{(due to the very soft SEDs)} and difficult to detect.

\begin{figure}[h] %%%[b]
\includegraphics[scale=.35]{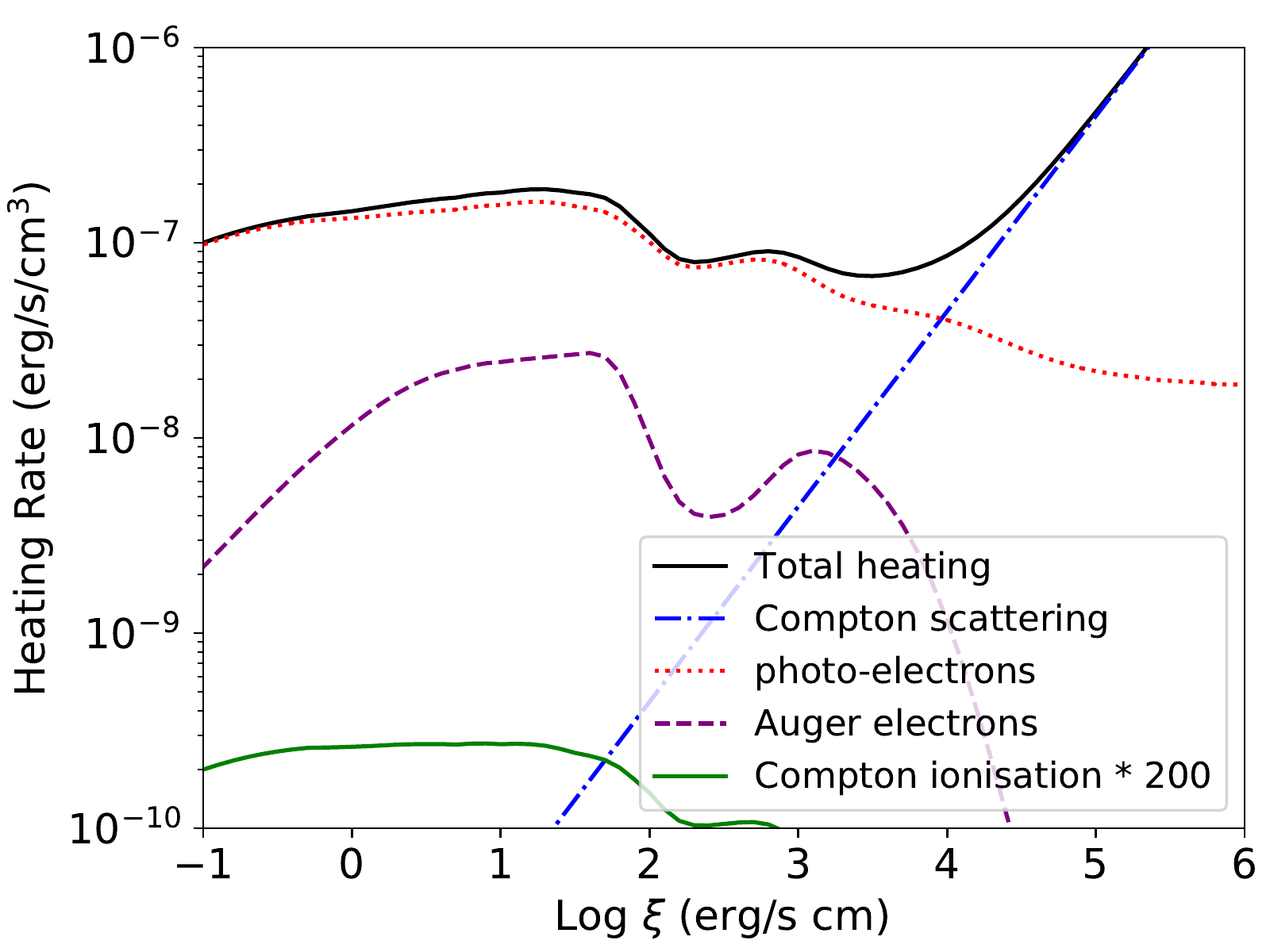}
\includegraphics[scale=.35]{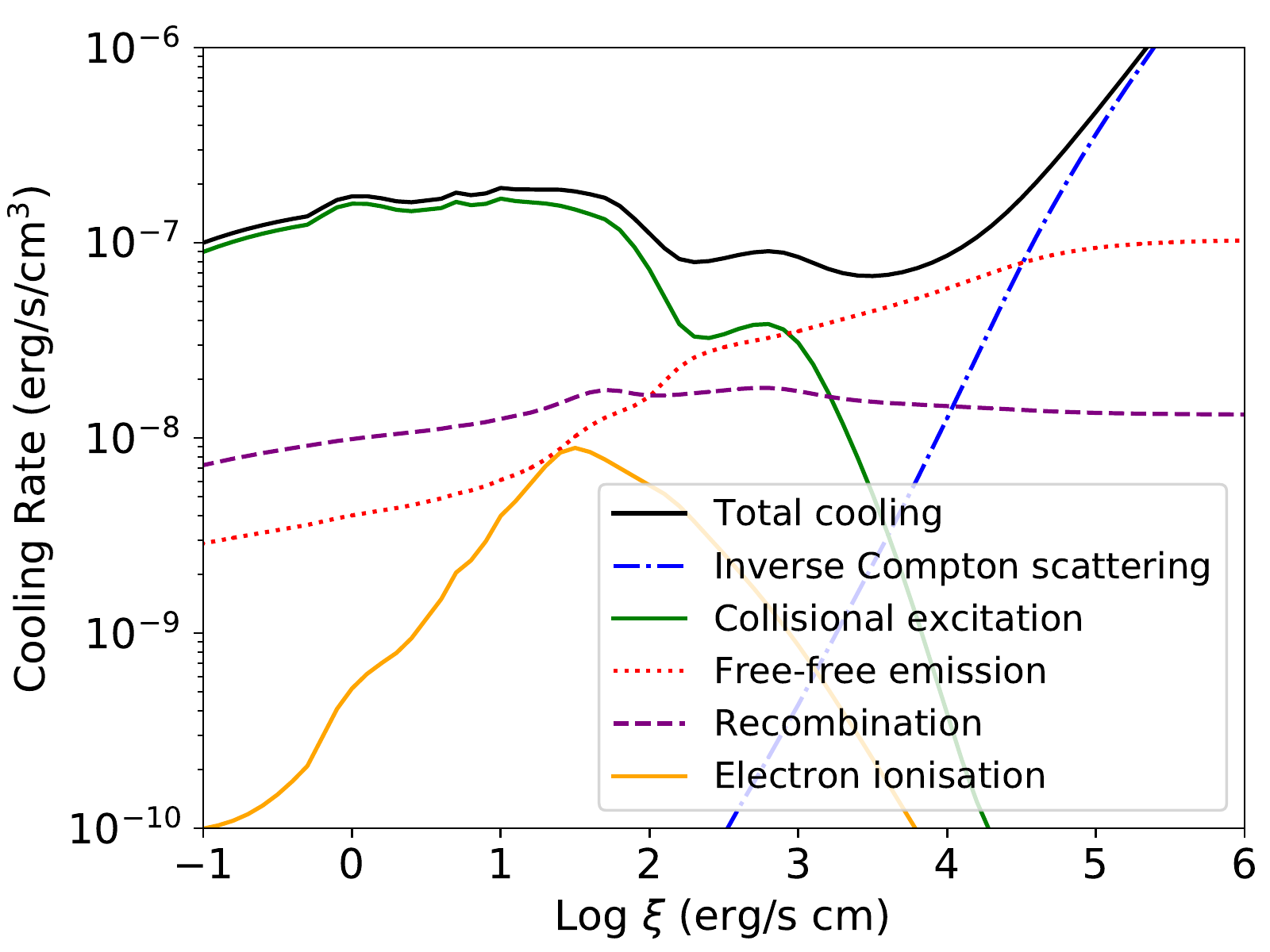}
\caption{Heating rates (left panel) and cooling rates (right panel) calculated for NGC 1313 X-1 (HUL) with the {\scriptsize{PION}} code available in {\sc{spex}} \cite{Mehdipour2016,Pinto2020a}.}
\label{fig:rates_heat_cool}       % Give a unique label
\end{figure}

Modern photoionisation codes provide details on the main heating and cooling processes that are involved in determining plasma equilibrium. In Fig. \ref{fig:rates_heat_cool} we show the heating and cooling rates as computed with {\scriptsize{PION}} for the NGC 1313 X-1 HUL spectrum assuming a gas density $n_{\rm H} = 10^{8} \, \rm cm^{-3}$ \cite{Pinto2020a}. The solid black lines corresponding to the total cooling and heating rates are almost indistinguishable, indicating that convergence to an equilibrium solution was found. At low $\xi$, the dominant processes are heating by photoelectrons and cooling by collisional excitation. At high $\xi$, the largest contribution to heating (cooling) is given by (inverse) Compton scattering. Recent comparisons between the different photoionisation codes {\sc{spex}}, {\sc{cloudy}} and {\sc{xstar}} showed a 10-30\,\% difference in the corresponding concentration of the ionic species. However, these differences are unlikely to significantly impact the results and implications \cite{Mehdipour2016}.

\begin{figure}[h] %%%[b]
\includegraphics[scale=.35]{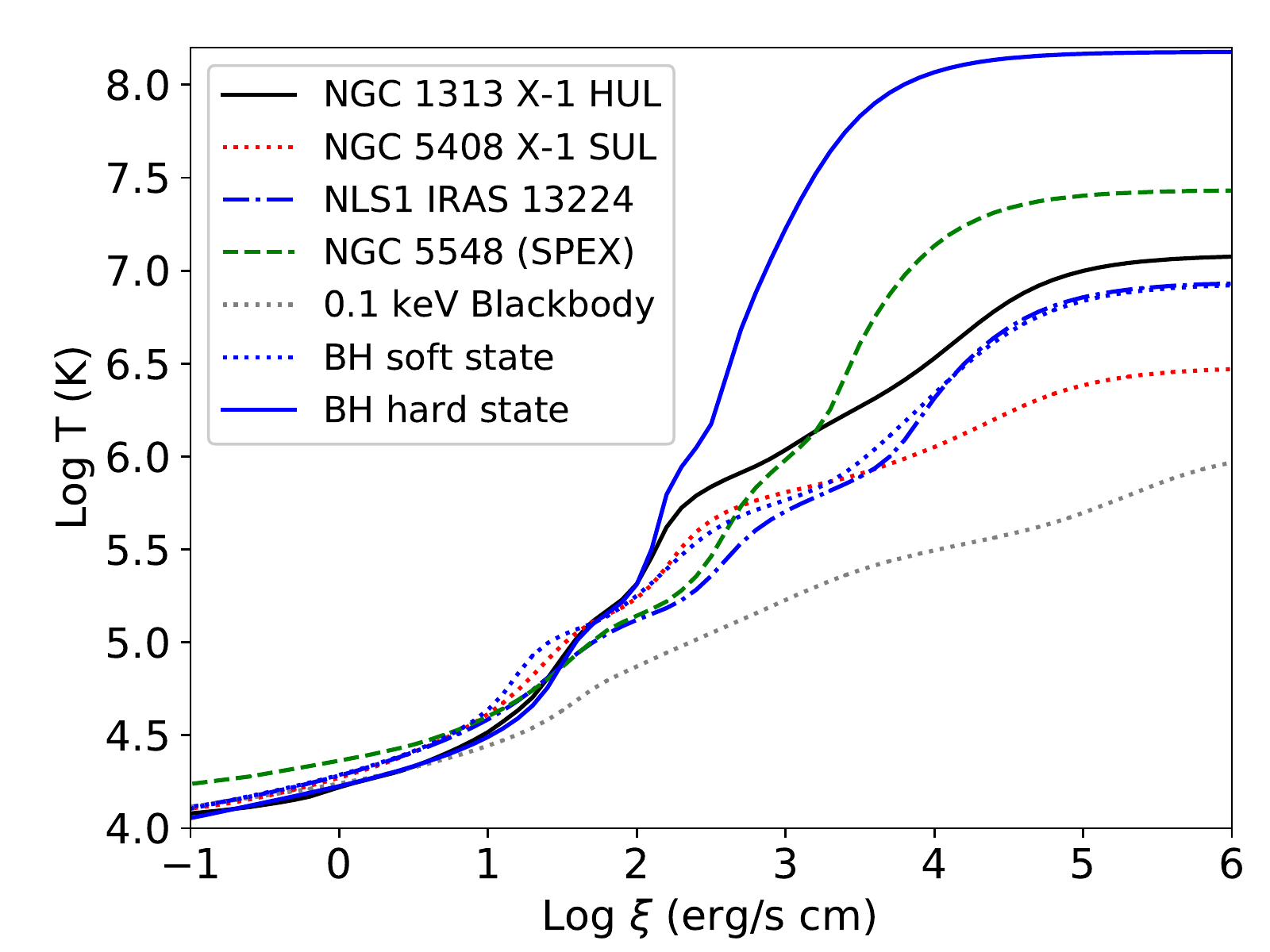}
\includegraphics[scale=.35]{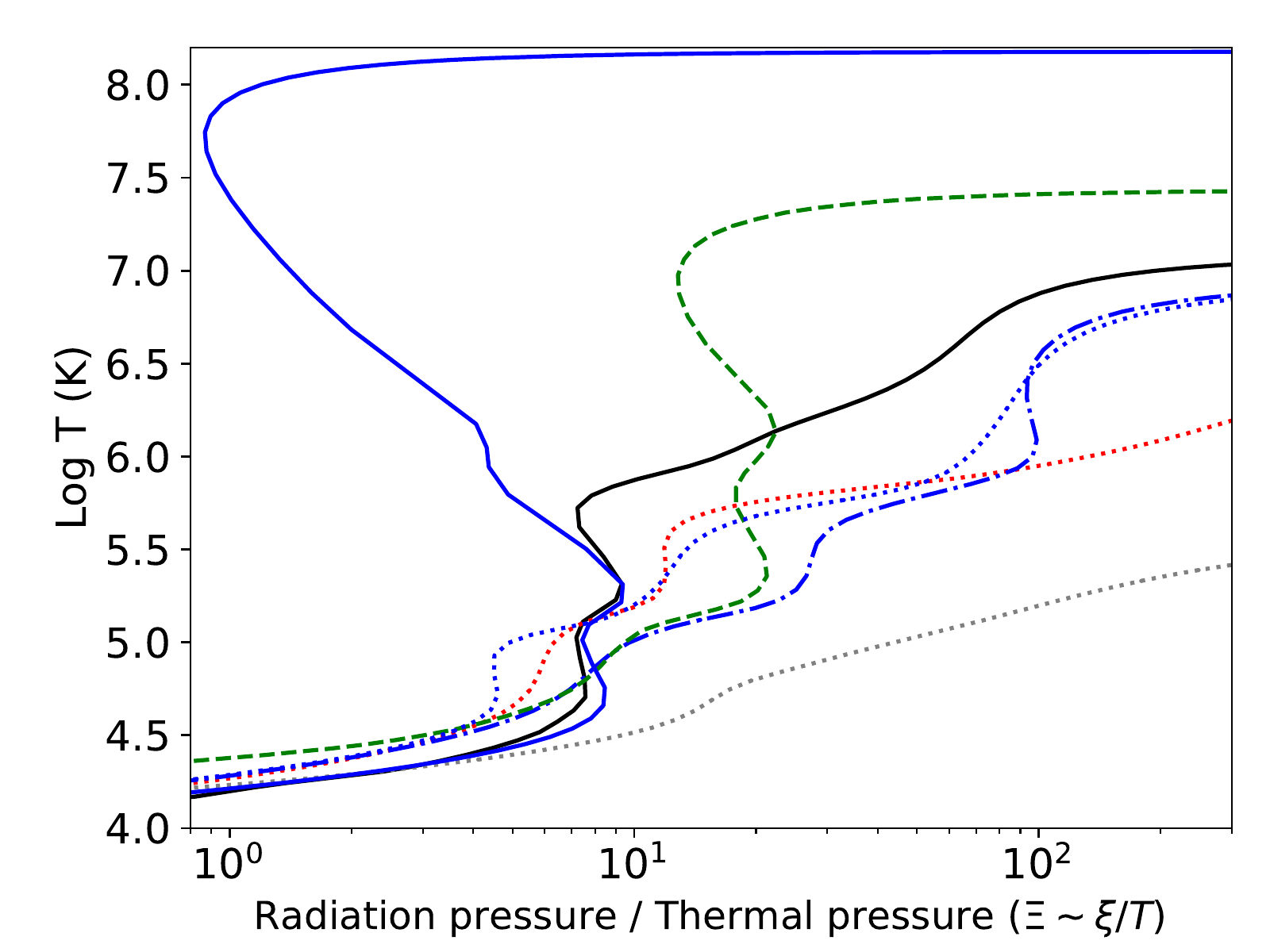}
\caption{Left panel: curves of ionisation balance (k$T-\xi$) for a variety of SEDs from Galactic XRBs, ULXs and AGN (see Fig. \ref{fig:SEDs} and \ref{fig:sed_columns}). Right panel: corresponding stability curves (k$T-\Xi$) with negative slopes indicating unstable branches \cite{Pinto2020a}.}
\label{fig:ion_bal}       % Give a unique label
\end{figure}

The ionisation balance of photoionised plasmas is usually described with the relationship between the temperature and the ionisation parameter (see Fig. \ref{fig:ion_bal}, left panel). These curves correspond to the computation for the SEDs shown in Fig. \ref{fig:sed_columns} (left panel) and the hard/soft XRB state in Fig. \ref{fig:SEDs} (left panel); the branches where the curves flatten indicate where Compton heating is important and show the temperature expected for a thermal wind. SEDs of sources with harder X-ray spectra, e.g. Sy1 NGC 5548 and the XRB hard state, are characterised by excess heating and high Compton temperature. A better visualisation of the ionisation balance is given by the stability curves (or `$S$ curves'). These show the temperature or the ionisation parameter as a function of the ratio between the radiation pressure ($F/c$) and thermal pressure  ($n_{\rm H}kT$), which is given by $\Xi = F / n_{\rm H} c kT = 19222 \, \xi / T$, with $F = L / 4 \pi R^2$ \cite{Krolik1981}. Along these curves heating equals cooling and, therefore, the gas is in thermal balance. On the left side of the $S$ curve, cooling dominates overheating, whilst on the right side, heating dominates over cooling. Where the $S$ curve has a positive gradient, the photoionised gas is thermally stable, i.e. small perturbations upwards (downwards) will be balanced by an increase of cooling (heating). Negative gradient branches are characterised by plasma that is thermally unstable to temperature perturbations. The stability curves computed for our representative SEDs are shown in Fig.\,\ref{fig:ion_bal} (right panel) and underline how hard X-ray spectra correspond to $S$ curves with large regions of instability. Photoionisation modelling of ULX X-ray spectra does indeed show that the wind is found preferentially along thermally stable branches \cite{Pinto2020a}. This occurs also for thermal winds in XRBs, which appear mainly in soft states with highly stable SEDs \cite{Ponti2012}, and AGN warm absorbers \cite{Kallman2019}. For more detail on ULX wind detections and properties see Sect. \ref{sec:line-detections} and \ref{sec:SEdd}.

\subsection{The quest for spectral lines in X-ray spectra}
\label{sec:line-detections}

The first well-exposed ($\sim 100$\,ks) RGS spectrum of Holmberg II X-1, one of the 5 brightest ULXs (in flux), showed an excess of counts at $569 \pm 9$ eV. With an equivalent width of $28 \pm 12$ eV, it was consistent with emission from a blend of lines from the He-like {\ovii} triplet (at
561-574 eV or 21.6-22.1 {\AA}), which would correspond to a luminosity of $\sim10^{38}$ erg\,s$^{-1}$\cite{Goad2006}. However, given the low significance of the feature, it was interpreted as being due to incorrect modelling of the neutral O edge in the ISM model caused by the assumption of Solar metallicity.

\begin{figure}[h] %%%[b]
\includegraphics[scale=.44]{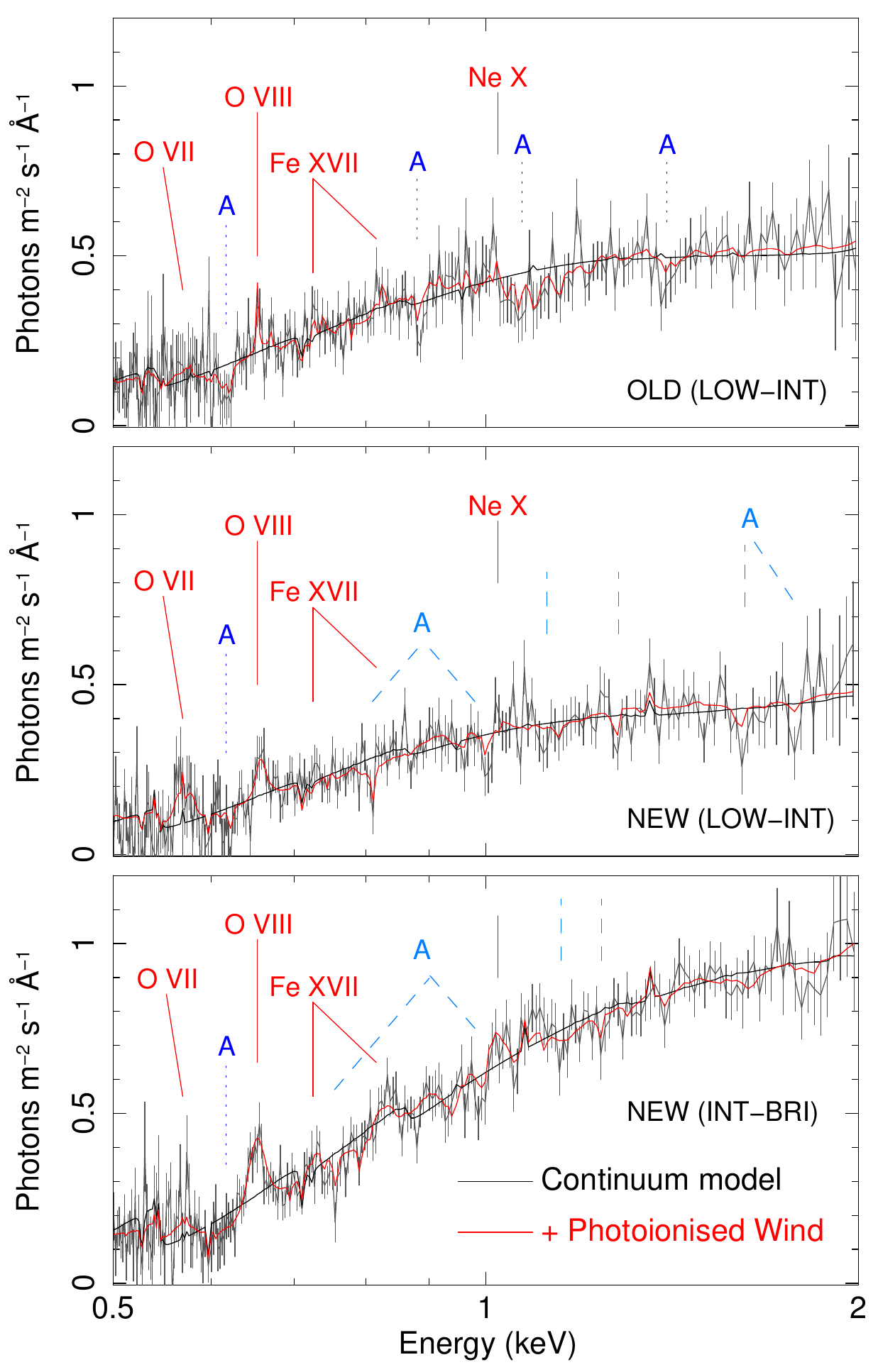}
\includegraphics[scale=.65]{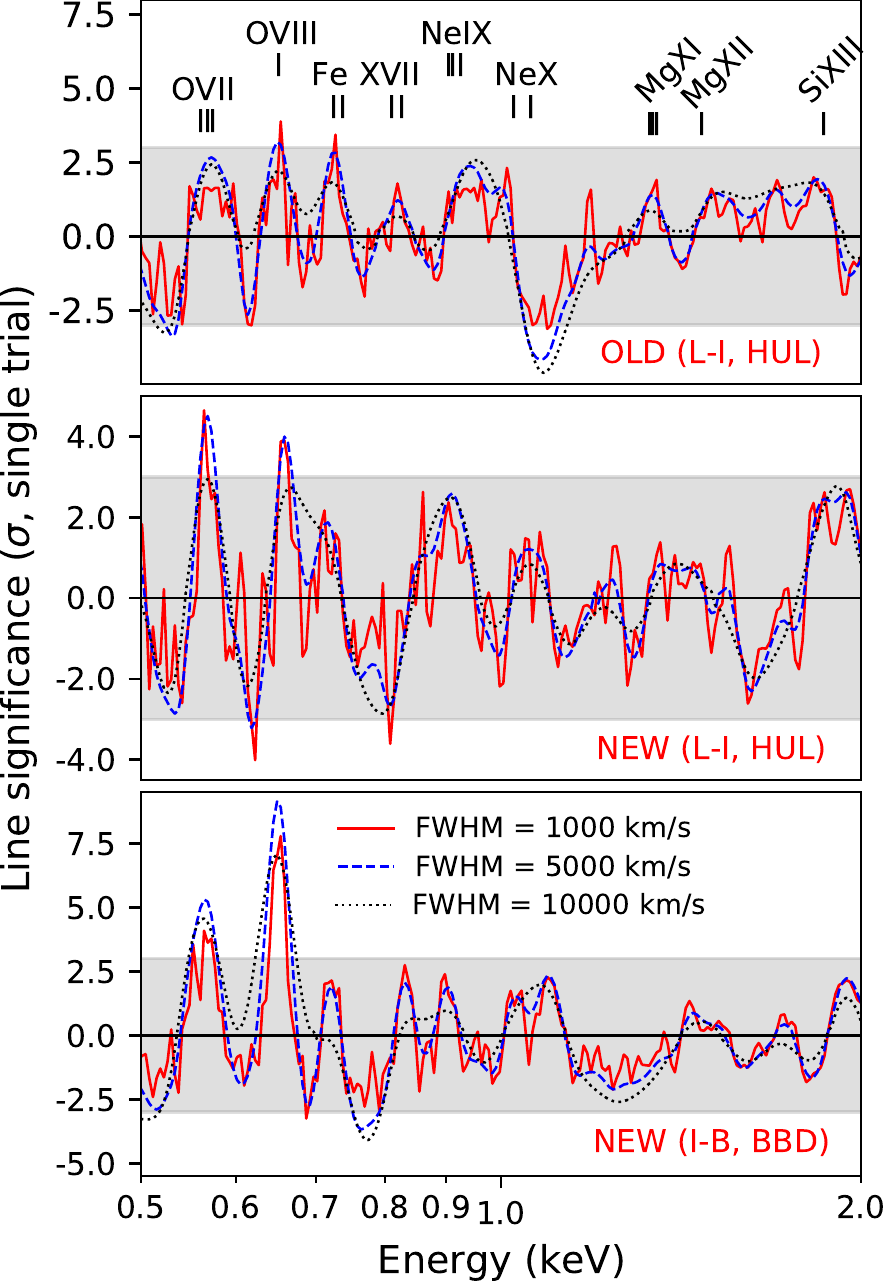}
\caption{\textcolor{black}{Left panel: Flux-resolved \textit{XMM-Newton} RGS spectra
%%%(0.45$-$2 keV), EPIC-pn (2$-$10 keV) and time-averaged \textit{NUSTAR} (10$-$20 keV) spectra of NGC 1313 X-1. From top to bottom: RGS and pn spectra
obtained by stacking the archival (2006-2012) HUL spectra \cite{Pinto2016}, low-to-intermediate flux observations from 2017 (also HUL) and intermediate-to-high flux observations from 2017 (BBD) \cite{Pinto2020b}. The black (red) lines show the continuum (best-fit photoionised wind) models. The strongest emission lines are indicated with solid red labels. Dotted blue (dashed light-blue) lines refer to previously known and confirmed (new) absorption features. Right panel: corresponding gaussian line scans performed on the three spectra for three different line widths (FWHM). The line significance is calculated as square root of the $\Delta C$-stat times the sign of the gaussian normalisation.}}
\label{fig:first_detection}  % Give a unique label
\end{figure}

The first major step in this field was achieved through the stacking of several observations totalling over 300 ks with the high spectral resolution RGS aboard \textit{XMM-Newton}. Such stacking was possible in two sources, NGC 1313 X-1 and NGC 5408 X-1, and enabled the 1 keV features in the CCD-resolution spectra of ULXs to be resolved into a forest of emission and absorption lines \cite{Pinto2016}. The emission lines are found near their laboratory wavelengths but their strength and position slightly vary with the source luminosity and spectral hardness \cite{Pinto2020b}, which indeed confirms they are associated with the ULX itself (see Fig. \ref{fig:first_detection}, left panel). When modelled with line-emitting plasma in either photoionisation or collisionally-ionisation equilibrium they yield X-ray luminosities of $L_{\rm X}\sim10^{38}$ erg\,s$^{-1}$, orders of magnitude brighter than the emission lines detected in the sub-Eddington Galactic XRBs \cite{Amato2021}. The strong absorption features above (below) 1 keV in NGC 1313 X-1 (NGC 5408 X-1) could not be identified with the laboratory energies of the dominant transitions. In order to find the exact line centroids and compute possible Doppler shifts, a spectral scan was performed using a moving gaussian line with a step of 250 km s$^{-1}$, about 1/3 of the RGS spectral resolution. Fig. \ref{fig:first_detection} (right panel) shows the results of this scan for three flux-resolved spectra of NGC 1313 X-1 adopting three different line widths. Apart from the {\ovii} He-like triplet, {\oviii} and {\nex} Ly-$\alpha$ emission lines, there is a strong absorption feature bluewards of the {\nex} line ($\gtrsim1$ keV) and other minor absorption-like features between 0.6-0.9 keV. Modelling with absorption from a photoionised plasma, e.g. {\scriptsize{PION}} in {\sc{spex}}, indicated ionic species identical to those seen in emission but requiring a huge blueshift of about 0.2\,$c$ in both ULXs \cite{Pinto2016}. Such velocities were never observed in Galactic Eddington-limited XRBs and likely reveal the long-sought powerful winds predicted by theoretical simulations of radiation pressure in super-Eddington accretion discs \cite{Ohsuga2005}. The combined significance of the absorption lines is between 4-5\,$\sigma$.

This discovery opened up a completely new avenue for studies of ULXs,
and a significant amount of work has since been undertaken to search for lines in other ULX spectra or with different detectors. An Fe K counterpart to the outflow in NGC 1313 X-1 was found in the form of an absorption feature at $8.77 \pm 0.06$ keV thanks to the combination of the imaging spectrometers aboard \textit{XMM-Newton} and \textit{NuSTAR} \cite{Walton2016a}. The outflow velocity inferred is somewhere in the range 0.2-0.25\,$c$, depending on whether {Fe \sc{xxv}} or {Fe \sc{xxvi}} is the dominant ion. Its equivalent width of $-61 \pm 24$ eV is comparable to the strongest Fe absorption line seen from a BH binary \cite{King2012}. Further work on low-to-moderate resolution detectors confirmed the presence of blueshifted absorption and emission features in other ULXs \cite{Wang2019, Brightman2022}. \textit{Chandra} gratings have a much lower effective area than the RGS, especially at $E \lesssim 1$ keV, requiring much longer and nearly uninterrupted exposures for ULX spectra (owing to the line variability \cite{Pinto2020b}). However, in 2017 the transient neutron star Swift J0243.6$+$6124 went through a super-Eddington outburst, shining above $10^{39} \rm \, erg \, s^{-1}$ in the 0.5-10 keV band, and thus became the first Galactic ULX pulsar. The proximity of the source meant it was an excellent target for \textit{Chandra} gratings and, indeed, observations with HETGS unveiled the presence of strong absorption features compatible with a 0.22\,$c$ outflow \cite{vdEijnden2019}. 

A thorough search for and identification of spectral lines in ULXs has been performed mainly with the \textit{XMM-Newton} / RGS spectrometers thanks to its combination of good effective area (in the context of grating spectrometers) and high spectral resolution. ULXs characterised by SUL / SSUL regimes, such as NGC 55 ULX-1, tend to show more lines, both in absorption and emission \cite{Kosec2021}. The first pulsating ULX to show a relativistic outflow was NGC 300 ULX-1 \cite{Kosec2018b}, which indicated that the dipole magnetic field of the central neutron star has to be sufficiently small ($B \lesssim 10^{12}$ G) to avoid dramatic truncation of the super-Eddington inner disc and decrease of the Thomson cross section \textcolor{black}{(i.e. a weaker the wind)}. For more detail on the structure of the super-Eddington disc see Sect. \ref{sec:super-Eddington}. NGC 55 ULX-1 and NGC 5204 ULX-1 are amongst the sources with the best example of blueshifted emission lines (0.1-0.3\,$c$) indicating either a conical (wind) or a collimated (jet) outflow \cite{Pinto2017,Kosec2018a}.

\subsubsection{\textcolor{black}{Line detection and significance estimation}}
\label{sec:line_developments}

%%% Spectral lines in high-resolution X-ray spectra are often searched for by visual inspection and by focusing on the well-known He- / H-like ionic transitions such as {\oviiviii}, {\neixx} and {\nvii} below 1 keV and heavier ions, e.g. {\mgxixii}, {\sixiiixiv} and {\fexxvxxvi} in the hard ($>1$ keV) band. These species are particularly favoured in the case of photoionised winds (see Fig.\,\ref{fig:sed_columns}, right panel), which are expected given the strong X-ray emission in ULX inner discs. 
The detected ULX winds reach very high velocities as expected from radiation pressure in super-Eddington accretion discs. However, these Doppler shifts can complicate the line detection and identification
%%% Although a spectral scan with a moving Gaussian line is a common procedure which indeed identified many lines in ULX RGS spectra (see Sect. \ref{sec:line-detections}), there is an important issue with 
owing to the chance of finding spurious features associated with Poisson noise (the so-called \textit{look-elsewhere} effect; L-E effect hereafter), imperfections in the instrumental calibration, etc. In particular, for lines that are not located at laboratory wavelengths a simple F-test is no longer a valid method to robustly estimate the detection significance \cite{Protassov2002}.

A standard procedure that is adopted to account for the L-E effect and estimate the false-alarm probability of any line detection is Monte Carlo (MC) method. MC consists in simulating a large number of spectra adopting a featureless continuum as a template model. The simulated spectra are then searched with a Gaussian line scan identical to the one used for the real data. This method is very accurate and has been used extensively, providing robust significance estimates with the best cases yielding confidence levels (CLs) between 3-4$\sigma$ for individual lines in ULXs with the best-quality data \cite{Pinto2020b,Kosec2018b}. This procedure is however expensive in terms of computational costs due to the need for a huge number of simulated spectra to scan for an individual observation of one source, e.g. 10,000 to estimate a CL $\gtrsim3\,\sigma$, which could easily last \textcolor{black}{a week} on one CPU (depending on the continuum complexity).

\begin{figure}[h] %%%[b]
\includegraphics[scale=.45]{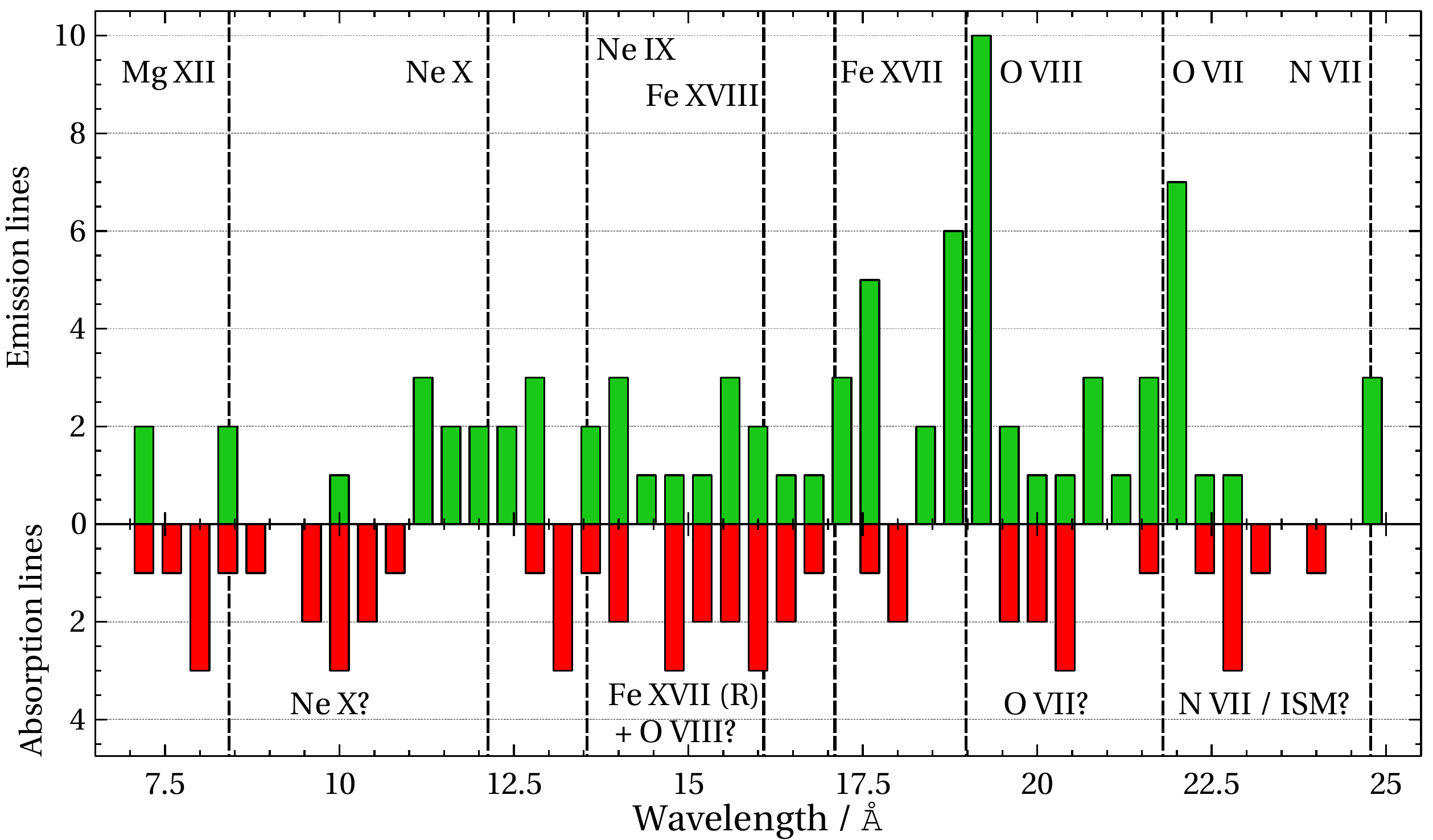}
\caption{Histograms of the emission (green) and absorption (red) lines detected in a ULX sample \cite{Kosec2021}. Labels show the identification of frequent emission lines, mainly detected at their laboratory wavelengths. The absorption lines are likely blueshifted.}
\label{fig:line_catalog}       % Give a unique label
\end{figure}

More recently a new technique has been developed which allows line searches in X-ray spectra to be undertaken with computing times that are 4 orders of magnitude shorter than spectral fitting. This consists of calculating the \textit{cross correlation} (CC) between a spectrum (the real data or a simulation) and a model, e.g. a Gaussian line model calculated using the same energy band and response matrix file for a grid of energy centroids. The false-alarm probability is then estimated in a similar manner to the MC technique, as the fraction of occurrences in which (re-normalised) CC values of the simulated spectra are equal to or higher than those in the real data. An easier version of this computation makes use of the spectral residuals of the observed spectrum (or the simulated spectrum) with respect to the continuum model, which can be cross-correlated with a grid of Gaussian models with energy centroids spanning the energy range of interest (0.3-2 keV for the RGS). With the dramatic improvement in the computing time, it became feasible to search for lines across the multi-epoch data from a reasonably large sample of 19 ULXs, which resulted in the detection of narrow (FWHM $\lesssim1,000$\,km s$^{-1}$) lines in $>60$\,\% of the sample \cite{Kosec2021}. Fig. \ref{fig:line_catalog} shows the histogram of the spectral lines detected in this sample of ULXs. The emission lines (green bars in the upper window) are mainly found near the laboratory wavelengths of dominant He- and H-like transitions. Most absorption lines (red bars in the lower window) are instead found away from the rest-frame wavelengths, which likely means that they are blueshifted.

\subsection{Physical models: parameter-space scan}
\label{sec:physical-scan}

The significance of individual spectral lines in the top-quality ULX high-resolution spectra is of about 4\,$\sigma$ (for nearby sources with exposure times above 300\,ks), but in most cases is $\lesssim$\,3\,$\sigma$. This is due to the moderate fluxes observed from ULXs (typically $\lesssim$ 10$^{-11}$ erg s$^{-1}$ cm$^{-2}$), which is due to their large distances ($\gtrsim$ 2 Mpc). 

The most obvious way to improve the detection and identification of winds in ULXs is to combine the statistics of multiple lines through the use of physical models of line emission or absorption. There has been a substantial investment of work in the last decade in order to improve atomic databases, particularly thanks to the availability of new laboratory measurements, theoretical calculations and high-resolution spectrometers (both gratings and microcalorimeters). The current uncertainties on the oscillator strength of the main spectral lines (typically from H- / He-like ions) are of about 10-20\,\% \cite{Mehdipour2016}. This is sufficiently small that it is possible to reasonably describe outflows with physical plasma models and, therefore, use these models in order to leverage the full range of emission/absorption lines predicted.
 
In the last 5 years a major improvement in the search for winds in ULXs was achieved thanks to the development of new techniques which perform a full and uniform search of the parameter space using physical models of plasmas in either collisional or photoionisation equilibrium. Typically, these scans involve creating a grid of line-of-sight velocities (e.g. Doppler shift) and key plasma properties: kT for collisional ionisation equilibrium (CIE) or $\xi$ for photoionisation equilibrium (PIE). These velocity and state parameters are fixed to each combination in the generated grid in turn, and the corresponding absorption model is then fit to the data allowing the column density of the absorbing or emitting plasma to vary. The scan of this parameter space is very accurate as it fits multiple lines simultaneously, and also avoids the risk of getting stuck in a local minimum, which is common to simple spectral fitting. This technique has been validated by applying it to well-known cases in bright Galactic XRBs with high-statistic line-rich high-resolution spectra, such as the CIE emission lines produced by the relativistic collimated jets in SS 433 and observed with \textit{XMM-Newton} / RGS \cite{Pinto2021}.

\subsubsection{Line-emitting gas}
\label{sec:physical-scan-emission}

In most cases, the emission lines in ULX spectra do not exhibit large (i.e. relativistic) Doppler shifts, although a shift of $\sim$1,000 km s$^{-1}$ is often detected. The exact position and strength of the emission lines can vary slightly over time and in response to the continuum variations. For NGC 1313 X-1 a deep observational campaign was undertaken in 2017 with the goal of understanding the line evolution with the spectral regime and the nature of the line-emitting plasma (see also Fig. \ref{fig:first_detection} middle and bottom panel). As previously discussed in Sect. \ref{sec:photoionisation}, the X-ray spectral lines from hot plasmas near the accretion discs of Galactic X-ray binaries and AGN exhibit diagnostic line ratios ($g$ and $r$) that are consistent with emission lines from photoionised plasmas. The RGS data available for NGC 1313 X-1 is so deep that the emission from the He-like {\ovii} triplet (resonance line at 21.6\,{\AA}, intercombination line at 21.8\,{\AA} and forbidden line at 22.1\,{\AA}) can be resolved. Accounting for uncertainties in the line widths, the continuum and the RGS calibration, the following ratios were estimated: $r = 0.6\pm_{0.4}^{1.0}$ and $g=1.3\pm_{0.8}^{1.6}$ \cite{Pinto2020b}. Despite the large uncertainties, it is clear that they are compatible with a high-density hybrid gas ($n_{\rm H} \sim 10^{10-12}$ cm$^{-3}$) where both recombination and collisional processes occur \cite{Porquet2000}, at least for the cool {\ovii} emitting plasma. 

\begin{figure}[h] %%%[b]
\includegraphics[scale=.325]{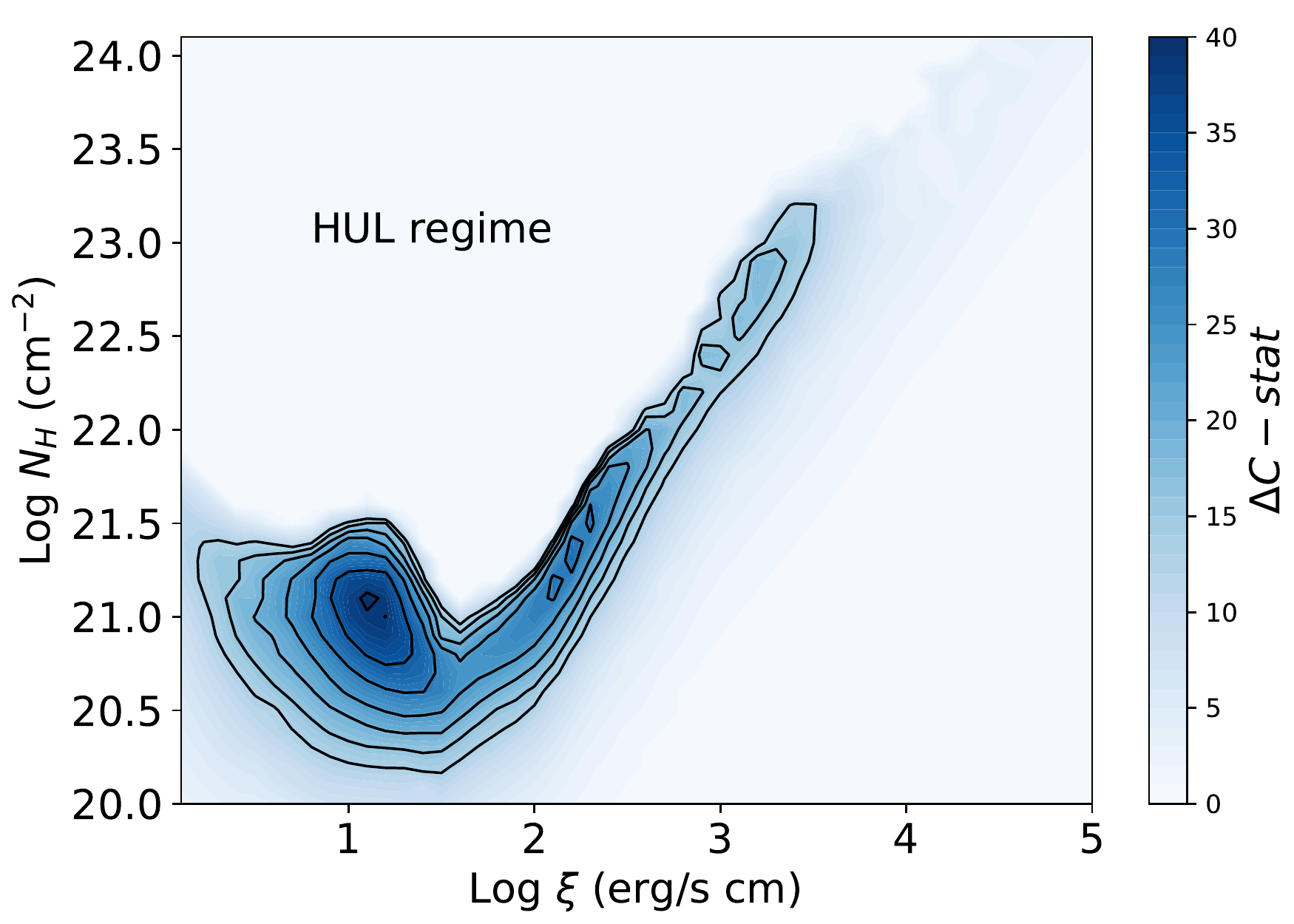}
\includegraphics[scale=.325]{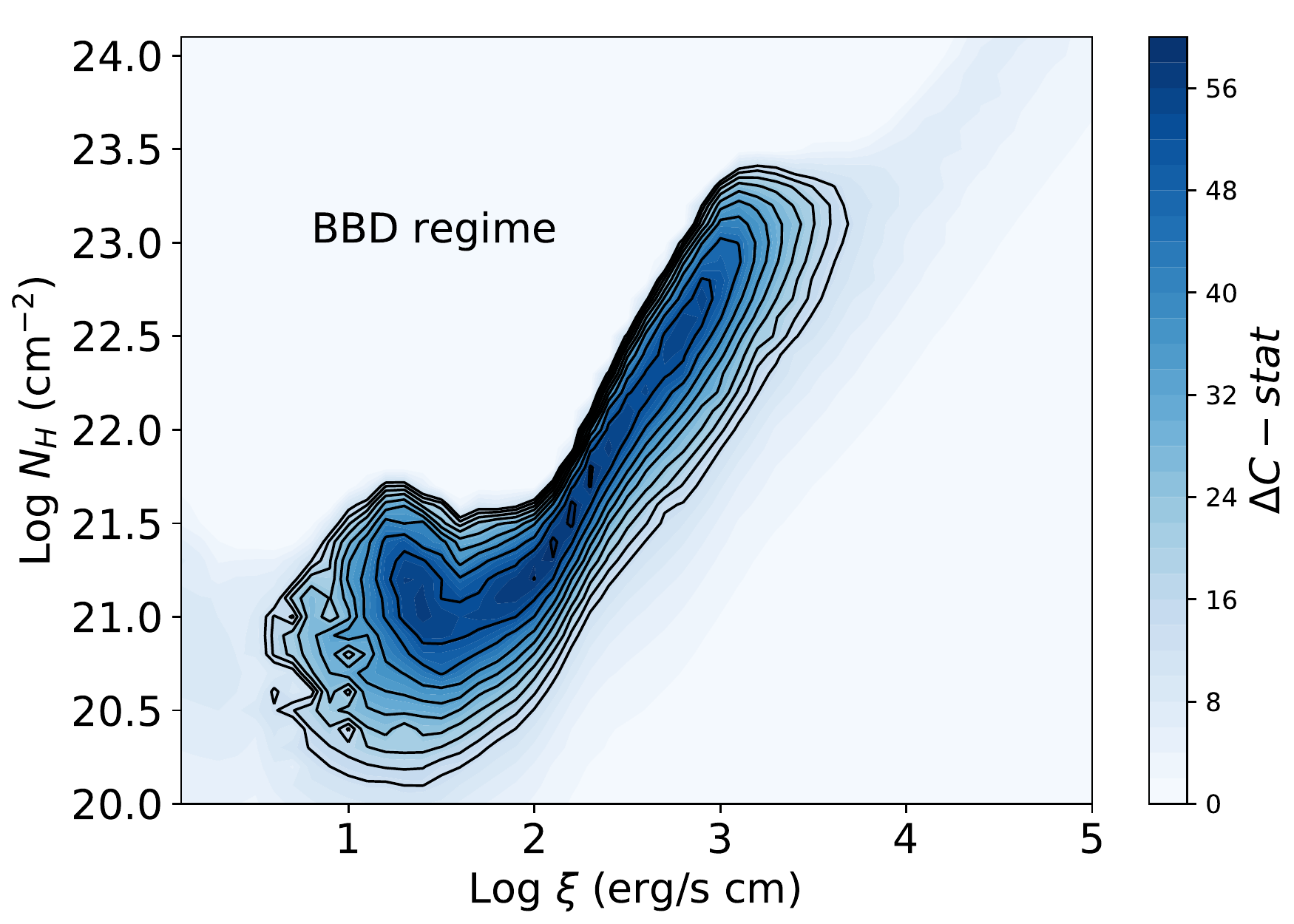}
\caption{Photoionisation emission model scan of the 2017 HUL (L-I, left) and BBD spectra (I-B, right panel) of NGC 1313 X-1 \textcolor{black}{(RGS $\sim$ 0.3-2 keV, EPIC $\sim$ 2-10 keV).} The color is coded according to the $\Delta C$-stat fit improvement to the continuum model. The black contours refer to the 2.5, 3.0, 3.5, 4.0\,$\sigma$, etc. confidence levels estimated through Monte Carlo simulations (see Sect.\,\ref{sec:significance}). Negative values of $\Delta C$-stat for high $N_{\rm H}$ at low-to-mild $\xi$ are set to zero for display purposes \cite{Pinto2020b}.}
\label{fig:line-emission}       % Give a unique label
\end{figure}

Using the {\scriptsize{PION}} code available in {\sc{spex}}, a scan of the $N_{\rm H}-\xi$ parameter space was performed \textcolor{black}{for NGC 1313 X-1 RGS spectra} adopting Solar metallicity, plasma density $n_{\rm H} = 10^{10}$ cm$^{-3}$, a full solid angle $\Omega=4\pi$, velocity dispersion of 1,000 km s$^{-1}$ and no Doppler shift ($v_{\rm LOS}=0$). Fig. \ref{fig:line-emission} shows the $\Delta C$-stat spectral improvements to a simple continuum model in the $N_{\rm H}-\xi$ space. The significance of the detections was constrained by running MC simulations (see Sect.\,\ref{sec:significance}). The plots show that the plasma is multi-phase, with a higher ionisation component that becomes dominant when the source is brighter (i.e. when the spectrum is in the bright broadened disc regime; right panel). This may indicate either a connection between the line emission and the accretion rate or that our view of the inner regions of the accretion disc changes with time (perhaps via disc precession \cite{Pinto2020b}).

For a few ULXs the RGS spectra show very strong emission lines with wavelengths that are incompatible with well-known transitions. A notable example are those from the intermediate spectral hardness HUL NGC 5204 X-1 and the soft-regime spectra of NGC 55 ULX-1 and NGC 247 ULX-1. In these cases, the scanning physical models require a deep exploration of a parameter space which includes at least the LOS velocity and a parameter describing the state of the plasma. For model grids of CIE line-emission, one can search through a grid of line of sight velocity ($v_{\rm LOS}$), velocity dispersion ($v_{\sigma}$) and electron temperature (k$T_{\rm e}$). The normalisation or emission measure ($n_{\rm H} n_{\rm e} V$) is a free parameter. In the case of line emission from PIE plasmas, instead of the electron temperature, the ionisation parameter $\xi$ is adopted, whilst the column density of the photoionised gas, $N_{\rm H}$, is a free parameter. 

The emission-like features in the time-averaged RGS spectrum of NGC 5204 X-1 can be best described with blueshifted (0.3\,$c$) {\fexvii} lines with the dominant resonance ($\lambda_0=15.0\,{\AA}$) and forbidden ($\lambda_0=17.1\,{\AA}$) lines describing the features found at $10.0\,{\AA}$ and $11.3\,{\AA}$ \cite{Kosec2018a}. This would suggest a collisionally-ionised rather than photoionised plasma with a temperature $\gtrsim0.5$\,keV, which is very similar to the cool component of the multiphase jet in the Galactic microquasar SS 433 \cite{Marshall2002}.

It is important to note that similar outflows producing strongly blueshifted emission lines around 1 keV have also been found, albeit at lower significance and only during particular epochs, in a few Galactic transients. For instance, the \textit{Chandra} / LETGS spectrum extracted during a very long burst of the NS LMXB SAX J1808.4-3658 exhibited a strange system of emission lines between 10-14\,{\AA} (mainly {\nex}) that can be best described with CIE emission blueshifted by about 0.1\,$c$ \cite{Pinto2014}. Another example is the very faint X-ray binary IGR J17062-6143 which showed a variable pattern of emission lines from the same ion but a lower blueshift ($\sim0.05\,c$) in a \textit{Chandra} / HETGS spectrum \cite{Degenaar2017}. In the former case, the outflow might have been driven by the powerful burst (amongst the brightest ever detected), whilst in the latter case the very low accretion rate might suggest an origin due to the propeller effect near the NS magnetosphere. Both scenarios might be relevant in ULXs since many of them are likely powered by super-Eddington magnetised neutron stars.

The emission lines detected in the SUL RGS spectra of NGC 55 ULX-1 \cite{Pinto2017} and NGC 247 ULX-1 \cite{Pinto2021} can be better described with emission from photoionised plasma as expected for a photoionised wind. The centroids of the lines are more blueshifted at higher fluxes which might indicate an enhanced radiation pressure at higher accretion rates. Further discussion is presented in Sect. \ref{sec:SEdd}.

\subsubsection{Absorption-line gas}
\label{sec:physical-scan-absorption}

As shown in Fig. \ref{fig:line_catalog}, most absorption lines are not consistent with the laboratory wavelengths of the dominant transitions expected in the soft X-ray band (0.3-2 keV). Therefore the scan of physical models must include $v_{\rm LOS}$ as a main parameter to account for the Doppler shift. The normalisation is represented by the column density for both CIE and PIE plasma models here. The ionisation balance is computed identically to the line-emitting gas but with the difference that here it can be done more quickly since the number of dominant transitions is smaller, which is mainly characterised by ground-state (often 1s--2p level) transitions. It is also well known that forbidden lines in absorption have typically a smaller oscillator strength than resonance lines. In {\sc{spex}} other than the aforementioned {\scriptsize{PION}} there is the faster (and more stable) {\scriptsize{XABS}} component which adopts a pre-calculated ionisation balance.

The first application of the {\sc{spex}} / {\scriptsize{XABS}} component to model the absorption lines in the RGS spectra of two ULXs, the HUL NGC 1313 X-1 and the SUL NGC 5408 X-1, revealed that they were compatible with absorption from mildly to highly ionised ionic species, the same as those seen in emission (such as {\oviiviii}, {\neixx} and {\fexvii}). The main difference was that the absorption component is blueshifted by about 0.2-0.25$\,c$ in both sources \cite{Pinto2016}. 
%%% The Fe K absorption feature detected in {\xmm} and \textit{NuSTAR} spectra was modelled with {\sc{xstar}} tables imported into {\sc{xspec}} and confirmed similar velocities with the exact solution depending on whether an {\fexxv} or {\fexxvi} is preferred \cite{Walton2016a}.
A grid of models of absorption spectra from photoionised gas, i.e. {\scriptsize{XABS}} components, provided a very good description of the absorption features in the RGS spectra of SUL NGC 55 ULX-1. The various lines required multiple components, which implied a complex dynamical structure with the ionisation state increasing with the outflow velocity. This may indicate the launching of plasmas from different regions of the accretion disc \cite{Pinto2017}. The comparison with the two ULXs mentioned above suggests that NGC 55 ULX-1 is being observed at a higher inclination (for more detail see Sect. \ref{sec:SEdd}). The wind partly absorbs the source flux above 1 keV, generating a spectral drop similar to that observed in ULS or SSUL sources that has been often modelled with an absorption edge in low-resolution CCD spectra \cite{Urquhart2016}. %%%% \textcolor{black}{The presence of this wind} may also explain the reprocessing of hard X-ray photons and the soft time lags detected in the light curves (see Fig. \ref{fig:pulsations}, right panel).

The first automatic scan that systematically explored a large parameter space ($v_{\rm LOS}, N_{\rm H}, v_{\sigma}, \xi $) at once was applied to the pulsating source NGC 300 ULX-1, whose {\xmm} / EPIC (2-10 keV) and RGS (0.3-2 keV) spectra were found to exhibit strong absorption features \cite{Kosec2018b}. The outflow is transient as it shows up only in one of the 2 observations. The velocity ($0.24\,c$) is comparable to the fast component in the other ULXs but the plasma is significantly more ionised (log $\xi \sim 4$ instead of 2-3 for the other ULXs). The detection of pulsations and the spectral hardness may indicate that the source is being seen closer to the disc axis (at a lower inclination) than other ULXs with wind detections and that, therefore, its wind plasma along our LOS is hotter. This was the first detection of a UFO in a NS ULX. The transient nature of the wind features can be explained by either a slight change of our LOS relative to the solid angle of the wind and/or most likely a clumpy wind.

\begin{figure}[h] %%%[b]
\includegraphics[scale=.325]{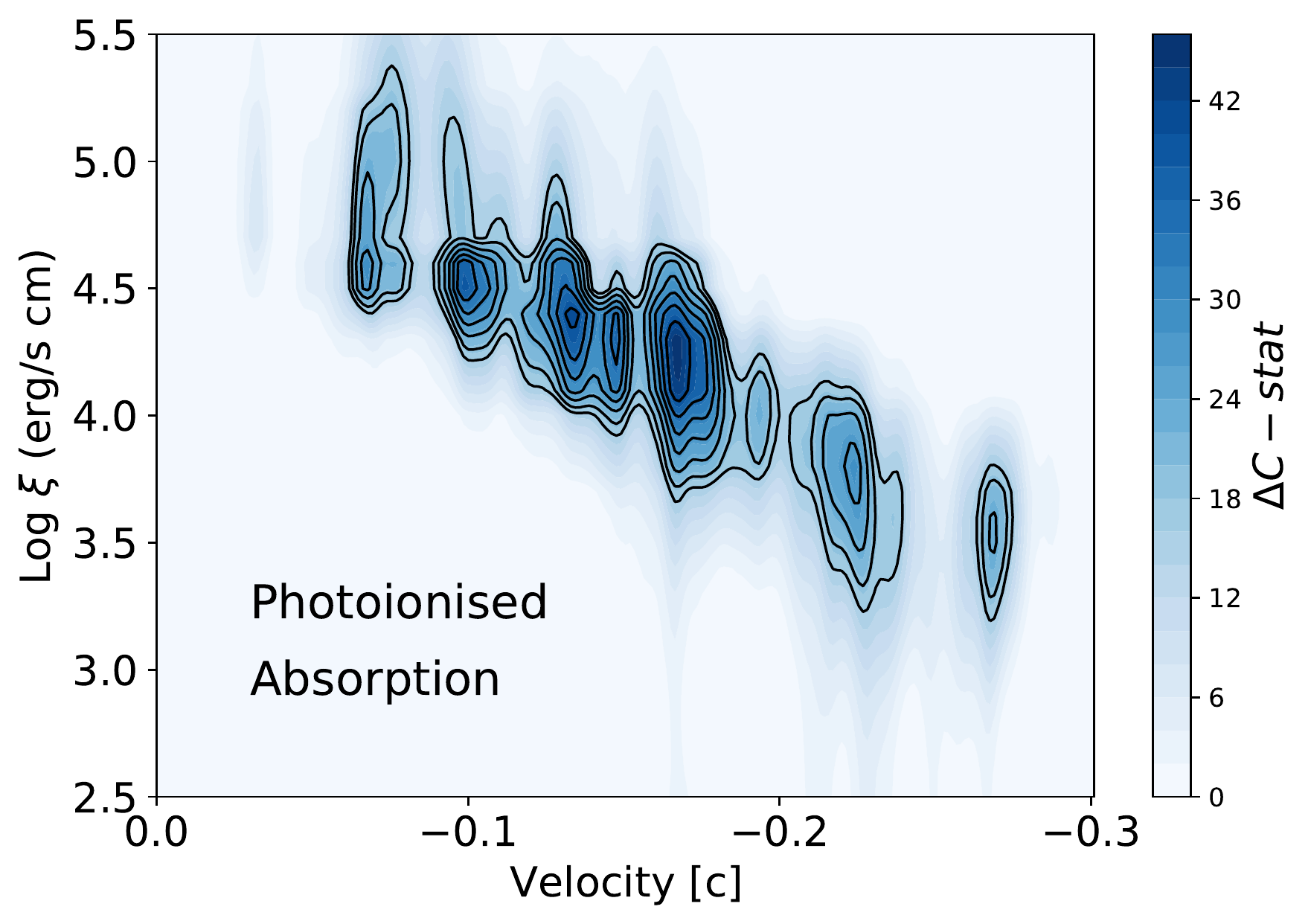}
\includegraphics[scale=.325]{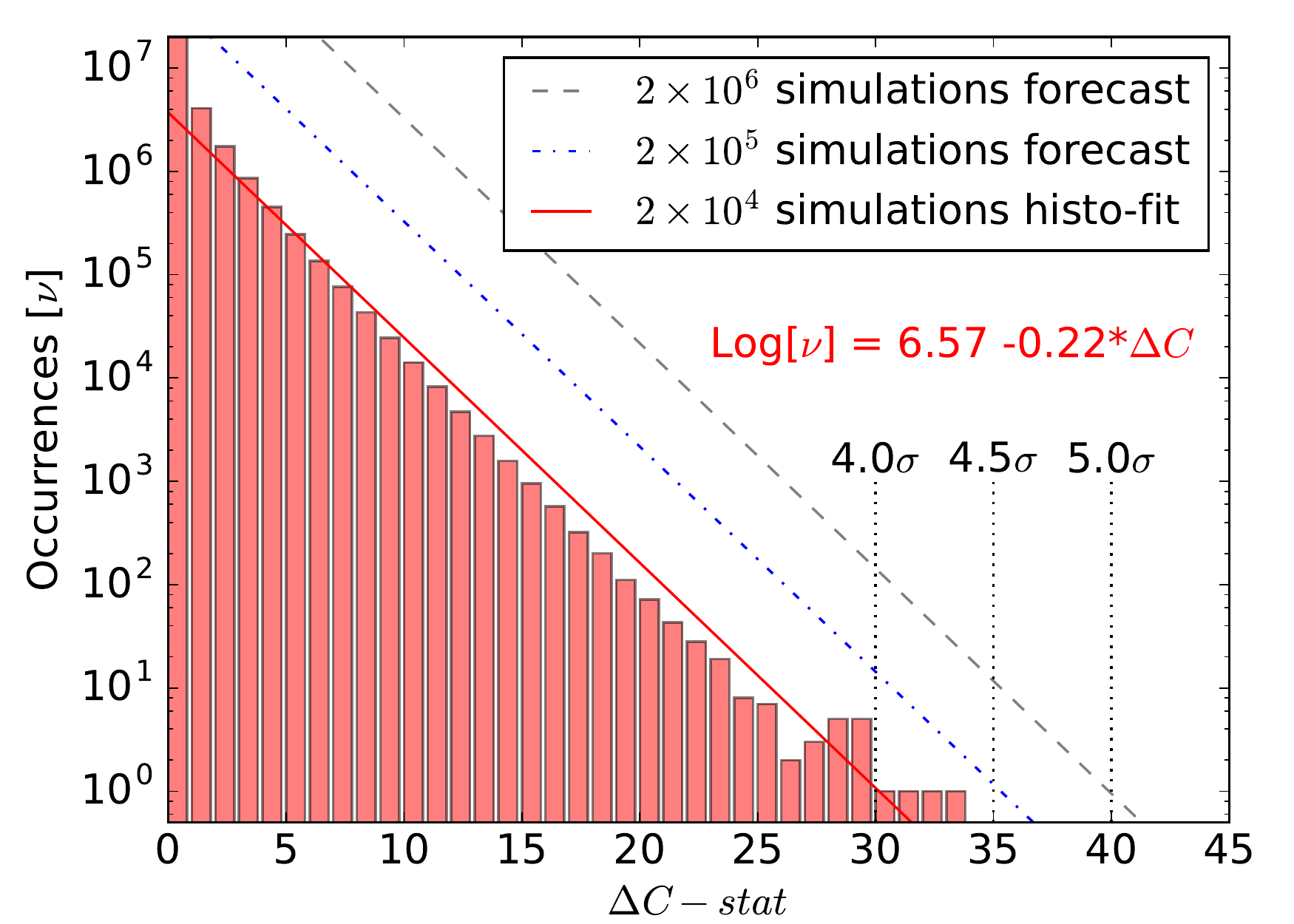}
%%%Left panel: automated scan of {\xmm} and {\nustar} spectra of pulsating NGC 300 ULX-1 with the {\sc{spex}} / {\scriptsize{XABS}} photoionised absorber and a velocity dispersion of 1000 km s$^{-1}$. The X-axis shows the systematic velocity of the absorber, whilst the Y-axis gives the $\Delta C$-stat fit improvement onto the broadband continuum model. Curves of different colours show results for different values of the ionisation parameter \cite{Kosec2018b}.
\caption{Left panel: $\Delta C$-stat contour plot of the scan using the {\sc{spex}} / {\scriptsize{XABS}} photoionised absorber models for the {\xmm} / EPIC and RGS time average spectra of SSUL NGC 247 ULX-1 \cite{Pinto2021}. Labels are same as in Fig. \ref{fig:line-emission}. The black contours refer to the 2.5, 3.0, 3.5, 4.0\,$\sigma$, etc. confidence levels. Right panel: Histogram and corresponding power-law fit of the 20,000 Monte Carlo simulations of NGC 247 ULX-1. Forecasts for 200,000 and 2 million simulations are also shown.}
\label{fig:line-absorption}       % Give a unique label
\end{figure}

These results confirmed that spectral searches with PIE grids unveiled winds in almost all ULXs with deep observations (each one lasting $t_{\rm exp}\gtrsim$\,100\,ks or a full {\xmm} orbit), as indicated by numerous line detections in the Gaussian line scans. In {Fig.\,\ref{fig:line-absorption}} (left panel) we show the example of the highly-significant detection of a photoionised absorber in the supersoft NGC 247 ULX-1 \cite{Pinto2021}, with a systematic velocity of about 0.17\,$c$. The comparison between the best-fitting $\xi$ solution and the stability curves showed that ULX winds are likely stable to thermal perturbations (see also Sect. \ref{sec:implications}). Many of the codes presented here, particularly those developed in the {\sc{spex}} framework, can be found online\,\footnote{https://github.com/ciropinto1982}. 

\subsubsection{\textcolor{black}{Significance of the detection for a grid of models}}
\label{sec:significance}

The significance of the detection of a line-emitting or absorption-line plasma through the use of a grid of physical models is estimated with simulations of spectra using featureless continuum models and searched with the same grid of physical models to compute the false-alarm probability, in a similar way as for the Gaussian line scan. Given that the absorption lines are blueshifted, they provide the best evidence of outflows but also suffer more from the \textit{look-elsewhere} effect. Therefore, an additional effort has been made in the estimate of their significance. The \textit{xabs} component within \textsc{spex} is the fastest of the available photoionisation codes, and is, therefore, best suited to scan the several thousands of simulated spectra involved in such work. Given the general similarity between the absorption spectra of photoionised and collisionally ionised gas (as well as between their emission spectra), the results with the \textit{xabs} component can still be used to assess the significance of other components.

In the case of NGC 247 ULX-1, 20,000 RGS and EPIC spectra were simulated by adopting as a template a multi-temperature three-blackbody continuum model \cite{Pinto2021}. Each simulated spectrum was scanned with the same \textit{xabs} grids shown in Fig.\,\ref{fig:line-absorption} (left panel). The results of these MC simulations are shown in Fig.\,\ref{fig:line-absorption} (right panel). No outlier was found with $\Delta C \geq \Delta C_{\rm max} = 46$, i.e. the value obtained in the real data, which suggests a significance $>4\sigma$ for the absorbing gas and a much higher significance for the line-emitting plasma. These are already high thresholds of significance but millions of simulations would be required to probe confidence levels above $5\sigma$. However, whilst running the simulations it was noticed that after several thousand had been completed the slope of the power-law relation describing the distribution or histogram of the $\Delta C$-stat (calculated with respect to the continuum alone) remained relatively constant, with a slope of around $-0.22$. Interestingly, a comparison of the various sets of these simulations performed in the literature -- the 20,000 simulations run for NGC 1313 X-1 (which were used to estimate the $\Delta C$ contours in Fig. \ref{fig:line-emission}) \cite{Pinto2020b}, the 2,000 simulations performed for NGC 5204 ULX-1 \cite{Kosec2018a}, and the 50,000 simulations performed with the new, faster, cross-correlation method highlighted above \cite{Kosec2021} -- found that the power-law slopes describing their $\Delta C$ distributions were all similar, with an average slope $\overline{\gamma}=-0.225\pm0.015$. The stability of these slopes has been therefore used to forecast the results of larger numbers of simulations, as in principle one can project the distribution a larger number of simulations would produce simply by re-scaling the normalisation of the best-fit power law. In Fig.\,\ref{fig:line-absorption} (right panel) we show the predictions for $2\times10^5$ (dash-dotted line) and $2\times10^6$ (dashed line) simulations for the wind in NGC 247 ULX-1. This would suggest that $\Delta C > 35$ and $>$ 40 correspond to detection significance of 4.5 and 5\,$\sigma$, respectively, with an uncertainty of $0.2\sigma$ based on the spread in the slopes of the other histograms. The confidence levels in $\sigma$ corresponding to the $\Delta C$-stat values for NGC 247 ULX-1 are shown as black contours in Fig.\,\ref{fig:line-absorption} (right panel).

% DJW: Note that I made the slope a lower-case gamma here, to avoid confusion with the photon index

As expected, the simultaneous modelling of multiple lines by means of physi\-cally-motivated models significantly improves the plasma detection significance in ULXs, reaching peaks of 5\,$\sigma$ in the deepest observations of ULXs with soft X-ray spectra ($\Gamma>2$, see e.g. Fig.\,\ref{fig:line-absorption} \cite{Kosec2021}). In HUL spectra the lines are notably fainter and more difficult to detect \cite{Kosec2021}, although deep ($\gtrsim$\,100\,ks) observations of nearby targets have confirmed highly significant detections in some cases (e.g. NGC 1313 X-1 and NGC 300 ULX-1 \cite{Pinto2020b,Kosec2018b}).

%%%It is important to note that evidence for winds in ULXs has also been found, albeit with more ambiguity, in optical spectra in the form of strong lines from H- and He- transitions. The lines are very similar to those observed in SS 433 and have been associated with the cool, outer, photosphere of the super-Eddington disc wind ($10^{5-6}$\,R$_{\rm G}$, see e.g. \cite{Vinokurov2018, Fabrika2021}). This agrees with the overall scenario in which SS 433 is a highly obscured ULX that is being seen almost edge-on, such that we can only see scattered emission, the jets and the outer photosphere. The presence of a wind in SS 433 is indeed confirmed by the detection of a possible blueshifted Fe K feature in the energy-dependent time lags seen in this source \cite{Middleton2021}.

\subsection{Cyclotron Resonant Scattering Features}

Highly-magnetised neutron stars are known to exhibit cyclotron resonant scattering features (CRSFs) which are produced by the interaction of X-ray photons with a strong magnetic field \cite{Truemper1978}. These features are often used to identify the accretor as a neutron star since black holes cannot produce such strong magnetic fields. 
Although formally resulting from a scattering process, since the scattering cross-section is strongly peaked at a specific energy (primarily set by the magnetic field of the neutron star and the nature of the scattering particles) CRSFs typically appear as broad absorption-like features in the X-ray spectra of accreting pulsars. In Galactic X-ray pulsars these features can be as broad as a few keV \cite{Tsygankov2006} and can sometimes be resolved even in spectra with a moderate resolution, such as those obtained by CCD in the hard X-ray band ($R = E / \Delta E \sim 50$ around 10 keV), although a broad bandpass is typically needed as these features are often seen at high energies ($>$10\,keV \cite{Caballero2012}).

Most observed CRSFs are generally expected to be electron CRSFs. However, in rare cases claims of narrow CRSFs (widths less than 0.4 keV) have been made. These have been attributed to proton CRSFs, given their much larger mass. Knowing the nature of the scattering particle is key for measuring neutron star magnetic fields, as for an electron $E_{\rm cyc,e} = 11.6 \, (1+z)^{-1} \rm (B/10^{12} \, G) \, keV$ whilst for a proton $E_{\rm cyc,p} = 6.3 \, (1+z)^{-1} \rm (B/10^{15} \, G) \, keV$. Here $z$ is the gravitational redshift of the line forming region, which is typically taken to be $z \sim 0.2$ assuming that CRSFs are formed close to the surface of the neutron star (such that $1+z_{\rm{cyc}} \simeq (1-2GM/R_{\rm NS} c^2)^{-1/2}$). None of these narrow CRSF candidates have been observed with high-resolution instruments to date, but such observations would be relevant in the future.

CRSF claims in ULXs are rare, as the sample of ULXs with high S/N broadband spectroscopy is still relatively small (and even in these cases the coverage with \textit{NuSTAR} only extends up to $\sim$40\,keV at most). However, \textcolor{black}{the detection of an isolated and narrow (width $\sim$0.1\,keV) absorption feature at $\sim$4.5\,keV in a spectrum of M 51 ULX-8 \cite{Brightman2018} suggest that this may be a proton CRSF, given the lack of other absorption features and the unusual energy,} implying an extreme, magnetar-level magnetic field of $B\sim 10^{15}$\,G. This source is not known to be a pulsar, but if this interpretation is correct it would still confirm the accretor to be a NS. Another potential narrow CRSF was suggested in the hyperluminous X-ray source NGC\,4045 X-1 \cite{Brightman2022}. In this case, the line energy was $\sim$8.6\,keV, which in principle could easily be reproduced by iron absorption in an ultrafast outflow with $v_{\rm{LOS}} \sim 0.2c$, similar to those seen in other ULXs. However, the equivalent width of this feature was extremely strong, $\sim 220$\,eV, much stronger than the equivalent width of the iron absorption from the outflow in NGC\,1313 X-1 (of about 60\,eV \cite{Walton2016a}), leading the authors to also discuss a CRSF interpretation. The only other notable claim is for a broad, electron CRSF at $\sim$13\,keV in the ULX pulsar NGC\,300 ULX-1. This would imply a more modest magnetic field of $\sim$10$^{12}$\,G \cite{Walton2018a}, similar to the magnetic fields seen in `normal' Galactic X-ray pulsars \cite{Caballero2012}. This is notable \textcolor{black}{as the presence of an ultrafast outflow is also claimed in this same source \cite{Kosec2018b}}. The more moderate field strength implied here is consistent with the idea that the super-Eddington inner regions of the accretion flow can form before the disc is truncated by the magnetic field of the NS, which is a condition likely required for the launching of such a fast outflow.

\section{Implications}
\label{sec:implications}

In this section we summarise the main results in the ULX field over the last decade, focusing on those obtained primarily through high-resolution X-ray spectroscopy. \textcolor{black}{We will first compare them with theoretical predictions, especially within the framework of super-Eddington accretion, and then discuss the implications on the properties of these systems.} Particular emphasis will be given to the role of ULXs in the understanding of galactic evolution and growth of supermassive black holes.

\subsection{Super-Eddington accretion}
 \label{sec:super-Eddington}

\begin{figure}[h] %%%[b]
\includegraphics[scale=.195]{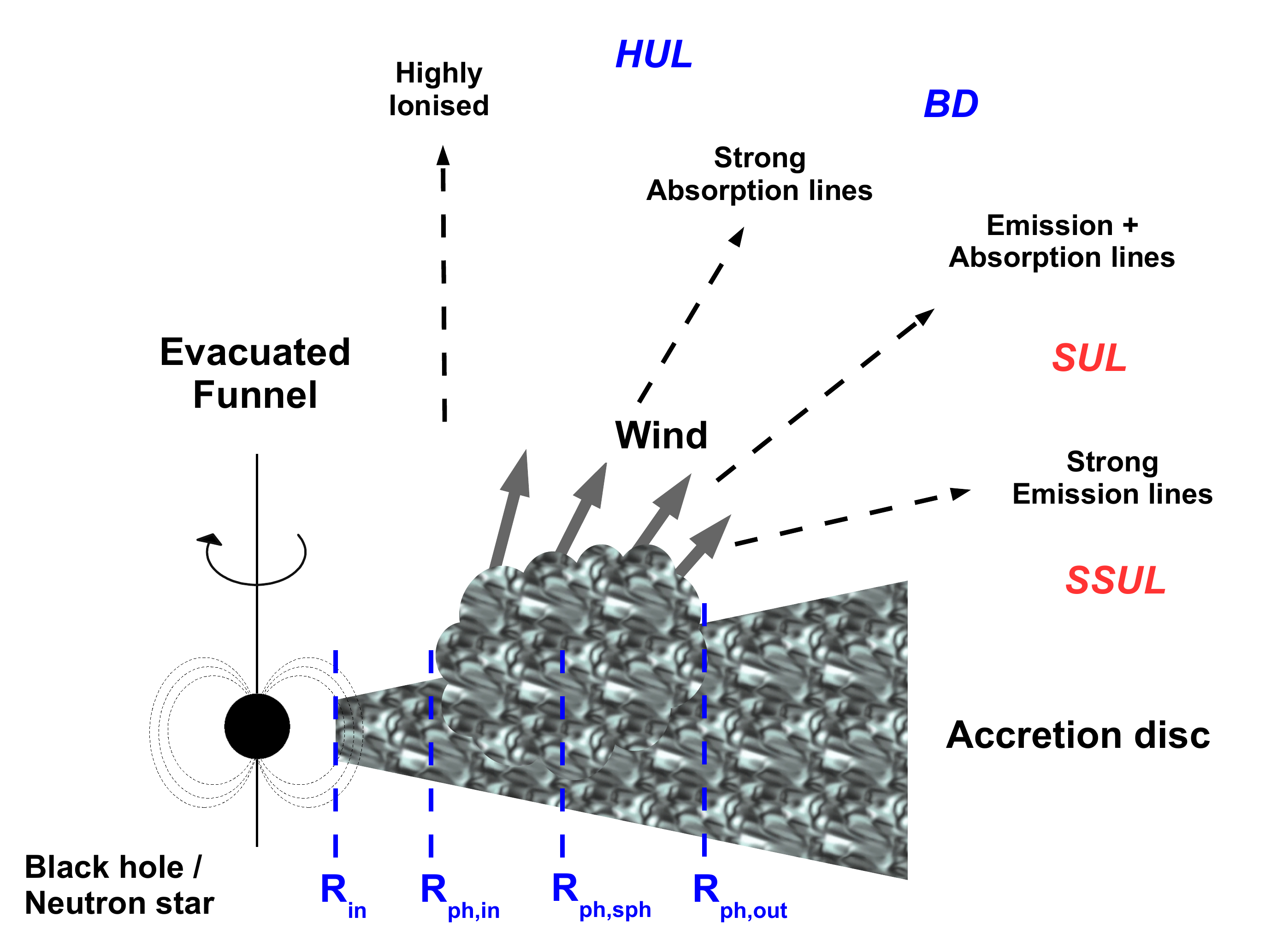}
\includegraphics[scale=.293]{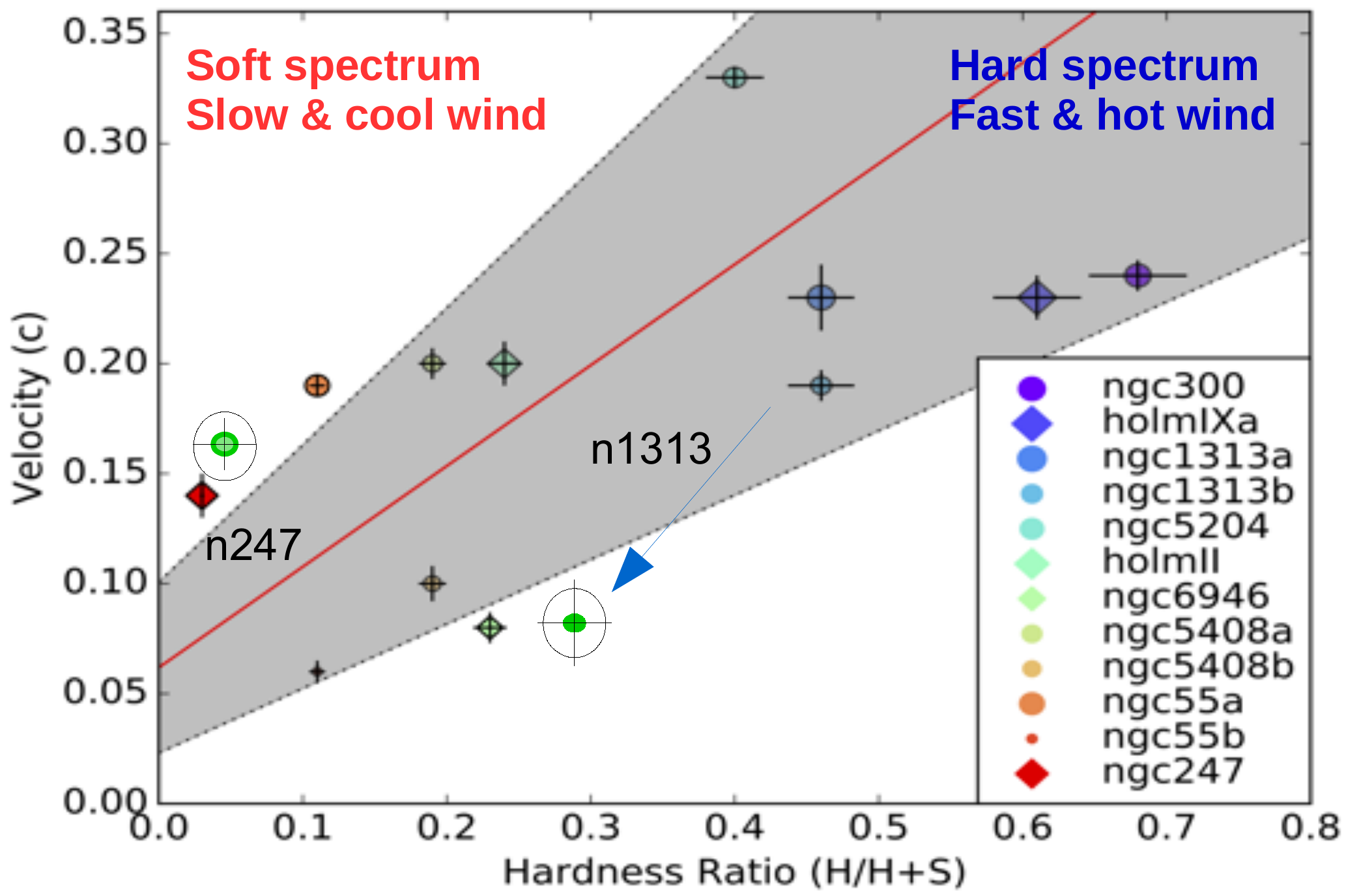}
\caption{Left panel: schematic cartoon of a super-Eddington accretion disc. The accretion disc is geometrically thick around the spherisation radius, $R_{\rm sph}$, where an optically-thick clumpy outflow is launched by radiation pressure. The wind becomes optically thin at $R_{\rm ph,out}$. Along LOS closer to the disc axis the spectra will appear hard. At moderate inclinations, the spectra will appear soft with strong absorption lines from the wind. At high inclinations the continuum is significantly absorbed and strong emission lines appear in the spectra with the latter becoming supersoft. Right panel: correlation between LOS velocity, ionisation parameter (colour) and hardness ratio (1-10 keV flux / 0.3-10 keV flux) for ULX wind detections (combining different sources and, where available, multiple epochs \cite{Pinto2020a}).}
\label{fig:disc-geometry}       % Give a unique label
\end{figure}

Under the approximation of spherical geometry and isotropic accretion of matter, the luminosity of an accreting compact object should not surpass a maximum defined as Eddington luminosity, $L_{\rm Edd}= 1.26\times10^{38} (M/M_{\odot}) \, {\rm erg \, s}^{-1}$, where $M$ is the mass of the accretor and $M_{\odot}$ is the Sun mass. This also implies a maximum accretion rate $\dot{M}_{\rm Edd}=L_{\rm Edd}/\eta c^2=1.86\times10^{18}(M/M_{\odot}) {\rm \, g \, s}^{-1} = 2 \times 10^{-7} \, M_{\odot} \, {\rm yr}^{-1}$ assuming a mass of $10\,M_{\odot}$ and a radiative efficiency \textcolor{black}{$\eta=10$\,\% typical for a weakly-spinning BH ($\eta$ depends on the BH spin, and varies from 6\,\% for a Schwarzschild BH to almost 40\,\% for a maximal Kerr BH)}. This also applies to more common sources e.g. main-sequence stars. At $L > L_{\rm Edd}$, the radiation pressure can surpass the gravitation pull and can thus drive powerful winds (conceptually similar to the radiation-pressure driven winds seen in bright, giant, stars and in outbursts of novae \cite{Pinto2012a}). The properties of the winds, e.g. velocity, outflow rate and temperature, depend on the accretion rate and the mass of the compact object. Magneto-hydrodynamic simulations of super-Eddington accretion onto black holes \cite{Ohsuga2005} have shown that above the supercritical accretion rate ($\dot{M}_{\rm cr} = 9/4 \, \dot{M}_{\rm Edd}$ \cite{Poutanen2007}) the radiation pressure inflates the inner accretion flow, which becomes geometrically thick, and launches fast winds; the escape velocity exceeds 0.1\,$c$ in the innermost regions ($R \lesssim 100 \, R_{\rm G}$). These super-Eddington winds are expected to be launched from inside the spherisation radius, $R_{\rm sph} = (\dot{M}/\dot{M}_{\rm Edd}) R_{\rm in}$, where $R_{\rm in}$ is the inner disc radius, which in absence of disc truncation is considered to be about $6 \, R_{\rm G}$ (see Fig. \ref{fig:disc-geometry}, left panel). The winds are also expected to have a mild ionisation state (log $\xi \sim 2-4$) owing to the relatively soft SEDs of super-Eddington accretion discs. Such ionisation parameters are fully consistent with the properties of the extreme outflows found in ULXs. The discovery of these outflows, together with their unusual broadband spectra and the discovery of pulsating ULXs, has helped establish the ULX population as the best local examples of super-Eddington \textcolor{black}{accretors}.

The presence of transient neutron stars that can reach luminosities in excess of the Eddington limit ($L_{\rm Edd, NS} \sim 2\times10^{38} \, {\rm erg \, s}^{-1}$) was already established 40 years ago \cite{Skinner1982}. However, the discovery of the first pulsating ULX in M 82 \cite{Bachetti2014} and, especially, the HLX NGC 5907 X-1 \cite{Israel2017a}, showed that these objects can reach highly super-Eddington luminosities; the latter reaches $10^{41} \, {\rm erg \, s}^{-1}$, i.e. 500 times $L_{\rm Edd}$. In principle, extreme magnetic field strengths (magnetar-level, $B \gg 10^{13}$\,G) can suppress the Thomson electron scattering cross-section. \textcolor{black}{This would reduce the impact of radiation pressure and, therefore, raise the effective Eddington limit, which may allow the NS to reach extreme rates \cite{Mushtukov2015}.}
However, the presence of ultrafast winds in many ULXs, including those with pulsations \cite{Kosec2018b}, would rather suggest that `classical' super-Eddington accretion flows do exist (i.e. such that the geometrically thick regions of the disc in which radiation pressure exceeds gravity do form) outside of the pulsar magnetosphere; this would correspond to the scenario in which $R_{\rm{sp}} > R_{\rm{M}}$, in turn suggesting that the magnetic field strengths are more moderate.

Some transient BH LMXBs, such as GRS 1915+105, sometimes reach mildly super-Eddington luminosities \cite{Motta2021}. SMBHs powering NLSy1 AGN can also shine up to a few times $L_{\rm Edd}$ \cite{Pinto2018a}. However, stellar-mass compact objects in the most extreme ULXs would require accretion rates (relative to Eddington) orders of magnitude higher. This can be achieved, for instance, through thermal-timescale mass transfer in high mass binaries \cite{King2001} when the companion star evolves, filling its Roche lobe. Binary evolution models have shown that the mass transfer can be very high during this phase, peaking at $10^{-3} ~ M_{\odot} \rm \, yr^{-1}$ for $\sim 10^4$ yr \cite{Rappaport2005,Wiktorowicz2015}. This can occur either for a BH with a companion star heavier than $\sim 10 \, M_{\odot}$ or a NS accreting from a Helium-burning secondary with mass $\sim$1--2\,$M_{\odot}$. These systems can often reach $1,000 \, \dot{M}_{\rm Edd}$ during the $10^5$~yr thermal-timescale phase and are prime candidates for explaining the presence of extreme ULXs.

\subsection{Super-Eddington disc-wind structure}
\label{sec:SEdd}

At $\dot{M} > \dot{M}_{\rm Edd}$, the accretion becomes radiatively inefficient due to strong advection, photon trapping and heat transfer within the disc. At higher accretion rates, the luminosity no longer increases linearly with $\dot{M}$, and should instead scale logarithmically. However, at the same time the disc is expected to become geometrically thick, assuming a funnel shape which, via scattering, can collimate photons produced in the inner regions into the evacuated cones away from the disc plane (geometrical beaming). This can help sources appear to reach high luminosities if spherically-symmetric emission is assumed. Accounting for all the effects, the observed or apparent luminosity calculated assuming spherical symmetry can be expressed as \textcolor{black}{$L \simeq L_{\rm{Edd}} [1+ \ln(\dot{M}/\dot{M}_{\rm{Edd}})/b]$}, where $b$ is the beaming factor. This is proposed to scale with the accretion rate as $b = 73 (\dot{M}/\dot{M}_{\rm Edd})^{-2}$ \cite{King2009}.

The current scenario \cite{Poutanen2007} suggests the presence of three main zones that characterise the structure of a super-Eddington accretion disc (see Fig. \ref{fig:disc-geometry}). Zone\,1 or the innermost region ($R<R_{\rm ph, in}$), where the wind is optically thin, the disc can be approximated with a hot accretion flow and produce the component with k$T\sim$few keV observed in ULXs. Zone\,2 between the inner edge of the photosphere and the spherisation radius ($R_{\rm ph, in}<R<R_{\rm sph}$) where the wind is opaque and forms a continuation of the accretion disc. As an expanding photosphere, this will have a $T \propto R^{-1/2}$ profile with typical temperatures of a few times 0.1 keV. Zone\,3 with $R_{\rm sph}<R<R_{\rm ph, out}$ where the optical depth of the wind falls as $1/R$. The luminosity of this region is Eddington limited and the temperature profile follows the standard $T \propto R^{-3/4}$ trend.

The funnel shape of the disc-wind structure implies that along the disc axis, the wind is either absent or highly ionised and the X-ray spectra would therefore appear harder (HUL) as the observed emission is dominated by the hot inner accretion flow. At larger viewing angles, the optically-thick clumps of the wind absorb the X-ray photons from the innermost hot regions and the spectra appear softer with strong absorption lines. At very high inclinations strong emission lines become visible against a highly-suppressed continuum. If the LOS is crossed by large wind clumps, or even by the thick disc itself, the spectra will appear supersoft. These general predictions are confirmed by the correlation between the LOS velocity and the ionisation parameter of the wind with the observed spectral hardness (characterised as a hardness ratio, defined as 1-10 keV flux / 0.3-10 keV flux) measured for a small sample of ULXs with high-quality RGS spectra (see Fig. \ref{fig:disc-geometry}, right panel \cite{Pinto2020a}). They are also supported by the anti-correlation between the spectral hardness and the number of lines detected in the RGS spectra of a sample of 19 ULXs \cite{Kosec2021}. \textcolor{black}{The transition from an intermediate-luminosity HUL ($\Gamma<2$) spectrum to a bright broadened disc (BBD) is likely due to a higher accretion rate as confirmed by the presence of a cooler and slower wind in the BBD spectrum of NGC 1313 X-1, which is expected for an expansion of $R_{\rm sph}$, i.e. a larger launching radius \cite{Pinto2020b}.} Here we assume that the wind speed is associated to the escape velocity at the wind launching radius $v_{\rm esc}=\sqrt{2GM/R}$.

Although there is not yet a large consensus, it is plausible that the ubiquitous low-velocity emission lines are produced by slow-moving thermal winds (see Sect. \ref{sec:wind-lines}) as these are unlikely to obscure the inner accretion flow but will emit as the gas cools and recombines \cite{Pinto2021,Middleton2022}. This would occur near the Compton radius $R_{IC}=GM \mu m_{\rm p}/(kT_{IC})$. For a $10\,M_{\odot}$ black hole and a Compton temperature of $T_{IC}\sim10^{6-7}$\,K (i.e. the temperature where the $T-\xi$ curve flattens for the ULXs in Fig.\,\ref{fig:ion_bal}, left panel), this would correspond to a radius of a few times $10^5 R_{\rm G}$. Alternative solutions would be reprocessing in the outer disc (especially for the cool {\ovii} component), or emission from a bulge near the magnetosphere \cite{Koliopanos2016} (particularly for the hotter {{\oviii} - {\nex}} component). In principle wind collisions with the CSM could contribute but they would vary on much longer timescales than what is observed, but \textcolor{black}{internal shocks of the ULX wind (if heated above 1 keV) or collisions with the stellar wind of the companion star} might also be a possible explanation.

\subsubsection{The case of magnetised neutron stars}
\label{sec:SEdd_NS}

As previously mentioned, some ULXs host neutron stars as shown by the discovery of pulsations in a handful of sources. M 82 X-2 \cite{Bachetti2014} and NGC 5907 X-1 \cite{Israel2017a} are very bright and could potentially host NS with high magnetic fields, i.e. $B \sim 10^{12-14}$\,G. Here, the Thomson electron scattering opacity can be significantly suppressed and facilitate high luminosities \cite{Mushtukov2015} with a significant truncation of the inner accretion disc. This might prevent the formation of the spherisation radius with the matter channeled along the magnetic field lines and forming of an accretion column. The presence of relativistic winds in many ULXs, including some with pulsations \cite{Kosec2018b,Kosec2021,vdEijnden2019}, would suggest small truncation and \textcolor{black}{low-to-moderate} magnetic fields.
%%%However, the high-resolution X-ray spectra of M 82 X-2 and NGC 5907 X-1 are of too low in quality and confused with X-ray emission from nearby sources and the local ISM to enable accurate wind searches. \textcolor{black}{@Peter / Dom, have you b.a.c looked at their CCD spectra near 1 keV? DJW: If you are asking about M82 X-2 and NGC5907 ULX-1 specifically, I think both are likely too absorbed to see much. For X-2 you would also only be able to use ACIS, and a lot of the observations have been taken more recently when the ACIS contamination build-up is pretty severe. I actually think you can get rid of most of this paragraph, as the main point is already made just a few sections earlier (1.5.1) isn't it?}

GR-MHD simulations of super-Eddington accreting NS with low magnetic fields have shown properties similar to black holes \cite{Ohsuga2007}. The \textcolor{black}{net} mass-accretion rate onto the NS exceeds the critical rate albeit being 20-30\,\% that of BHs for the same input $\dot{M}$. This means that the mass-outflow rate is a few times larger in flows around non-magnetic NSs than in flows around BHs. 
%%% The outflow velocities are 0.2-0.3\,$c$, which implies that \textcolor{black}{the X-ray line-emitting, outflowing, plasma in SS 433 may be explained as a shocked outflow from} the supercritical accretion onto a NS \cite{Middleton2021}. \textcolor{black}{DJW: Aren't you confusing the jet and the disc wind here? I'm also not sure why we're discussing SS433 in a section about winds in magnetic NS ULXs.} CP: because I am not sure the X-ray line-emitting plasma seen in SS 433 is a jet, it could be something else, e.g. shocks within a hot phase of the wind. Although in order to avoid confusion better not discussing it in this review.

At intermediate magnetic fields ($B \sim 10^{10}$\,G) the supercritical disc is truncated at a few NS radii, where the magnetic pressure associated with the magnetic field of the central NS balances with the radiation pressure in the disc.  The transport of angular momentum spins up the NS of about $-10^{-11}$ s s$^{-1}$ \textcolor{black}{or even more}. The ejecta consists of collimated outflows (jets moving at 0.4\,$c$) and winds with typical velocities of 0.1\,$c$, the latter with a much higher density \cite{Takahashi2017}. Similar work on NS with a magnetic field $B \sim 10^{12}$\,G has shown that copious photons are generated at the shock near the NS surface and escape from the sidewall of the accretion columns but high-density inflow and low-density outflows appear within the column. The matter significantly slows down and much of it does not reach the surface due to a strong radiation force \cite{Kawashima2020}. At high accretion rates with the aid of advection, the matter can broaden the column, mass-load the field lines, and produce radiation-driven, mildly relativistic ejecta with properties similar to those seen in ULX winds \cite{Abolmasov2022}.

The predicted spin up rates and outflows are remarkably consistent with the observations \cite{Bachetti2014,Pinto2016} which would provide strong evidence for ULXs to be mainly powered by neutron stars with magnetic fields $B \sim 10^{10-13}$\,G, $\dot{M}\sim10-100\,\dot{M}_{\rm Edd}$, and magnetospheric radii $R_{M} \lesssim R_{\rm sph} \sim 10^{7-8}$\,cm or $10-100$ NS radii \cite{Mushtukov2019,King2020}. We adopted the standard definition $R_{M} \sim 7 \times 10^7 \Lambda m^{1/7} R_{6}^{10/7} B_{12}^{4/7} L_{39}^{-2/7}$ cm, where m = $M_{\rm NS}/M_{\odot}$, $B_{12}= B/10^{12}$ G, $ L_{39} = L/10^{39}$ erg s$^{-1}$, $R_{6} = R_{\rm NS}/10^6$ cm and $\Lambda=0.5$ is a constant corresponding to a specific accretion disc geometry \cite{Mushtukov2015}. The discovery of pulsations in ULXs characterised by harder (typically HUL) spectra would confirm the funnel shape of the system although this could also be a result of the accretion column structure. Moreover, given that the magnetic field might be buried by accretion over time, a large fraction of the ULX population likely consists of NS that do not exhibit pulsations. This is suggested by the remarkable similarity between the spectra of ULX with/without pulsations \cite{Walton2018a}.

\subsection{Feedback and growth rate}
\label{sec:feedback}

ULX winds have extreme velocities and are therefore expected to have a significant impact on their surrounding environments. Many ULXs, including NGC 1313 X-1, are indeed surrounded by huge interstellar cavities. Some of them have a supersonic expansion rate (80-250 km s$^{-1}$) and optical line ratios which suggest a mechanical expansion driven by outflows \cite{Gurpide2022}. In these cases, they are called bubbles and show typical ages of $10^{5-6}$ yr and H\,$\alpha$ luminosities that require a mechanical power of $10^{39-40}$\,erg\,s$^{-1}$ \cite{Pakull2003}. In other cases the optical lines seem to be more consistent with photoionisation \cite{Berghea2010}, which could be caused by the ULX radiation field (in this case they are known as nebulae). Regardless of the exact driving mechanism, it is clear that ULXs can deposit huge amounts of energy into the surrounding ISM.

It is possible to estimate the mechanical power of ULX winds and the amount of matter that is carried by the wind. The mass outflow rate is taken from the mass conservation law: $\dot{M}_w = 4 \, \pi \, R^2 \, \rho \, v_w \, \Omega \, C$, where $\Omega$ and $C$ are the solid angle and the volume filling factor (or \textit{clumpiness}), respectively, $\rho=n_{\rm H}  \, m_p \, \mu$ is the gas density, and $R$ is the distance from the ionising source. From this, an expression for the kinetic luminosity can be derived by substituting the $R^2 \rho$ factor from the definition of the ionisation parameter ($\xi=L_{\rm ion}/n_{\rm H}R^2$), giving $L_w = 0.5 \, \dot{M}_w \, v_w^2 =  2 \, \pi \, m_p \, \mu \, \Omega \, C \, L_{\rm ion} \, v_w^3 \, / \, \xi$.

According to GR-MHD simulations \cite{Kobayashi2018}, radiatively-driven winds in super-Eddington systems are highly porous ($C\sim0.01-0.1$) and have solid angles of about $\Omega/4\pi\gtrsim0.3$. Inserting the typical values of ionisation parameters found in ULX winds (log $\xi\sim$ 2-4) in Eq. (23) of \cite{Kobayashi2018} we estimate comparable values of clumpiness. The fraction of ULXs with detection of emission lines is very high ($>60\,\%$ \cite{Kosec2021}), whilst absorption lines (i.e. fast winds) are detected in a smaller fraction of the sample (e.g. $\sim30\,\%$ \cite{Pinto2020a}). Assuming the detection fraction as a proxy for the solid angle implies values broadly consistent with the simulations \cite{Kobayashi2018}.

The $\Omega$ and $C$ parameters along with $\xi$ and $v_{\rm LOS}$ estimated for the wind through the photoionisation models (see Sect. \ref{sec:physical-scan-absorption}) yield a kinetic power $L_w = 10^{39-40}$ erg s$^{-1}$, which is comparable to the ULXs X-ray luminosity. This means that about 50\,\% of the total energy budget is expended powering the winds observed, and their mechanical power is sufficiently high to inflate the bubbles. Using the estimates of $L_w$, the age ($\tau$) of the bubbles and adopting a typical ISM density $\rho_{\rm \footnotesize{\, ISM}} \sim 10^{-24}$ g cm$^{-3}$, it is possible to predict a bubble size $R_{\rm bubble} = 0.76 ( {\dot{M}_w \, v_w^2}/ {2 \, \rho_{\rm \footnotesize{\, ISM}}} )^{1/5}  \tau^{3/5} \sim 100$ pc \cite{Castor1975}, which is in excellent agreement with the observations \cite{Pakull2003}.\,\footnote{\textcolor{black}{Here, a lifetime of $10^6$ yr is adopted from the current expansion velocities of $\sim100$\,km s$^{-1}$ and the expansion law $R \sim t^{3/5}$. The long recombination time of He {\sc iii} (see Sect.\,\ref{sec:multiwavelength} \cite{Berghea2010}) and, alternatively, arguments related to the maximum amount of matter that can be accreted onto a NS without becoming a BH \cite{Pinto2020a} agree with such long timescales.}}

Regarding the wind mass outflow rate, assuming a standard accretion efficiency, i.e. $\eta=0.1$, we estimate $\dot{M}^{\rm max}_w\sim$\,90\% of the total inflow rate through the outer disc. However, if we take into account advection and photon trapping we estimate a more realistic mass loss rate of 10-50\% \cite{Mushtukov2019}. This means that the compact object would still grow rapidly despite the presence of these powerful outflows.

The total energy output of a ULX integrated over its time of activity can be as high as $10^{53}$ erg, and released within a region of $\sim1$\,kpc$^2$. The present population of ULXs in nearby galaxies is rather small (1-2 per galaxy). However, in the local Universe the number of ULXs per galaxy \textcolor{black}{increases in galaxies with lower metallicity \cite{Lehmer2021} and, especially, with higher star formation rates \cite{Mineo2012, Lehmer2019}}. It is therefore possible that at $z=2-3$, near the peak of star formation, ULXs were much more abundant and had a significant effect on the evolution of their host galaxies. The effects, perhaps comparable to those of supernovae, may help to explain the inconsistencies between the observed amount of galaxies and the (higher) theoretical predictions in the low-mass end of the mass function of galaxies \cite{Silk2012}.

\subsection{ULX as probes of the primordial black holes}
\label{sec:ULXs_vs_AGN}

Winds in active galactic nuclei are common features and include a variety of phases from the slow ($v_{\rm LOS} \lesssim$ 5,000 km s$^{-1}$) warm absorbers to the ultrafast outflows reaching $0.4\,c$ (see Sect. \ref{sec:wind-lines}). The fraction of AGN exhibiting UFOs is of about 40\,\% \cite{Tombesi2010} \textcolor{black}{but the fraction might increase at higher accretion rates} as shown by the frequent detections in quasars and NLSy1 \cite{Reeves2003, Kosec2018c, Walton2019, Pinto2018a, Xu2022}. This could be an indication that radiation pressure starts to be important and to contribute to the launch of the winds which at low $\dot{M}$ might be driven by magnetic fields instead. The main difference between ULX and AGN UFOs is that in the latter case they exhibit higher values of ionisation parameter (on average) due to the harder SEDs of AGN. However, amongst the AGN UFOs, the highly-accreting NLSy1 show cooler winds with properties more similar to those seen in ULXs.

The discovery of AGN powered by SMBHs with masses $\sim 10^{9} M_{\odot}$ at redshift $\gtrsim7$ when the Universe was less than a Gyr old \cite{Fan2003, Banados2018} indicated that some channels of the fast growth of SMBHs are required.  Possible channels for the formation of supermassive black holes are the merging of many smaller black holes, Eddington-limited accretion onto IMBH seeds or super-Eddington accretion. Presently, all these channels provide viable solutions: recent gravitational wave discoveries have shown that merging amongst compact objects does occur \cite{Abbott2016}, although it might be not straightforward to merge a million BHs due to the system instability. Of course, \textcolor{black}{BH-BH mergers may have helped} to reach large black hole masses of e.g. $\sim 10^{3-4} M_{\odot}$ in the very early Universe \cite{Volonteri2013} but then a substantial accretion rate has to be sustained in order to reach the SMBH masses in a few hundred million years \cite{Jin2016}. Accretion is a mass-scale invariant, even if in the case of SMBHs they primarily accrete from the ISM (e.g. during galactic mergers) rather than a binary companion (although in rare cases they can still accrete from stars during tidal disruption events; TDEs).

TDEs are transient flares of activity caused by partial or total disruptions of stars that get too close to a SMBH. Typical BH masses are $\lesssim 10^{8}\,M_{\odot}$, and the peak accretion rates during these events are widely expected to be highly super-Eddington. They are typically characterised by supersoft X-ray spectra ($kT_{\rm BB}\sim0.1$ keV) and are believed to reach fallback luminosities $\sim 100 \, L_{\rm Edd}$ for a short period \cite{Wu2018}. One \textcolor{black}{very bright TDE that occurred in a nearby galaxy provided a high-quality, high-resolution spectrum which} showed evidence of radiation-pressure driven winds at low and high velocities \cite{Miller2015,Kara2018}. For these reasons, TDEs and NLSy1 (both characterised by SMBHs with moderate masses and high accretion rates) could be considered the supermassive counterparts of ULXs \cite{Dai2018}. GR-MHD simulations confirm that the black hole can rapidly grow with a growth time scale given by \textcolor{black}{$\tau_{\rm growth} = M/\dot{M}=4.5\times10^6(\dot{m}/100)^{-1}$ yr \cite{Ohsuga2005}.} This is a fairly short timescale and indicates that SMBHs can be formed by super-Eddington accretion provided that such a high accretion rate is sustained for long enough given that at most about $50$\,\% of the matter is lost into the wind \cite{Pinto2020a}. 
%%%systems.%%%\textcolor{black}{DJW: Need to define "growth time".} The equation defines it...

Finally, even though it now seems like they would be in the minority, it is still possible that some ULXs could host IMBHs ($10^{2-5} M_{\odot}$) and shed light onto another of the potential SMBH formation pathways, with ESO 243-49 HLX-1 ($L^{\rm max} = 2 \times 10^{42}$ erg s$^{-1}$) and M 82 X-1 ($L^{\rm max} \sim 10^{41}$ erg s$^{-1}$) amongst the best candidates. They both follow an $L \propto T^4$ relationship which, \textcolor{black}{together with their X-ray variability and radio properties (the latter for HLX-1), would suggest masses of a few 10,000 $M_{\odot}$ and 400 $M_{\odot}$, respectively \cite{Webb2012,Pasham2014}.} However, there are large uncertainties in these measurements and some results would instead suggest masses of $\sim 10^5 M_{\odot}$ for the former \cite{Titarchuk2016} and $20-118 \, M_{\odot}$ for the latter \cite{Brightman2016a}. This would place ESO 243-49 HLX-1 in the low end of the SMBHs and M82 X-1 in the high end of the stellar-mass BHs. HLX-1 is indeed found in a cluster of stars and which might be the remnant of a tidally stripped dwarf satellite galaxy \cite{Soria2013} with the SMBH accreting from a captured, \textcolor{black}{partially-disrupted}, companion star \cite{Lin2020}.

%%%\textcolor{black}{Not sure about the link between ULXs and gravitational waves a-la Wiktoriwicz, Mondal and Belczinski as Hi-Res work has nothing to do with it. Maybe we can just leave a comment in the introduction where we say that a (small) fraction of ULXs will go through mergers, whilst many of the mergers of compact objects (40\%) have gone through a ULX super-Eddington phase... although their simulations also provide useful prediction for BH/NS ULX fraction in different Z-environment and phases of the SFR outburst). Maybe in the next section with Athena and Lisa...}

\section{Future prospects}
\label{sec:prospects}

\textcolor{black}{Novel techniques in X-ray timing and spectroscopy, combined with theoretical simulations, have provided a remarkable improvement in our knowledge of the astrophysics of ULXs.  
%%%We now know that most of them are powered by super-Eddington accretion onto stellar-mass compact objects. High-resolution X-ray spectroscopy of ULXs provides unambiguous evidence of extremely fast and powerful winds which deposit a significant amount of energy in the surrounding medium, and may influence galactic evolution at certain cosmological epochs. Given their super-Eddington nature, understanding the accretion/outflows in ULXs may also be important for understanding the rapid growth of supermassive black holes and their formation in the early Universe. 
However, despite the significant recent progress,} several important questions remain unanswered. \textit{What is the relative fraction of BH-ULXs and NS-ULXs? Are there any spectroscopic differences between them? Are the spectral transitions triggered by orbital variations in the accretion rate or by stochastic variations in the wind? How does ULXs feedback evolve over time and affect the evolution of their host galaxies?} To solve these issues a combination of accurate plasma modelling, spectral-timing studies and more sensitive observations is required.

\subsection{Current limitations}
\label{sec:limitations}

%%%\textcolor{black}{DJW: A general thought about this section is that it comes across a little as though it's saying that we've basically exhausted all we can do for ULXs with our current generation of facilities, and I don't think we want to give that impression.}

It is difficult to distinguish BHs from NSs in absence of pulsations, cyclotron lines and dynamical mass measurements (see Sect. \ref{sec:timing-pulsations} and \ref{sec:multiwavelength}). As mentioned in Sect. \ref{sec:SEdd_NS}, some differences are expected to be found in wind properties depending on the accretor. In principle QPOs, sharp and high-amplitude flux drops in the light curves and $L-T$ trends may place further constraints, but these strongly depend on theoretical models and their interpretations. Of course, having two or more methods pointing towards the same solution e.g. a NS or a BH would be encouraging but in many cases we face signal-to-noise limitations. With the current X-ray instruments (such as \textit{XMM-Newton}) only about 20 ULXs have statistics sufficiently good to search for winds through individual lines in grating spectra. Only about 30 sources have time series with sufficient statistics to detect pulsations. CRSFs are difficult to detect as they require high sensitivity across a very broad energy band, which is possible to achieve only for a handful of sources. This means we lack the statistics that are necessary to search for putative ULX sub-samples with different properties, \textcolor{black}{although new deep observations could serve this purpose.}

One way to study in detail the size of the accretion disc, and the wind launching region and place constraints on the accretor nature would be using timing arguments such as energy-dependent lags and the response of the wind to continuum variability. However, we lack both the spectral resolution and the sensitivity in ULX X-ray time series. A significant detection of spectral lines requires observations longer than about 100 ks with the grating spectrometers which prevent us from resolving the timescales of continuum variability and the 100\,-\,1,000 s delays between different energy bands \cite{Alston2021,Kara2020,Kobayashi2018}. This also creates ambiguity in the variability process, the structure and the size of the emitting region.

The emission line plasma is clearly multiphase but the current sensitivity is not sufficient to distinguish between CIE and PIE equilibrium \textcolor{black}{in most observations}, which prevents a full understanding of the nature of the X-ray line emission. An accurate estimate of the thermal structure and plasma response to continuum changes (which seems to occur in time-integrated flux-resolved spectra \cite{Pinto2020b,Pinto2021,DAi2021}) would allow us to distinguish amongst different interpretations of the absorption component \textcolor{black}{(e.g. from close to the NS accretion column \cite{Abolmasov2022} or at $R_{\rm sph}$ \cite{Poutanen2007})} and provide further detail of the ULX effects on the binary system and the surrounding medium.

Another issue is the lack of sensitivity and spectral resolution in the Fe K band (6-10 keV) which limits our capability of detecting the hottest components of the wind launched from the inner disc. These are expected to be detected in absorption primarily in ULXs with hard spectra (HUL), and would fill an observational gap that remains present in the outflow structure. Currently, the only Fe K detections are those for NGC 1313 X-1, NGC 300 ULX-1 and NGC 4045 HLX-1 \cite{Walton2016a, Kosec2018b, Brightman2022}.

\subsection{Future missions and technologies}
\label{sec:future-missions}

There are a number of future X-ray missions/mission concepts (both approved and under study) that will enhance our ability to study hot and energetic plasmas, both in the next few years and on longer timescales. Some will carry instruments with a larger collecting area and increased spectral resolution, which will improve our sensitivity to atomic lines, others will extend the X-ray energy band over which sensitive observations can be taken, which will improve our knowledge of the spectral continuum and our ability to detect of broad features such as CRSFs.

\subsubsection{High spectral resolution detectors}
\label{sec:future-missions-resolution}

The sensitivity to weak lines scales with the so-called \textit{figure of merit} which is typically defined as the product of the resolving power ($R=E / \Delta E$) and the effective area: $FoM=\sqrt{A_{eff} \, R}$ (but sometimes as $\sqrt{A_{eff} / \Delta E}$ instead). A remarkable improvement with respect to the gratings aboard {\xmm} (Chapter 2) and {\chandra} (Chapter 3) was achieved thanks to the launch of \textit{Hitomi} (previously known as \textit{ASTRO-H}, see Chapter 5) in 2016. This satellite carried a micro-calorimeter detector which measured the energy of an incoming photon through a thermometer cooled down to 50 mK. This device achieved an excellent resolution of 5 eV in the Fe K band and was able to cover the 0.3-12 keV band. The effective area of the calorimeter was better than all of the high-resolution instruments currently operating above 1 keV. Although an observation of NGC 1313 X-1 was approved for the performance verification phase, the satellite was unfortunately lost after 5 weeks, but not before showing its capabilities thanks to a deep observation of the Perseus galaxy cluster \cite{Hitomi2016}. The \textit{X-ray Imaging and Spectroscopy mission (XRISM)}, planned for launch in 2023, will once again provide the high-resolution capabilities of its predecessor through the \textit{Resolve} micro-calorimeter spectrometer. \textcolor{black}{Similarly to \textit{Hitomi}, \textit{XRISM / Resolve} is expected to yield a $FoM$ improvement by a factor 5 around 1 keV and above 10 in the Fe K band with respect to present instruments}, albeit with a limited spatial resolution ($PSF \sim 1'$). The \textit{Arcus} mission concept \cite{Smith2020} has complementary capabilities. It consists of multiple grating detectors with a high effective area and a spectral resolution that provides an order of magnitude improvement in the $FoM$ below 1 keV. Finally, the largest planned X-ray mission relevant for high-resolution studies of ULXs will be the \textit{Advanced Telescope for High-ENergy Astrophysics (Athena)}, which is currently under ESA study. This facility is planned for launch after 2035 and will bear the most powerful high-resolution X-ray instrument so-far conceived: the \textit{X-ray Integral Field Unit (X-IFU)} \cite{Barret2022}. At the time of writing, \textit{X-IFU} is designed to achieve a 2.5 eV spectral resolution coupled with a high effective area (1 m$^2$) and a good spatial resolution (5-10$''$). With an $FoM$ that is a factor 10 better than any of the current or planned instruments at $\gtrsim 1$ keV, and coverage of the 0.2-12 keV band, \textit{Athena} will revolutionise high-resolution studies of ULXs enabling us to 1) probe shorter variability timescales, 2) resolve ULXs in crowded fields and 3) observe more distant and fainter sources thereby boosting the sample of ULXs with detected spectral lines.

\begin{figure}[h] %%%[b]
%\sidecaption
% For example, with the graphicx style use
\includegraphics[scale=.325]{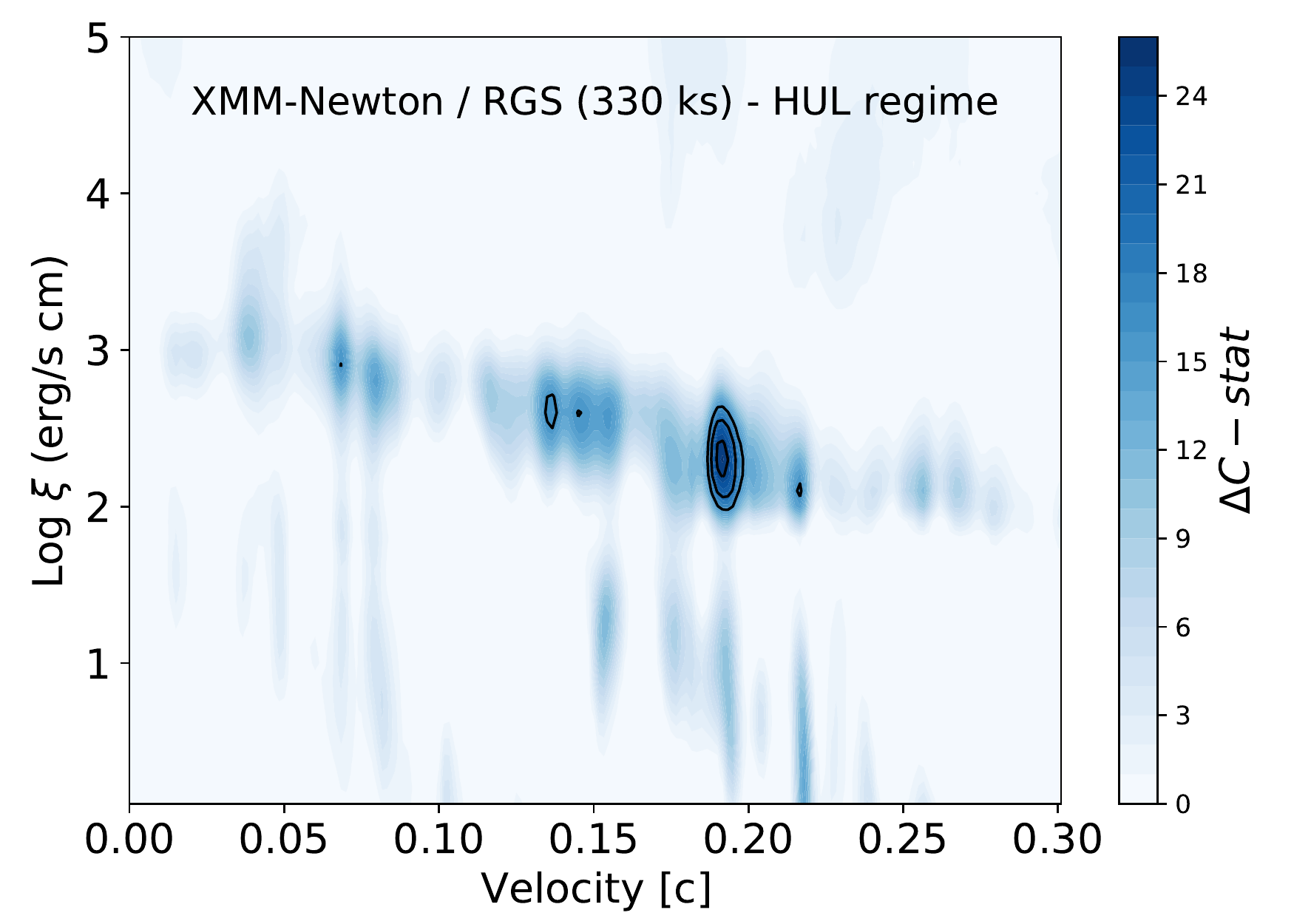}
\includegraphics[scale=.325]{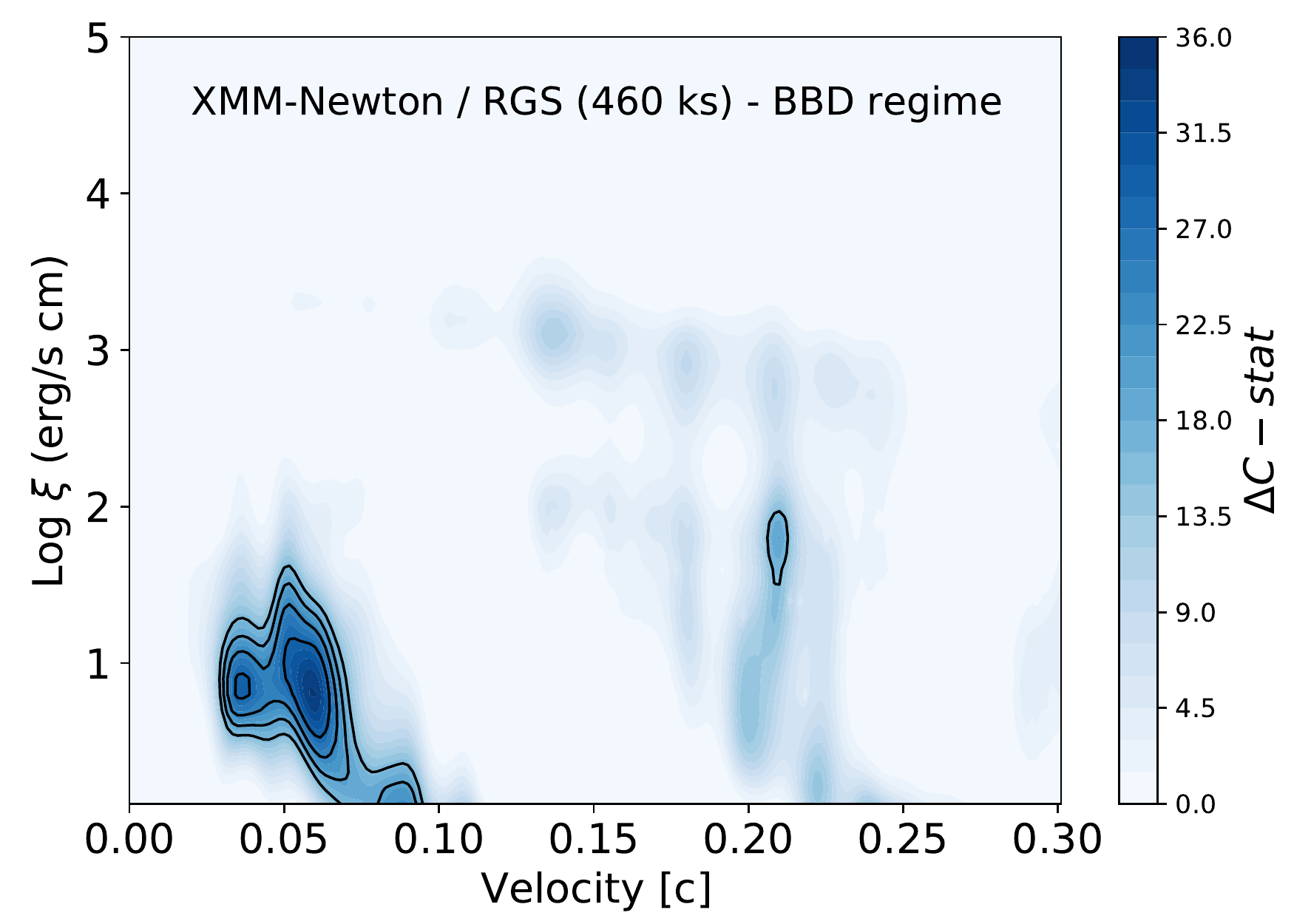}
\includegraphics[scale=.325]{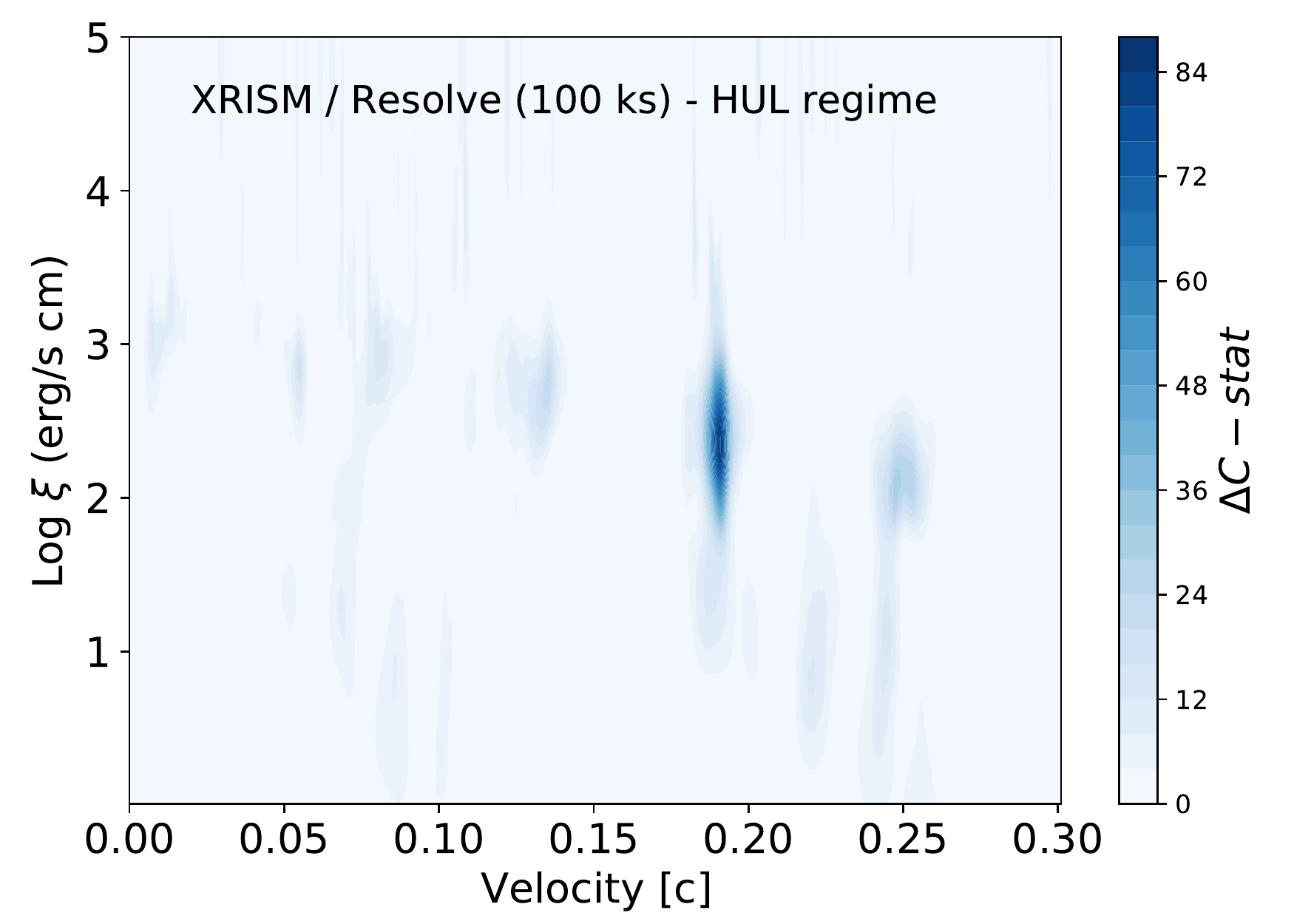}
\includegraphics[scale=.325]{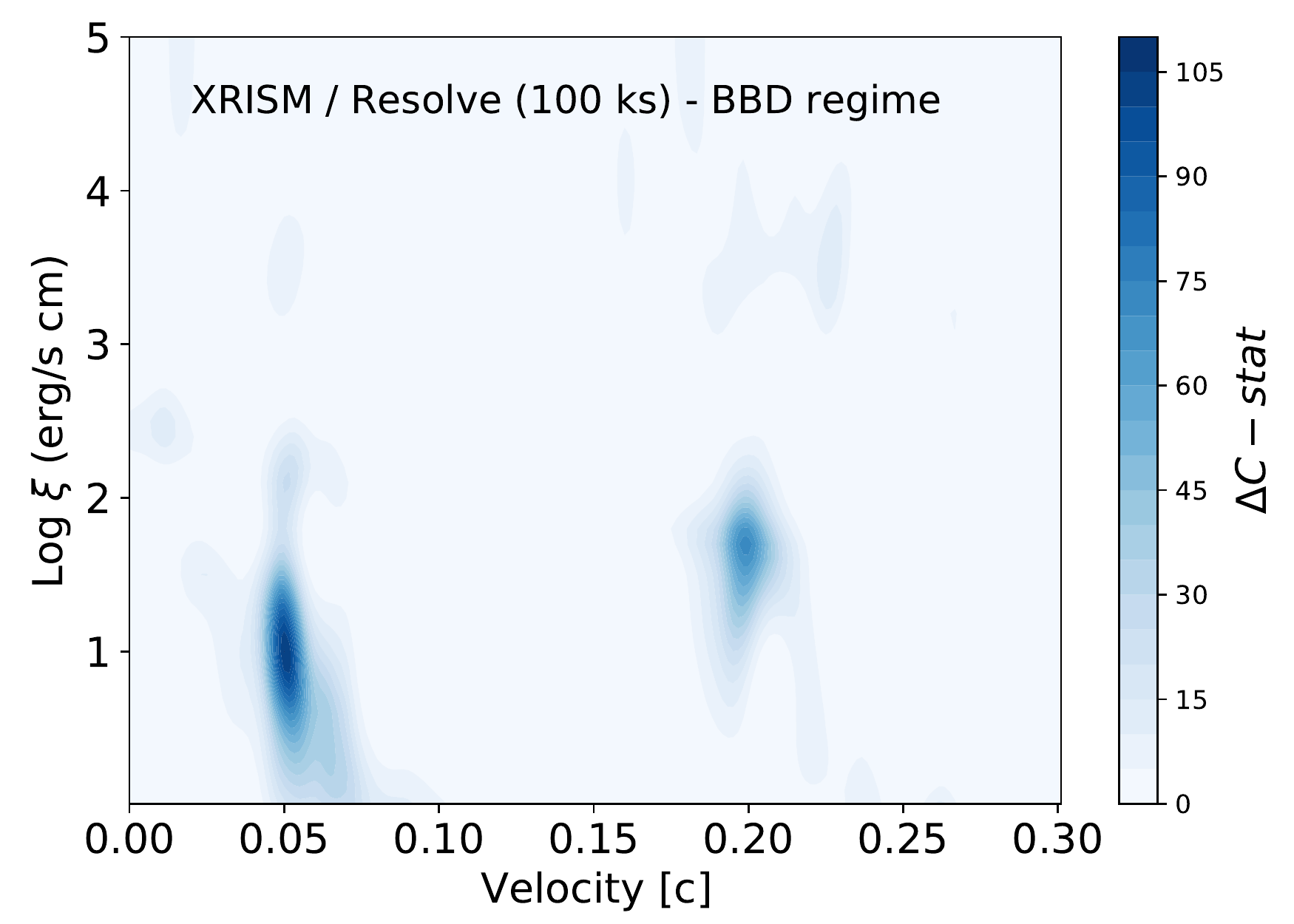}
\includegraphics[scale=.325]{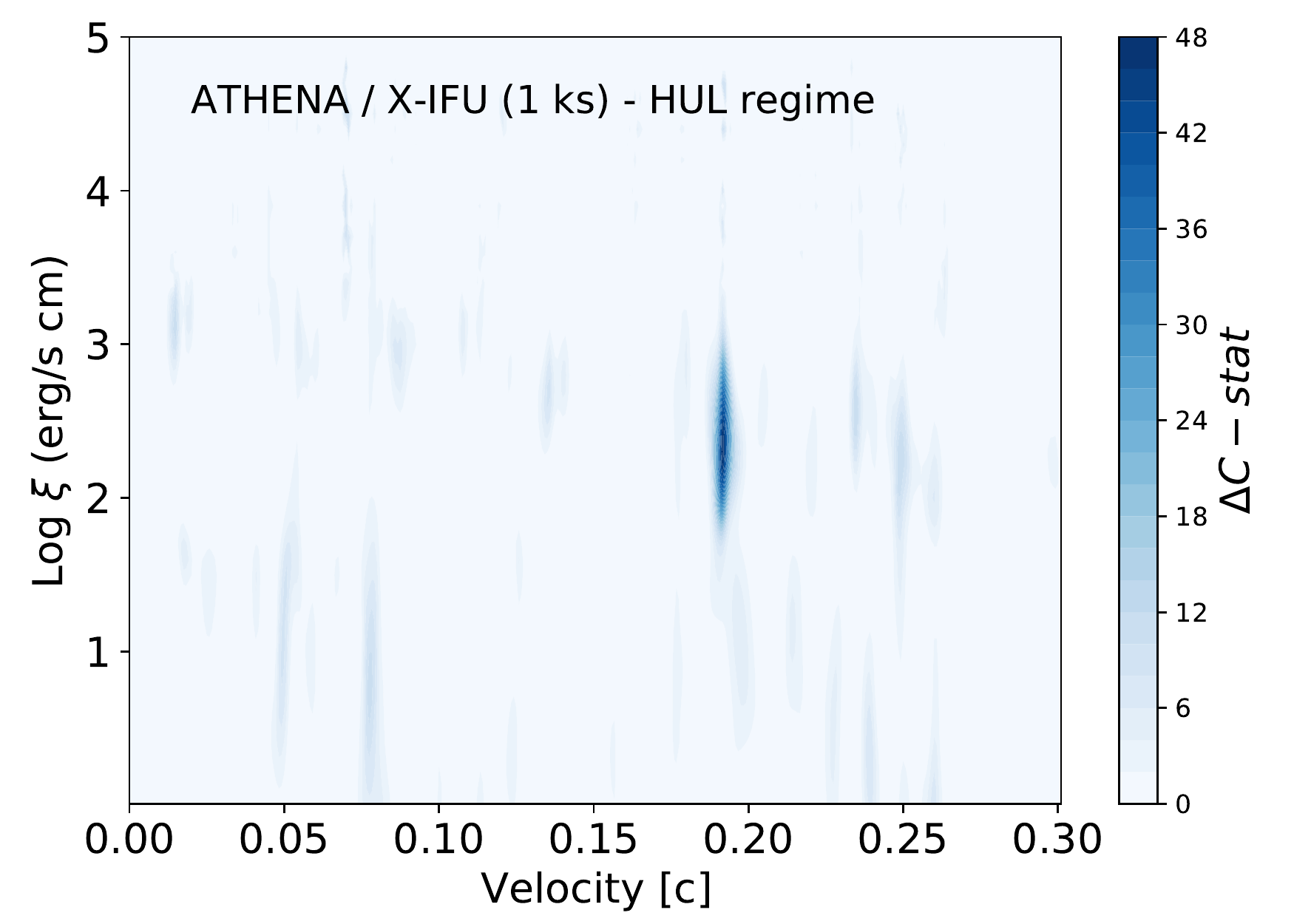}
\includegraphics[scale=.325]{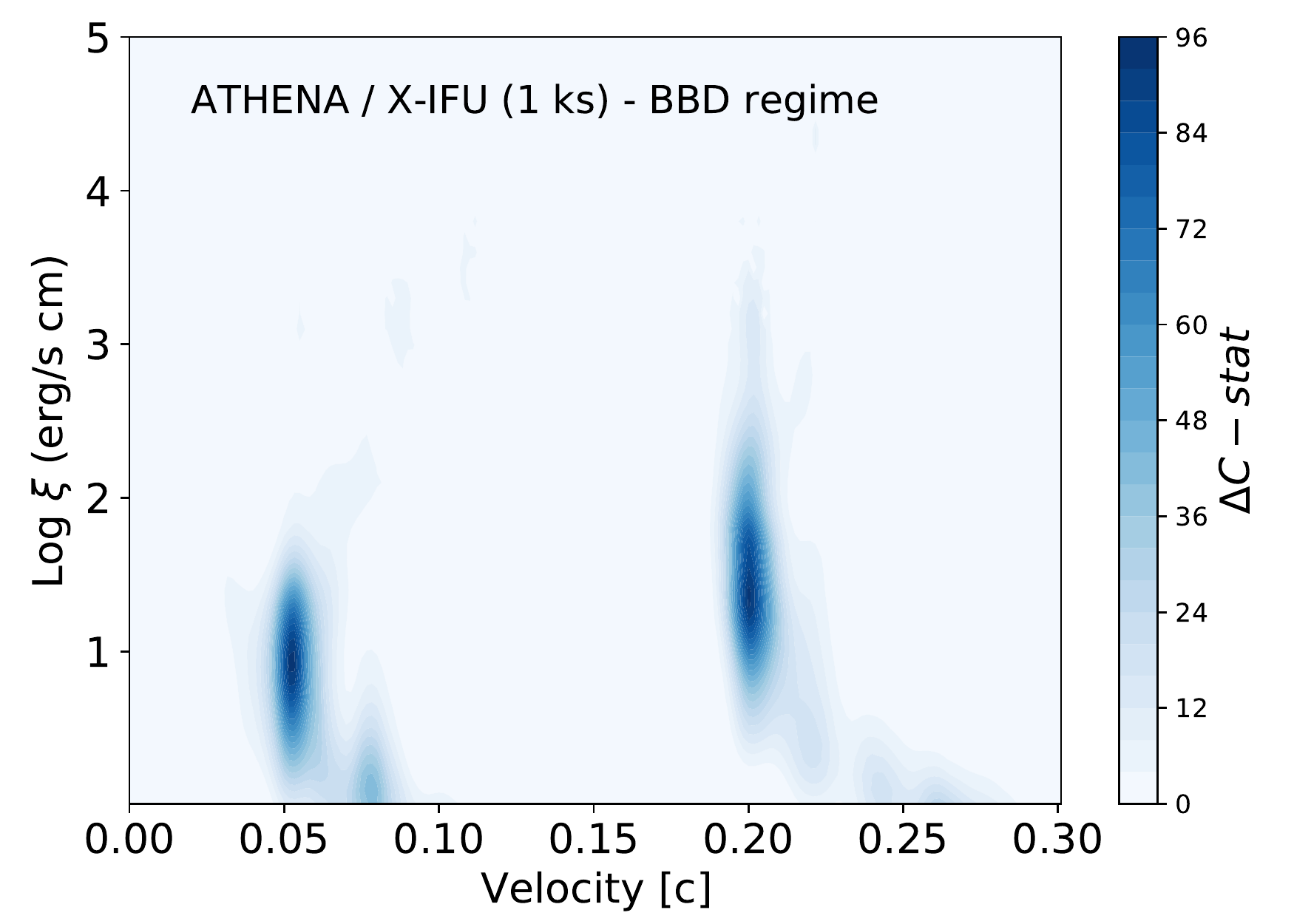}
%
% If no graphics program available, insert a blank space i.e. use
%\picplace{5cm}{2cm} % Give the correct figure height and width in cm
%
\caption{Top panels: photoionisation absorption model scan of the 2006-2012 low-flux HUL spectrum (left, 330 ks exposure) and the 2017 high-flux BBD spectrum (right, 460 ks) of NGC 1313 X-1 taken with {\xmm} (see also Fig. \ref{fig:first_detection}). Colour scale and labels are same as in Fig. \ref{fig:line-absorption} \cite{Pinto2020b}. Middle and bottom panels: corresponding simulations with the \textit{XRISM / Resolve} (100 ks) and \textit{Athena / X-IFU} (1 ks) micro-calorimeters adopting the same the best-fit continuum + wind model.}
\label{fig:Simulations}       % Give a unique label
\end{figure}

Here we showcase the capabilities of \textit{XRISM} and \textit{Athena} by presenting some simulated observations. In Fig. \ref{fig:Simulations} top panels we show the results of photoionisation absorption model scans of the HUL and BBD spectra of NGC 1313 X-1 (see also Fig. \ref{fig:first_detection} top and bottom panels). Over 300 ks of exposure time is required by \textit{XMM-Newton / RGS} in order to achieve significant detections (up to $4\,\sigma$ accounting for the L-E effect with MC simulations \cite{Pinto2020b}). Starting from the best-fitting continuum + wind models, we simulated \textit{Resolve} and \textit{X-IFU} observations for the HUL and BBD spectra of NGC 1313 X-1, then we removed the ionised wind from the spectral model and performed an identical routine (as applied to the \textit{XMM-Newton} datasets) to search for the evidence of outflows in the simulated spectra. The simulations performed with \textit{Resolve} (middle panels) used a clean exposure of 100 ks and those with \textit{X-IFU} (bottom panel in Fig. \ref{fig:first_detection}) an exposure of just 1 ks. It is evident that \textit{XRISM} will significantly reduce the present issues with degeneracy between different $(\xi,v_{\rm LOS})$ solutions, and also improve the detection of multiple components (see the case for the more complex outflow in the BBD spectrum). Given the values of $\Delta C-stat$ for a 100 ks \textit{XRISM} exposure, it is obvious that even exposures shorter by a factor 2 would produce excellent results. The results obtained with \textit{X-IFU} will be transformative, given that small snapshots of just 1 ks are required to achieve significant wind detections in the brightest ULXs. This will enable us to probe the timescales of the X-ray variability of the wind, and the nature of the soft X-ray lags \cite{Heil2010,Pinto2017,Kara2020} and the overall disc structure. Longer exposures will instead allow us to search for winds in much more distant sources, significantly increasing the sample of wind detections in ULXs. Similar conclusions might be derived about our future capabilities to search for further putative narrow proton CRSFs in ULXs.

\subsubsection{Broadband \textcolor{black}{and other} X-ray detectors}
\label{sec:future-missions-broadband}

In Sect. \ref{sec:broadband} we have discussed the importance of an accurate knowledge of the continuum shape of ULX X-ray spectra and the complex multi-component modelling. 
%%%The spectral curvature above $\sim$2\,keV provided a key indication that ULXs were super-Eddington accretors even before the discovery of winds and pulsations. In this respect, the simultaneous availability of missions such as {\xmm}, {\chandra}, {\suzaku} and {\nustar}, providing a broad coverage of the 0.3-80 keV band has been decisive for those studies along with the sensitive \textit{HST} observations at long wavelengths. 
A correct determination of the continuum has also proved to be mandatory in order to search for narrow and broad spectral features which provided evidence for winds and CRSFs (see Sect. \ref{sec:hrxs}). Broadband X-ray detectors are also equipped with the highest effective area resulting in the best counting statistics and are, therefore, the ideal instruments to search for pulsations and other timing features.

%%% There are several new or planned missions that should be considered for their important features. Searches of optical-IR counterparts of nearby ULXs can be significantly improved with \textit{JWST} thanks to its remarkable sensitivity (Sect. \ref{sec:ULXs_vs_AGN}). New facilities in the radio and $\gamma$-ray band, e.g. \textit{SKA} and \textit{CTA}, may enhance the currently small sample of ULXs with detections of jets (see Sect. \ref{sec:multiwavelength}). 

\textcolor{black}{Amongst the best facilities} for ULX-related science there is the \textit{enhanced X-ray Timing and Polarimetry} ({\extp}) mission (2027-, \cite{Zhang2019}). It will provide a high effective area in the hard X-ray band ($>$\,2\,keV) which together with a low background due to the low Earth orbit will dramatically improve the detection of highly-ionised wind components in the Fe K band and cyclotron lines with respect to {\xmm} and {\nustar}. We have performed a simple simulation of an 80\,ks observation of PULX NGC 300 ULX-1 using the \textit{Spectroscopic Focusing Array (SFA)} aboard {\extp} and the best-fit continuum + wind model \cite{Kosec2018b}. We have found that the {\fexxv} absorption line complex may be detected at $5\,\sigma$. The use of physical model scans (see Sect. \ref{sec:physical-scan}) will allow us to fit multiple lines at once and achieve such significance at much shorter exposure times, thereby probing variability timescales of a few hours. Moreover, for the brightest sources, {\extp} might provide the first estimates of polarised flux as an alternative probe of the system geometry.

\textcolor{black}{Although not formally a new mission, a facility that still has to show its potential is \textit{eROSITA}.} Currently on hold, the mission scanned the whole sky for four years every six months. It is therefore plausible that, as soon as the data become public, many new transient ULXs will be discovered and followed up with dedicated observations and other facilities. Amongst the proposed X-ray missions, the \textit{High Energy X-ray Probe (HEX-P)} \cite{Madsen2018} is expected to achieve an effective area 10 times better than the present detector ({\nustar}) at energies around 10 keV, along with the first focused coverage up to 200 keV. \textcolor{black}{Unlike {\nustar}, this facility will also cover the soft X-ray band down to 0.2 keV.} This will be ideal to estimate the high-energy curvature of ULX spectra, study the structure of accretion columns, and detect pulsations and CRSFs in ULXs. We have run another simulation for PULX NGC 300 ULX-1 adopting the CRSF best-fit model, i.e. a broad ($\sigma\sim3$ keV) feature at 13 keV \cite{Walton2018a}. A \textcolor{black}{100 ks} observation with \textit{HEX-P} will not only achieve a result comparable to that obtained with 200 ks of observations each with {\xmm} and {\nustar}, but the broader band coverage will also break the degeneracy between the line energy and width. Moreover, \textit{HEX-P} will provide the necessary statistics to distinguish between a CRSF model and alternative continuum models that do not require CRSFs \cite{Koliopanos2019}. \textcolor{black}{Other concept missions with great potential thanks to their high effective area, spatial and (for the latter) spectral resolution are \textit{AXIS} and \textit{Lynx}. \textit{AXIS} \cite{Mushotzky2019} would boost our capabilities to detect ULXs, particularly at high distances and in crowded fields. \textit{Lynx} \cite{Gaskin2019} should also be equipped with both micro-calorimeter and grating spectrometers, providing for the first time high spatial ($0.5''$) and spectral ($R>2,000$) resolution in the whole $0.2-10$ keV bandpass, which can boost ULX wind studies at greater distances and significantly enlarge the sample.}

\begin{acknowledgement}
-- We acknowledge support from the European Union’s Horizon 2020 Programme under the AHEAD2020 project (grant agreement n. 871158). The authors are grateful to Peter Kosec for reading the draft and providing useful insights for the discussion.
\end{acknowledgement}

%%%%%%%%%%%%%%%%%%%%%%%%%%%%%%%%%%%%%%%%%%%%%%%%%%%%%%%%%

%\biblstarthook{References should be \textit{cited} in the text by number. The reference list should be \textit{sorted} in alphabetical order. All works by the author alone, ordered chronologically by year of publication. All works by the author with a coauthor, ordered alphabetically by coauthor. All works by the author with several coauthors, ordered chronologically by year of publication. For the reference style, we suggest to use \textit{LaTeX (US)} from INSPIRE.}

%%%%%%%%%%%%%%%%%%%%%%%%%%%%%%%%%%%%%%%%%%%%%%%%%%%%%%%%%

%%%% \tableofcontents

%%%%%%%%%%%%%%%%%%%%%%%%%%%%%%%%%%%%%%%%%%%%%%%%%%%%%%%%%

\end{document}